\documentclass[twocolumn,showpacs,preprintnumbers, amsmath,amssymb,amsfonts, final, a4paper,aps, citeautoscript,footinbib,prb,superscriptaddress]{revtex4-1}
\usepackage[pdftex]{graphicx,color}
\usepackage[latin1]{inputenc}
\usepackage{mathrsfs}
\usepackage{epstopdf}
\usepackage{times}
\usepackage{float}
\usepackage[vcentermath]{youngtab}
\usepackage[sort&compress]{natbib}
\usepackage[colorlinks,bookmarks=true,citecolor=blue,linkcolor=blue,urlcolor=blue, breaklinks=true]{hyperref}
\graphicspath{{Figures/}}
\newcommand{\bolS}{\mathrm{\bf S}}
\newcommand{\bolT}{\mathrm{\bf T}}

\newcommand{\bolr}{\mathrm{\bf r}}

\newcommand{\boln}{\mathrm{\bf n}}

\newcommand{\be}{\mathrm{e}}
\newcommand{\ket}[1]{| #1 \rangle}  % ket-vector
\newcommand{\VEV}[1]{\langle #1 \rangle}  % Vacuum Expectation Value

\begin{document}
%%%%%%%%%%%%%%%%%%%%%%%%%%%%%%%%%%%%%%%%%%%%%%%%%%%%%%%
\title{Phase diagrams of one-dimensional half-filled two-orbital SU($N$) cold fermions systems}
%%%%%%%%%%%%%%%%%%%%%%%%%%%%%%%%%%%%%%%%%%%%%%%%%%%%%%%
\author{V. Bois}
\affiliation{Laboratoire de Physique Th\'eorique et Mod\'elisation, CNRS UMR 8089,
Universit\'e de Cergy-Pontoise, Site de Saint-Martin,
F-95300 Cergy-Pontoise Cedex, France.}
\author{S.\ Capponi} \affiliation{Laboratoire de Physique Th\'eorique, CNRS UMR 5152, 
Universit\'e Paul Sabatier, F-31062 Toulouse, France.}
\author{P. Lecheminant}
\affiliation{Laboratoire de Physique Th\'eorique et
 Mod\'elisation, CNRS UMR 8089,
Universit\'e de Cergy-Pontoise, Site de Saint-Martin,
F-95300 Cergy-Pontoise Cedex, France.}
\author{M. Moliner}
\affiliation{IPCMS (UMR 7504) and ISIS (UMR 7006), Universit\'e de Strasbourg et CNRS, Strasbourg, France.}
\author{K. Totsuka}
\affiliation{Yukawa Institute for Theoretical Physics, 
Kyoto University, Kitashirakawa Oiwake-Cho, Kyoto 606-8502, Japan.}
%%%%%%%%%%%%%%%%%%%%%%%%%%%%%%%%%%%%%%%%%%%%%%%%%%%%%%%
\date{\today}
\pacs{71.10.Pm, 75.10.Pq}

% PACS used:
% 71.10.Pm Fermions in reduced dimensions 
% 75.10.Pq Spin chain models
%%%%%%%%%%%%%%%%%%%%%%%%%%%%%%%%%%%%%%%%%%%%%%%%%%%%%%%
\begin{abstract}
We investigate possible realizations of exotic SU($N$) symmetry-protected topological (SPT) phases with alkaline-earth cold fermionic atoms loaded into one-dimensional optical lattices. 
A thorough study of two-orbital generalizations of the standard SU($N$) Fermi-Hubbard model, directly relevant to recent experiments, is performed. 
Using state-of-the-art analytical and numerical techniques, we map out the zero-temperature phase diagrams at half-filling and identify several Mott-insulating phases. 
While some of them are rather conventional (non-degenerate, charge-density-wave or spin-Peierls like), we also identify, for even-$N$, two distinct types of SPT phases: an orbital-Haldane phase, analogous to a spin-$N/2$ Haldane phase, and a topological SU($N$) phase, which we fully characterize by its entanglement properties. 
We also propose sets of non-local order parameters that characterize the SU($N$) topological phases found here. 
\end{abstract}
%%%%%%%%%%%%%%%%%%%%%%%%%%%%%%%%%%%%%%%%%%%%%%%%%%%%%%%
\maketitle
%%%%%%%%%%%%%%%%%%%%%%%%%%%%%%%%%%%%%%%%%%%%%%%%%%%%%%%
\section{Introduction}
%%%%%%%%%%%%%%%%%%%%%%%%%%%%%%%%%%%%%%%%%%%%%%%%%%%%%%%
High continuous symmetry based on the SU($N$) unitary group with $N>2$ plays a fundamental role in the standard model
of particle physics. The description of hadrons stems from an approximate SU($N$) symmetry where $N$ is 
the number of species of quarks, or flavors. 
In contrast, the SU($N$) symmetry was originally introduced in condensed matter physics 
as a mathematical convenience to investigate the phases of strongly correlated systems.  
For instance, we enlarge the physically relevant spin-SU(2) symmetry to SU($N$) 
and use the $N$ as a control parameter that makes various mean-field descriptions possible 
in the large-$N$ limit.  We then carry out the systematic $1/N$-expansion 
to recover the original $N=2$ case.\cite{Auerbach,Sachdev}

Extended continuous symmetries have been also used to unify several seemingly different 
competing orders in such a way that the corresponding order parameters can be transformed to 
each other under the symmetries. \cite{Zhang-97,HermeleSF05}   
A paradigmatic example is the SO(5) theory\cite{Zhang-97,DemlerHZ04} 
for the competition between $d$-wave superconductivity and 
antiferromagnetism, where the underlying order parameters are combined 
to form a unified order parameter quintet.  The high continuous symmetry often emerges 
from a quantum critical point unless it is simply introduced phenomenologically.
In this respect, for instance, the consideration of SU(4) symmetry might be a good starting point 
to study strongly correlated electrons with  
orbital degeneracy. \cite{Li-M-S-Z-98,Yamashita-S-U-98,Pati-S-K-98,Frischmuth-M-T-99,Azaria-G-L-N-99}

At the experimental level, realizations in condensed matter systems of enhanced 
continuous symmetry (in stark contrast to the SU(2) case) are very rare since they usually require 
substantial fine-tuning of parameters.
Semiconductor quantum dots technology provides a notable exception as it enables the realization of an SU(4) Kondo effect resulting from the interplay between spin and orbital degrees of freedom. \cite{Keller14}

Due to their exceptional control over experimental parameters, 
ultracold fermions loaded into optical lattices might be ideal systems
to investigate strongly correlated electrons with a high symmetry.
While ultracold atomic gases with alkali atoms can, in principle, explore the physics with
SO(5) and SU(3) symmetries, \cite{Wu2003,HonerkampH04,Lecheminant-B-A-05,Wu06,RappZHH07,AzariaCL09}
alkaline-earth atoms are likely to be the best candidates for experimental realizations  of exotic
SU($N$) many-body physics. \cite{Gorshkov-et-al-10,Cazalilla-H-U-09,Cazalilla-R-14}
These atoms and related ones, like ytterbium atoms, have a peculiar energy spectrum associated with
the two-valence outer electrons. 
The ground state (``$g$'' state) is a long-lived singlet state $^1S_0$ and the spectrum exhibits a metastable triplet excited state (``$e$'' state) $^3P_0$. Due to the existence of an ultranarrow optical transition 
$^1S_0$-$^3P_0$  between these states, alkaline-earth-like atoms appear to be excellent candidates for atomic clocks and quantum simulation applications. \cite{daley11}
Moreover, the $g$ and $e$ states have zero 
electronic angular momentum, so that the nuclear spin $I$ is almost decoupled from the
electronic spin. The nuclear spin-dependent variation of the scattering lengths
is expected to be smaller than $\sim 10^{-9}$ for the $g$ state and $\sim 10^{-3}$ for the $e$  
state. \cite{Gorshkov-et-al-10}
This decoupling of the electronic spin from the nuclear one in atomic collisions
 paves the way to the experimental realization of fermions with
an SU($N$) symmetry where $N= 2 I +1$ ($I$ being the nuclear spin) is the number of nuclear states.

The cooling of fermionic isotopes of these atoms below the quantum degeneracy has been
achieved for strontium atoms $^{87}$Sr  with $I= 9/2$ \cite{DeSalvo-Y-M-M-K-10,Tey-S-G-S-10} 
and ytterbium atoms $^{171}$Yb, $^{173}$Yb  
with $I= 1/2,5/2$. \cite{Fukuhara-T-K-T-07,Taie-etal-PRL-10}
These atoms enable the experimental exploration of the physics of fermions with an emergent SU($N$) symmetry where $N$
can be as large as 10.
In this respect, experiments on  $^{173}$Yb  atoms loaded into a three-dimensional (3D) optical lattice 
have stabilized an SU(6) Mott insulator \cite{Taie-Y-S-T-12} 
while the one-dimensional (1D) regime has also been investigated. \cite{Pagano-et-al-14}
Very recent experiments on $^{87}$Sr (respectively $^{173}$Yb)  atoms in a two-dimensional (2D) 
(respectively 3D) optical lattice have directly observed the existence of the SU($N$) symmetry 
and determined the specific form of the interactions
between the $g$ and $e$ states. \cite{Zhang-et-al-14,Scazza-et-al-14} 
All these results and future experiments might lead to the investigation 
of the rich exotic physics of SU($N$)  fermions as for instance the 
realization of a chiral spin liquid phase with non-Abelian statistics. \cite{Hermele-G-11,Cazalilla-R-14} 

The simplest effective Hamiltonian to describe an $N$-component Fermi gas with an SU($N$) symmetry
loaded into an 1D optical lattice is the SU($N$) generalization of the famous Fermi-Hubbard model:
\begin{equation}
{\cal H}_{\text{SU($N$)}} = - t \sum_{i, \alpha}  \left(c_{\alpha,\,i}^\dag c_{\alpha,\,i+1}  + \text{H.c.}\right)
+ \frac{U}{2}  \sum_{i} n_{i}^2 ,
\label{HubbardSUN}
\end{equation} 
$c_{\alpha,\,i}^\dag$ being the fermionic creation operator for site $i$ and nuclear spin
states $\alpha = 1, \ldots, N$, and $n_{i} = \sum_{ \alpha}  c_{\alpha,\,i}^\dag c_{\alpha,\,i}$ is the
density operator. All parameters in model (\ref{HubbardSUN}) are independent from the nuclear states
which express the existence of an  global SU($N$) symmetry: $c_{\alpha,\,i} \mapsto \sum_{\beta} U_{\alpha \beta}  c_{\beta,\,i}$, $U$ being an SU($N$) matrix. 
Model (\ref{HubbardSUN}) describes alkaline-earth atoms in the $g$ state loaded into the lowest band of the optical
lattice. The interacting coupling constant $U$ is directly related to the scattering length associated with
the collision between two atoms in the $g$ state. In stark contrast to the $N=2$ case, the  SU($N$)  Hubbard model
(\ref{HubbardSUN})  is not integrable by means of the Bethe ansatz approach. However, most of its physical properties are 
well understood thanks to field theoretical and numerical approaches. For a commensurate filling of one atom per site,
which best avoids issues of three-body loss, a Mott-transition occurs for a repulsive interaction when $N>2$ between
a multicomponent Luttinger phase and a Mott-insulating phase with $N-1$ gapless degrees of freedom.
\cite{Assaraf-A-C-L-99,Manmana-H-C-F-R-11}
In addition, the fully gapped Mott-insulating phases of model (\ref{HubbardSUN}) are  known to 
be spatially nonuniform for commensurate fillings.\cite{Szirmai-L-S-2008}

The search for exotic 1D Mott-insulating phases with SU($N$) symmetry requires thus to go beyond
the simple SU($N$) Fermi-Hubbard model (\ref{HubbardSUN}).
One possible generalization is to exploit the existence of the $e$ state in the spectrum of alkaline-earth atoms
and to consider a two-orbital extension of the  SU($N$) Fermi-Hubbard model which 
is directly relevant to recent experiments.  \cite{Zhang-et-al-14,Scazza-et-al-14} 
The interplay between orbital and SU($N$) nuclear spin degrees of freedom is then expected to give rise to several
interesting phases, including symmetry-protected topological (SPT) phases. \cite{Gu-W-09,Chen-G-W-10}
The latter refer to non-degenerate fully gapped phases which do not break any symmetry
and cannot be characterized by local order parameters.  
Since any gapful phases in one dimension have short-range entanglement,
 the presence of a symmetry is necessary  to protect the properties of that 1D topological phase, in particular the existence of non-trivial edge states. \cite{Chen-G-W-10,Chen-G-L-W-12}

In this paper, we will map out the zero-temperature phase diagrams of several two-orbital SU($N$) lattice models 
 at half-filling by means of complementary use of analytical and numerical approaches.  
 A special emphasis will be laid on 
 the description of SU($N$) SPT phases which can be stabilized in these systems.
 In this respect, as it will be shown here, several distinct SPT phases will be found.
 In the particular $N=2$ case, i.e. atoms with nuclear spin $I=1/2$,  the paradigmatic example of 1D SPT 
 phase, i.e. the spin-1 Haldane phase \cite{Haldane-PLA-83,Haldane-PRL-83}, 
 will be found for charge, orbital, and nuclear spin degrees of freedom. 
 This phase is a non-degenerate gapful phase with spin-1/2 edges states which are protected by
 the presence of at least one of the three discrete symmetries: 
the dihedral group of $\pi$ rotations along the $x,y,z$ axes, time-reversal, 
and inversion symmetries.\cite{Pollmann-B-T-O-12}
In the general $N$ case, we will show that the spin-$N/2$ Haldane phase emerges only for the
orbital degrees of freedom in the phase diagram of the two-orbital SU($N$) model.
The resulting phase will be called orbital Haldane (OH) phase and is an SPT phase
when $N/2$ is an odd integer.
On top of these phases, new 1D SPT phases will be found which stem from the higher SU($N$) continuous
symmetry of these alkaline-earth atoms. These phases are the generalization of the Haldane phase
for SU($N$) degrees of freedom with $N>2$. As will be argued in the following, 
these topological phases for general $N$ are protected by the presence of PSU($N$) $=$ SU($N$)/$\mathbb{Z}_N$  symmetry.  
Even in the absence of the latter symmetry, SU($N$) topological phases may remain topological in the presence
of other symmetries.  
For instance, with the (link-)inversion symmetry present, 
our SU($N$) topological phase when $N/2$ is odd (i.e., $I=1/2,5/2,9/2,\ldots$ which is directly relevant to ytterbium and strontium atoms) crosses over to the topological Haldane phase. 
A brief summary of these results has already been given in a recent paper \cite{Nonne-M-C-L-T-13}
where we have found these SU($N$) topological phases for a particular 1D two-orbital SU($N$) model.

The rest of the paper is organized as follows.
In Sec.~\ref{sec:models-strong-coupling}, we introduce two different lattice models 
of two-orbital SU($N$) fermions and discuss their symmetries.  
Then, strong-coupling analysis is performed which gives some clues
about the possible Mott-insulating phases and the global phase structure.  
We also establish the notations and terminologies used in the following sections, 
and characterize the main phases 
that are summarized in Table~\ref{tab:abbreviation}.
 
The basic properties of the SU($N$) SPT phase identified in the previous section   
are then discussed in detail in Sec.~\ref{sec:SUN-topological-phase} paying particular attention to 
the entanglement properties.  The use of non-local (string) order parameters to detect the SU($N$) SPT phases 
will be discussed, too.  
In Sec.~\ref{sec:weak-coupling}, a low-energy approach of the two-orbital SU($N$) lattice models is
developed to explore the weak-coupling regime of the lattice models.   
The main results of this section are summarized in the phase diagrams in Sec.~\ref{sec:RG-phase-diag}.  
As this section is rather technical,  
those who are not familiar with field-theory techniques may skip Secs.~\ref{sec:continuum_description} and 
\ref{sec:RG_analysis} for the first reading. 

In order to complement the low-energy and the strong-coupling analyses,  
we present, in Sec.~\ref{sec:DMRG}, our numerical results for $N=2$ and $4$ obtained by the density
matrix renormalization group (DMRG) simulations. \cite{White-92}  
Readers who want to quickly know the ground-state phase structure may read Sec.~\ref{sec:models-strong-coupling} 
first and then proceed to Sec.~\ref{sec:DMRG}.    
Finally, our concluding remarks are given in Sec.~\ref{sec:conclusion} and the paper is supplied with four appendices which
provide some technical details and additional information.

%%%%%%%%%%%%%%%%%%%%%%%%%%%%%%%%%%%%%%%%%%%%%%%%%%%%%%%
\section{Models and their strong-coupling limits} 
\label{sec:models-strong-coupling}
%%%%%%%%%%%%%%%%%%%%%%%%%%%%%%%%%%%%%%%%%%%%%%%%%%%%%%%

In this section, we present the lattice models related to the physics of the 1D two-orbital SU($N$) model
that we will investigate in this paper. In addition, the different strong-coupling limits of the models will be discussed 
to reveal the existence of SPT phases in their phase diagrams.

%%%%%%%%%%%%%%%%%%%%%%%%%%%%%%%%%%%%%%%%%%%%%%%%%%
\subsection{Alkaline-earth Hamiltonian}
\label{sec:Gorshkov-Hamiltonian}
%%%%%%%%%%%%%%%%%%%%%%%%%%%%%%%%%%%%%%%%%%%%%%%%%%
Let us first consider alkaline-earth cold atoms where the atoms can occupy the ground state $g$ and excited metastable state $e$.
In this case, four different elastic scattering lengths can be defined due to the two-body collisions between
two atoms in the $g$ state ($a_{gg}$), in the $e$ state ($a_{ee}$), and finally between the $g$ and $e$
states ($a^{\pm}_{ge}$). \cite{Gorshkov-et-al-10}
On general grounds, four different interacting coupling constants are then 
expected from these scattering properties and a rich physics might emerge from this complexity.
The model Hamiltonian, derived by  Gorshkov {\em et al}. \cite{Gorshkov-et-al-10}, 
which  governs the low-energy properties of these atoms loaded into a 1D optical 
reads as follows ({\em $g$-$e$ model}): 
\begin{equation}
\begin{split}
&  \mathcal{H}_{g\text{-}e} = 
  - \sum_{m=g,e}  t_{m} \sum_{i} \sum_{\alpha=1}^{N} 
   \left(c_{m\alpha,\,i}^\dag c_{m\alpha,\,i+1}  + \text{H.c.}\right) \\
&  -\sum_{m=g,e}\mu^{(m)} \sum_i n_{m,i}  
 + \sum_{m=g,e} \frac{U_{mm}}{2} \sum_{i} n_{m,\,i}(n_{m,\,i}-1) \\
&  +V \sum_i n_{g,\,i} n_{e,\,i} 
  + V_{\text{ex}}^{g\text{-}e} \sum_{i,\alpha \beta} 
  c_{g\alpha,\,i}^\dag c_{e\beta,\,i}^\dag 
  c_{g\beta ,\,i} c_{e\alpha,\,i} ,
  \end{split}
\label{eqn:Gorshkov-Ham}
\end{equation} 
where the index $\alpha$ labels the nuclear-spin multiplet ($I^{z}=-I,\ldots,+I$, $N=2I+1$, 
$\alpha=1,\ldots,N$) and the orbital indices $m=g$ and $e$ label the two atomic states 
${}^{1}S_{0}$ and ${}^{3}P_{0}$, respectively.  The fermionic creation operator with quantum numbers 
$m,\alpha$ on the site $i$ is denoted by $c_{m\alpha,\,i}^\dag$.
The local fermion numbers of the species $m=g,e$ are defined by
\begin{equation}
n_{m,i} = \sum_{\alpha=1}^{N}c^{\dagger}_{m\alpha,i}c_{m\alpha,i} 
= \sum_{\alpha=1}^{N} n_{m\alpha,i} \; .
\end{equation}
We also introduce the total fermion number at the site $i$:
\begin{equation}
n_{i} = \sum_{m=g,e} n_{m,i} \; .
\end{equation}

\begin{figure}[htb]
\begin{center}
\includegraphics[scale=0.5]{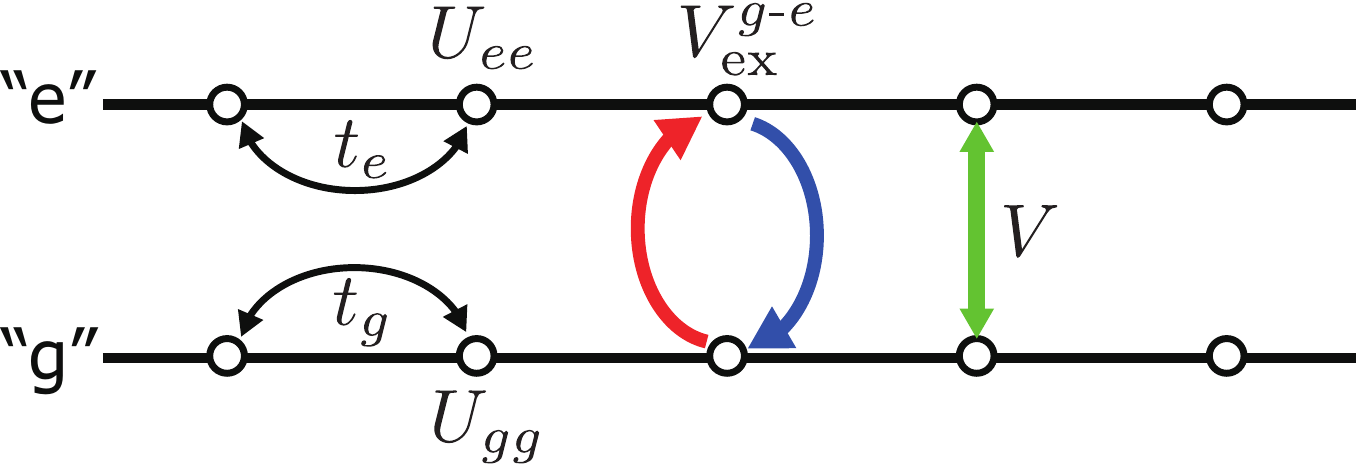}
\caption{(Color online) The two-leg ladder representation of the $g\text{-}e$ model \eqref{eqn:Gorshkov-Ham}. 
Two single-band SU($N$) Hubbard chains are coupled to each other only by 
the inter-chain particle exchange ($V_{\text{ex}}^{g\text{-}e}$) and 
the interchain density-density interaction ($V$).   
Note that splitting of a single physical chain into two is fictitious. 
\label{fig:alkaline-2leg}}
\end{center}
\end{figure}

In order to understand the processes contained in this Hamiltonian, it is helpful to 
represent it as two coupled (single-band) SU($N$) Hubbard chains (see Fig.~\ref{fig:alkaline-2leg}). 
On each chain, we have the standard hopping $t$ along each chain (which may be different for $g$ 
and $e$) and the Hubbard-type 
interaction $U$, and the two are coupled to each other by the $g$-$e$ 
contact interaction $V$ and the $g$-$e$ exchange process $V_{\text{ex}}^{g\text{-}e}$.
Model (\ref{eqn:Gorshkov-Ham}) is invariant under continuous U(1)$_{\text{c}}$  and SU($N$) symmetries:
\begin{equation}
c_{m\alpha,\,i} \mapsto \be^{i \theta} c_{m\alpha,\,i} \, , \; 
c_{m\alpha,\,i} \mapsto \sum_{\beta} \mathcal{U}_{\alpha \beta}  c_{m\beta,\,i} ,
\label{eqn:U(N)}
\end{equation}
with $\mathcal{U}$ being an SU($N$) matrix. The two transformations (\ref{eqn:U(N)}) 
respectively refer to the conservation of the total number of atoms and the 
SU($N$) symmetry in the nuclear-spin sector.
On top of these obvious symmetries, the Hamiltonian is also invariant under 
\begin{equation}
c_{g\alpha,\,i} \mapsto \be^{i \theta_{\text{o}}} c_{g\alpha,\,i} \, , \; 
c_{e\alpha,\,i} \mapsto \be^{-i \theta_{\text{o}}} c_{e\alpha,\,i}  \; .
\label{eqn:orbital-U1}
\end{equation}
This is a consequence of the fact that the total fermion numbers for $g$ and $e$ are 
conserved {\em separately}.\footnote{%
This breaks down when there is a hopping (transition) between $g$ and $e$.}  
%%%%%%%%%%%%%%%%%%%%%%%%%%%%%%%%%%%%%%%%%%%%%%%%%%%%%%%%%%
%%%%%%%%%%%%%%%%%%%%%%%%%%%%%%%%%%%%%%%%%%%%%%%%%%%%%%%

In the case of SU(2), it is well-known that the orbital ($g$, $e$) exchange process 
can be written in the form of the Hund coupling.   
Let us write down such expressions in two ways.  
First, we introduce the second-quantized SU($N$) generators of each orbital
\begin{equation}
\hat{S}_{m,i}^{A} = c^{\dagger}_{m\alpha,i}(S^{A})_{\alpha\beta}c_{m\beta,i}  \quad (m=g,e, \; 
A=1,\ldots, N^{2}-1), 
\label{eqn:second-quant-SUN}
\end{equation} 
as well as the {\em orbital pseudo spin} $T^{a}_{i}$ ($a=x,y,z$):
\begin{equation}
T_i^a = \frac{1}{2} c_{m \alpha,\,i}^\dag \sigma^a_{m n} c_{n \alpha,\,i} 
= \sum_{\alpha=1}^{N}T_{\alpha,i}^{a}
\quad (m,n=g,e)  \; ,
\label{eqn:1pseudospinoperator}
\end{equation}
where a summation over repeated indices is implied in the following and $\sigma^a$ denotes the Pauli matrices.  
If we normalize the SU($N$) generators $S^{A}$ as\footnote{%
This corresponds to, e.g., using the SU(2) generators $\sigma^{a}/\sqrt{2}$ ($a=x,y,z$) instead of 
the standard ones $\sigma^{a}/2$.}
\begin{equation}
\text{Tr}\,(S^{A}S^{B}) = \delta_{AB} \; ,
\end{equation}
the generators $S^{A}$ satisfy the following identity:
\begin{equation}
\sum_{A=1}^{N^{2}-1} (S^A)_{\alpha\beta} (S^A)_{\gamma \delta}  =  \left(
\delta_{\alpha \delta} \delta_{\beta \gamma} - \frac{1}{N} 
\delta_{\alpha \beta} \delta_{\gamma \delta} \right)  \; .
\label{eqn:SASA-tensor}
\end{equation}
The above U(1)$_{\text{o}}$ transformation \eqref{eqn:orbital-U1} amounts to the rotation along 
the $z$-axis:
\begin{equation}
\begin{split}
& T_i^{\pm} \mapsto \be^{\mp2i \theta_{\text{o}}} T_i^{\pm} \\
& T_i^{z} \mapsto T_i^{z}
\end{split}
\end{equation}
generated by 
\begin{equation}
T_i^z = \frac{1}{2}(n_{g,\,i}-n_{e,\,i}) \; .
\end{equation}

Then, it is straightforward to show that the orbital-exchange ($g \leftrightarrow e$) can be written as 
the Hund coupling for the SU($N$) `spins' or that for the orbital pseudo spins:
\begin{equation}
\begin{split}
& \sum_{i} 
  c_{g\alpha,\,i}^\dag c_{e\beta,\,i}^\dag 
  c_{g\beta ,\,i} c_{e\alpha,\,i}   \\
& = - \sum_{i}
\hat{S}_{g,i}^{A}\hat{S}_{e,i}^{A} 
- \frac{1}{N} \sum_{i}n_{g,i}n_{e,i} \\
& = \sum_{i}( \bolT_{i})^{2}
- \frac{1}{4}\sum_{i} n_{m,i}(n_{m,i}-1) 
\\
& \phantom{=} 
-\frac{3}{4} \sum_{i} n_{i}
+ \frac{1}{2}\sum_{i}n_{g,i}n_{e,i}  \; .
\end{split}
\label{eqn:exchange-to-Hund}
\end{equation}
The fermionic anti-commutation is crucial in obtaining the two opposite signs in front of 
the Hund couplings.  
The above expression enables us to rewrite the original alkaline-earth Hamiltonian \eqref{eqn:Gorshkov-Ham} 
in two different ways
\begin{widetext}
\begin{equation}
\begin{split}
\mathcal{H}_{g\text{-}e} =& 
  -  \sum_{i}\sum_{m=g,e} t_{m}
   \left(c_{m\alpha,\,i}^\dag c_{m\alpha,\,i+1}  + \text{H.c.}\right)
  -  \sum_i  \sum_{m=g,e} \mu^{(m)}  n_{m,i}  \\
   & \quad +\sum_{i}\sum_{m=g,e} \frac{U_{mm}}{2} n_{m,\,i}(n_{m,\,i}-1)
+\left(V - \frac{1}{N}V_{\text{ex}}^{g\text{-}e}\right) \sum_i n_{g,\,i} n_{e,\,i} 
- V_{\text{ex}}^{g\text{-}e}
\sum_{i}
\hat{S}_{g,i}^{A}\hat{S}_{e,i}^{A} 
\\
=& 
 -  \sum_{i}\sum_{m=g,e} t_{m} 
   \left(c_{m\alpha,\,i}^\dag c_{m\alpha,\,i+1}  + \text{H.c.}\right)
  -  \sum_i \sum_{m=g,e}\left(\mu^{(m)} + \frac{3}{4} V_{\text{ex}}^{g\text{-}e}\right) n_{m,i}  \\
   & \quad 
+\sum_{i}\sum_{m=g,e} \frac{U_{m m} - V_{\text{ex}}^{g\text{-}e}/2}{2} n_{m,\,i}(n_{m,\,i}-1)
+\left(V + V_{\text{ex}}^{g\text{-}e}/2 \right) \sum_i n_{g,\,i} n_{e,\,i} 
+ V_{\text{ex}}^{g\text{-}e} \sum_{i}  ( \bolT_{i})^{2} 
 \; .
\end{split}
\label{eqn:Gorshkov-Ham-Hund}
\end{equation} 
\end{widetext}
From this, one readily sees that positive (negative) $V_{\text{ex}}^{g\text{-}e}$ tends to 
quench (maximize) orbital pseudo spin $\bolT$ and maximize (quench) the SU($N$) spin.  
This dual nature of the orbital and SU($N$) is the key to understand the global structure of 
the phase diagram.  

Using the orbital pseudo spin $T^{a}$, we can rewrite the original $g$-$e$ Hamiltonian \eqref{eqn:Gorshkov-Ham} 
as
\begin{equation}
\begin{split}
\mathcal{H}_{g\text{-}e} = & - \, \sum_{i} \sum_{m=g,e}
t_{m}  \left( c_{m\alpha,\,i}^\dag c_{m\alpha,\,i+1}+  \text{H.c.}  \right) \\
& -\frac{1}{2}\left(\mu _e + \mu _g\right) \sum_{i} n_{i} - \left(\mu _g -\mu _e\right)\sum_{i} T^{z}_{i}  \\
& +\frac{U}{2} \sum_i  n_i^2 + U_{\text{diff}} \sum_i  T^{z}_{i} n_{i}  \\
&  +J \sum_i \left\{ (T_i^x)^2 + (T_i^y)^2\right\} + J_z \sum_i (T_i^z)^2  ,
\end{split}
\label{alkamodel-2b}
\end{equation}
with 
\begin{equation}
\begin{split}
& U=\frac{1}{4} (U_{g g}+ U_{ee}  
+2 V), \;\; 
U_{\text{diff}} = \frac{1}{2}(U_{gg} - U_{ee}) , \\
& J=V^{g\text{-}e}_{\text{ex}} , \;\;
J_{z} = \frac{1}{2}(U_{e e} + U_{g g} -2 V), \\
& \mu_{g} = \frac{1}{2} (2 \mu^{(g)}+ U_{g g} + V^{g\text{-}e}_{\text{ex}}), \\
& \mu_{e} = \frac{1}{2} (2 \mu^{(e)}+ U_{e e} + V^{g\text{-}e}_{\text{ex}})
\; .
\end{split}
\label{eqn:HG-parameters}
\end{equation}
The site-local part of the above Hamiltonian \eqref{alkamodel-2b} gives the starting point for the strong-coupling 
expansion:
\begin{equation}
\begin{split}
\mathcal{H}_{\text{atomic}} = & 
-\frac{1}{2}\left(\mu _e + \mu _g\right) \sum_{i} n_{i} - \left(\mu _g -\mu _e\right)\sum_{i} T^{z}_{i}  \\
& +\frac{U}{2} \sum_i  n_i^2 
+ U_{\text{diff}} \sum_i  T^{z}_{i} n_{i} \\
&  +J \sum_i \left\{ (T_i^x)^2 + (T_i^y)^2\right\} + J_z \sum_i (T_i^z)^2 
\; .
\end{split}
\label{eqn:atomic-limit-Ham}
\end{equation}
Since the model contains many coupling constants, it is highly desirable to consider a simpler
effective Hamiltonian which encodes the most interesting quantum phases of the problem.
In this respect, for the DMRG calculations of Sec. \ref{sec:DMRG}, we will set 
$t_g = t_e=t$, $U_{g g} = U_{ee} =U_{mm}$, and 
$\mu_{g}=\mu_{e}$ to get the following  Hamiltonian 
({\em generalized Hund model})\cite{Nonne-B-C-L-10}:
\begin{equation}
\begin{split}
\mathcal{H}_{\text{Hund}} = & -t \, \sum_{i}
  \left( c_{m\alpha,\,i}^\dag c_{m\alpha,\,i+1}+  \text{H.c.}  \right) \\
&  -\mu\sum_i n_i
  +\frac{U}{2} \sum_i  n_i^2  \\
&  +J \sum_i \left\{ (T_i^x)^2 + (T_i^y)^2\right\} + J_z \sum_i (T_i^z)^2 \; .
\end{split}
  \label{alkaourmodel}
\end{equation}
Now, the equivalence mapping between the models \eqref{eqn:Gorshkov-Ham} 
and \eqref{alkaourmodel} reads as
\begin{equation}
\begin{split}
&J=V_{\text{ex}}^{g\text{-}e},  \quad J_z=U_{mm} - V , \\
&U=\frac{U_{mm}+V}{2}, \quad
\mu=\frac{U_{mm}+V^{g\text{-}e}_{\text{ex}}}{2}+\mu_{g}  \; .
\end{split}
\label{eqn:Gorshkov-to-Hund}
\end{equation}
It is obvious that the first three terms in Eq. \eqref{alkaourmodel} 
are U($2N$)-invariant and the remaining orbital part ($J$ and $J_{z}$) breaks it down to 
\begin{equation}
\begin{split}
\text{U($2N$)} = 
\text{U(1)}_{\text{c}}{\times}\text{SU($2N$)} &\xrightarrow{J= J_{z}(\neq 0)} 
\text{U(1)}_{\text{c}}{\times}\text{SU($N$)}_{\text{s}}{\times}\text{SU(2)}_{\text{o}} \\
& \xrightarrow{J\neq J_{z}} \text{U(1)}_{\text{c}}{\times}\text{SU($N$)}_{\text{s}}{\times}\text{U(1)}_{\text{o}} \;.
\end{split}
\label{eqn:symmetry-change}
\end{equation} 
Therefore, the generic continuous symmetry of this model is 
$\text{U}(1)_{\text{c}}\times\text{SU}(N)_{\text{s}}\times \text{U}(1)_{\text{o}}$.   
Physically, the orbital-$\text{U}(1)_{\text{o}}$ symmetry of $\mathcal{H}_{g\text{-}e}$ \eqref{alkamodel-2b} may be 
traced back to the vanishingly weak $g \leftrightarrow e$ transition.\cite{Gorshkov-et-al-10}   
%%%%%%%%%%%%%%%%%%%%%%%%%%%%%%%%%%%%%%%%%%%%%%%%%%
\subsection{\texorpdfstring{$\boldsymbol{p}$}{p}-band Hamiltonian}
\label{sec:p-band-definition}
%%%%%%%%%%%%%%%%%%%%%%%%%%%%%%%%%%%%%%%%%%%%%%%%%%
There is yet another way to realize the two orbitals using a simple setting.  
Let us consider a one-dimensional optical lattice (running in the $z$-direction) 
with moderate strength of (harmonic) confining potential $V_{\perp}(x,y)=\frac{1}{2}m\omega_{xy}^{2}(x^{2}+y^{2})$ 
in the direction (i.e. $xy$) 
perpendicular to the chain.  Then, the single-particle part of the Hamiltonian reads as
\begin{equation}
\begin{split}
\mathcal{H}_{0} &= \left\{
- \frac{\hbar^{2}}{2m}\partial_{z}^{2} + V_{\text{per}}(z) 
\right\}  +
\left\{ 
- \frac{\hbar^{2}}{2m}\left( \partial_{x}^{2} + \partial_{y}^{2} \right) 
 + V_{\perp}(x,y) 
 \right\}  \\
& \equiv \mathcal{H}_{\perp}(x,y) + \mathcal{H}_{/\!/}(z)
 \; ,
 \end{split}
 \label{eqn:single-particle-Ham}
 \end{equation}
 where $V_{\text{per}}(z)$ is a periodic potential that introduces a lattice structure in the chain 
 (i.e. $z$) direction.  
If the chain is infinite in the $z$-direction, we can assume the Bloch function in 
the following form:
\begin{equation} 
\psi^{(n)}_{n_x,n_y,k_z}(x,y,z) = \phi_{n_x,n_y}(x,y) \varphi^{(n)}_{k_z}(z) \; .
\end{equation}
The two functions $\varphi^{(n)}_{k_z}(z)$ and $\phi_{n_x,n_y}(x,y)$ respectively satisfy
\begin{subequations}
\begin{equation}
\mathcal{H}_{/\!/}(z) \varphi^{(n)}_{k_z}(z) = \epsilon^{(n)}(k_z) \varphi^{(n)}_{k_z}(z)  
\end{equation}
and 
\begin{equation}
\mathcal{H}_{\perp}(x,y) \phi_{n_x,n_y}(x,y)  = \epsilon_{n_x,n_y}  \phi_{n_x,n_y}(x,y)  \; .
\label{eqn:Schroedinger-transverse-part}
\end{equation}
\end{subequations}
Since the second equation is the Schr\"{o}dinger equation of the two-dimensional harmonic 
oscillator, the eigenvalues $\epsilon_{n_x,n_y}$ are given by 
\begin{equation}
\epsilon_{n_x,n_y} = \left( n_x + n_y + 1 \right) \hbar \omega_{xy}  \quad 
(n_x, n_y = 0,1,2, \ldots ) \; .
\end{equation}
The full spectrum of $\mathcal{H}_{0}$ is given by
\begin{equation}
E^{(n)}_{n_x,n_y}(k_z) =  \epsilon^{(n)}(k_z) +  \epsilon_{n_x,n_y}  
\end{equation}
and each Bloch band specified by $n$ splits into the sub-bands labeled by $(n_x,n_y)$.   
We call the subbands with $(n_x,n_y)=(0,0)$, $(1,0)$, and $(0,1)$ as `$s$', `$p_x$' and `$p_y$', 
respectively.   
The shape of the bands depends only on the band index $n$ and the set of integers $(n_x,n_y)$ 
determines the $k_z$-independent splitting of the sub-bands.  

Now let us consider the situation where only the $n=0$ bands are occupied,  
and, among them, the lowest one (the $s$-band) is completely filled.  
Then, it is legitimate to keep only the next two bands $p_x$ and $p_y$ in the effective Hamiltonian.%
\cite{Kobayashi-O-O-Y-M-12,Kobayashi-O-O-Y-M-14}    
To derive a Hubbard-type Hamiltonian, we follow the standard strategy\cite{Jaksch-Z-05} 
and move from the Bloch basis $\psi^{(n)}_{n_x,n_y,k_z}(x,y,z)$ to the Wannier basis
\begin{equation}
W^{(n)}_{n_x,n_y;R}(x,y,z) \equiv 
\frac{1}{\sqrt{N_{\text{cell}}}} \phi_{n_x,n_y}(x,y) \sum_{k_z}
\be^{- i k_z R} \varphi^{(n)}_{k_z}(z)
\label{eqn:Wannier-pxpy}
\end{equation}
($R$ labels the center of the Wannier function and $N_{\text{cell}}$ is then number of unit cells).   
Expanding the creation/annihilation operators in terms of the Wannier basis and keeping only 
the terms with $n=0$ and $(n_x,n_y)=(1,0)$ or $(0,1)$, we obtain the following Hamiltonian 
(see Appendix \ref{sec:p-band-hamiltonian})
\begin{equation}
\begin{split}
& \mathcal{H}_{p\text{-band}} \\
& = - t \sum_{i}
( c_{m\alpha,i}^{\dagger}c_{m\alpha,i+1}  + \text{H.c.} ) \\
& + \sum_{i} \sum_{m=p_x,p_y} 
(\epsilon_{m} - t_{0}) \, n_{m,i} \\
&+  \frac{1}{2}U_{1}  \sum_{i}
n_{m,i}(n_{m,i}-1) +  U_{2}\,  \sum_{i}n_{p_x,i}n_{p_y,i} \\
& + U_{2} \sum_{i}
c_{p_x\alpha,i}^{\dagger} c_{p_y\beta,i}^{\dagger}c_{p_x\beta,i}c_{p_y\alpha,i} \\
& + U_{2} \sum_{i}\left\{ (T^{x}_{i})^{2} - (T^{y}_{i})^{2} \right\}  \; .
\end{split}
\label{eqn:p-band}
\end{equation}
In the above, we have introduced a short-hand notation $m=p_{x},p_{y}$ with 
$p_{x}=(n_x,n_y)=(1,0)$ and $p_{y}=(n_x,n_y)=(0,1)$.  
The last term comes from the pair-hopping between the two orbitals 
(see Appendix \ref{sec:p-band-hamiltonian}) and breaks U(1)$_{\text{o}}$-symmetry in general.  
Since the Wannier functions are real and the two orbitals $W_{p_x/p_y;R}^{(0)}(\bolr)$ 
are related by $C_4$-symmetry, there are only two independent couplings $U_1$ and $U_2$ 
[see Eq. \eqref{eqn:def-U1-U2}].   
In fact, due to the axial symmetry of the potential $V_{\perp}(x,y)$, even the ratio $U_{1}=3 U_{2}$ 
is fixed and we are left with a single coupling constant.   

Except for the last term, $\mathcal{H}_{p\text{-band}}$ coincides with the Hamiltonian 
\eqref{alkaourmodel} after the identification
\begin{equation}
\begin{split}
& U=\frac{1}{2}(U_1+U_2) \, , \; U_{\text{diff}}=0 \, , \; J=U_2 \, , \; J_{z}=U_1 - U_2 \\
& \mu = -(\epsilon_{m}-t_{0})+ \frac{1}{2}(U_1 + U_2)  \; .
\end{split}
\end{equation}
Incorporating the last term, we obtain the following (orbital) anisotropic model
\begin{equation}
\begin{split}
\mathcal{H}_{p\text{-band}} = & -t \,  \sum_{i}
  \left( c_{m\alpha,\,i}^\dag c_{m\alpha,\,i+1}+  \text{H.c.}  \right)  \\
& -\mu\sum_i  n_i
+\frac{1}{4}(U_1 + U_2) \sum_i  n_i^2  \\
&  +\sum_i \left\{ 2U_2 (T_i^x)^2 + (U_1 - U_2)(T_i^z)^2\right\}  \; .
\end{split}
\label{eqn:p-band-simple}
\end{equation}
One may think that the last term breaks $\text{U(1)}_{\text{o}}$.   
However, as $U_{1}=3 U_{2}$ for {\em any} axially-symmetric $V_{\perp}(x,y)$, it has in fact a {\em hidden} 
U(1)$_{\text{o}}$-symmetry: $2U_2\left\{  (T_j^x)^2 +(T_j^z)^2\right\}$ and 
$\mathcal{H}_{p\text{-band}}$ reduces to $\mathcal{H}_{\text{Hund}}$ [Eq. \eqref{alkaourmodel}] 
after the due redefinition of $\mathrm{\bf T}$.\footnote{%
This is in a sense an artifact of the choice of the basis ($p_x$ and $p_y$).   
In fact, if we choose the angular-momentum (along the $z$-axis) basis, the U(1)-symmetry is obvious.}
 Higher continuous symmetries may also 
appear in model (\ref{eqn:p-band-simple}) when $U_2 =0$ since it decouples into two independent
U($N$) Hubbard chains, as it can be easily seen from Eq. (\ref{eqn:p-band}). Moreover, along the line $U_1 =U_2$,
the $p$-band model  (\ref{eqn:p-band-simple}) is equivalent to the $U_2 =0$ case after a redefinition
of $\mathrm{\bf T}$. Finally, as we will see in the next section,  
the $p$-band model for $N=2$ at half-filling enjoys an enlarged 
SU(2) $\times$ SU(2) $\sim$ SO(4) symmetry for all $U_1,U_2$ 
which stems from an additional SU(2) symmetry for the charge degrees 
of freedom at half-filling.\cite{Kobayashi-O-O-Y-M-14}
%%%%%%%%%%%%%%%%%%%%%%%%%%%%%%%%%%%%%%%%%%%%%%%%%%%%%%%%%%
\begin{figure}[hbt]
\begin{center}
\includegraphics[scale=0.5]{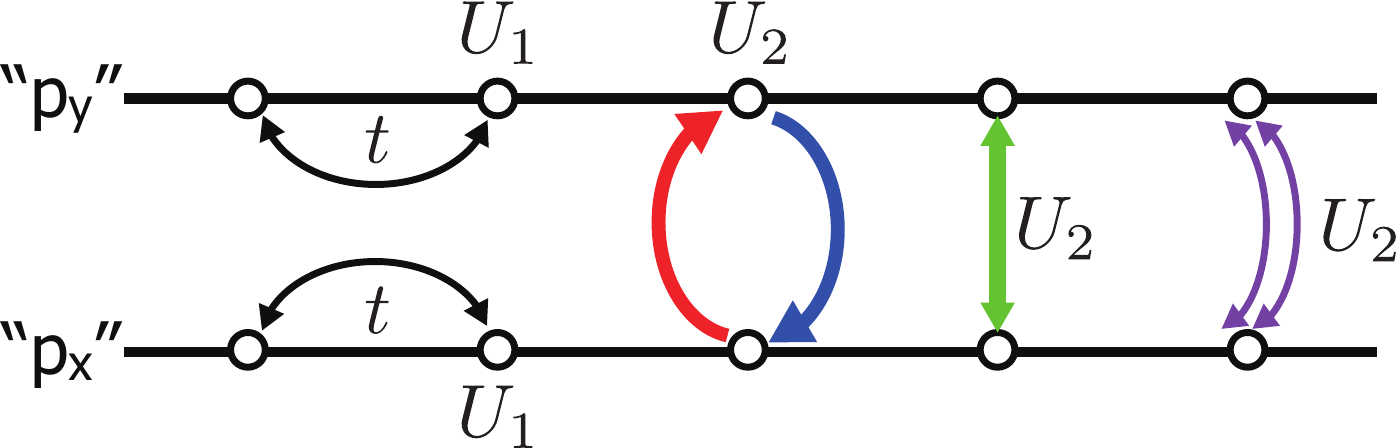}
\caption{(Color online) The two-leg ladder representation of the $p$-band model \eqref{eqn:p-band}. 
On top of the interactions included already in the $g$-$e$ model, pair-hopping processes between 
the two orbitals are allowed.   
\label{fig:p-band-2leg}}
\end{center}
\end{figure}
%%%%%%%%%%%%%%%%%%%%%%%%%%%%%%%%%%%%%%%%%%%%%%%%%%%%%%%

\begin{figure}[H]
\begin{center}
\includegraphics[scale=0.6]{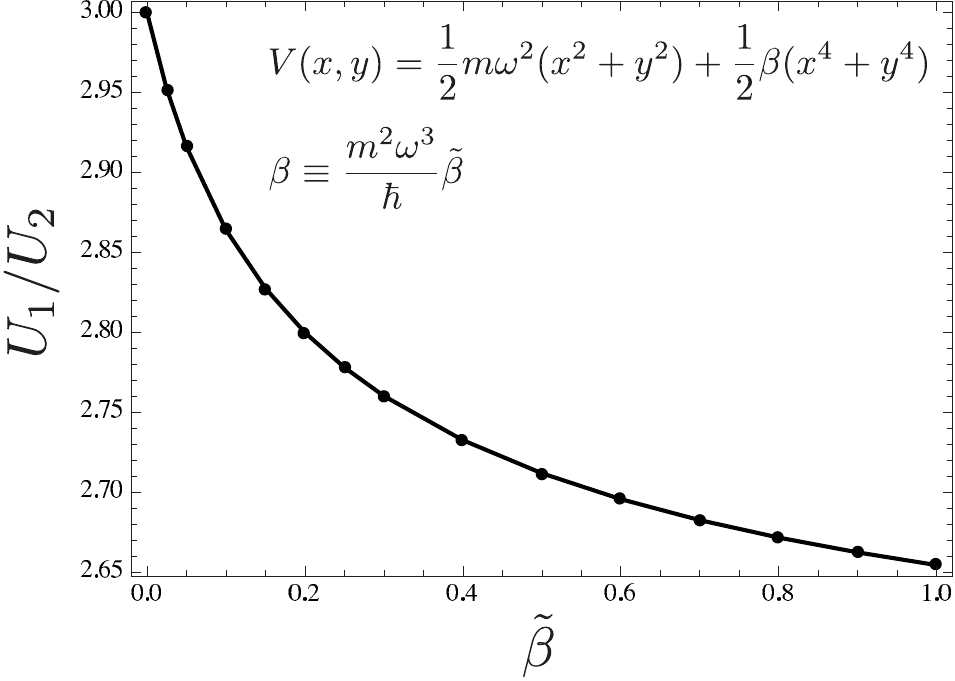}
\caption{The ratio $U_{1}/U_{2}$ for anharmonic potential \eqref{eqn:anharmonic-potential} 
obtained by solving the Schr\"{o}dinger equation \eqref{eqn:Schroedinger-transverse-part} 
numerically.     
\label{fig:U1overU2}}
\end{center}
\end{figure}

The $p$-band model is convenient since the axial symmetry guarantees that 
the parameters are fully symmetric for the two orbitals $p_x$ and $p_y$.  
However, the same symmetry locks the ratio $U_{1}/U_{2}(=3)$ and we cannot control it 
as far as $V_{\perp}(x,y)$ is axially symmetric. 
One simplest way of changing the ratio is to break the axial symmetry and 
consider the following anharmonic potential:
\begin{equation}
V_{\perp}(x,y) = \frac{1}{2}m\omega_{xy}^{2} (x^{2}+y^{2}) 
+ \frac{1}{2}\beta (x^{4}+y^{4}) \quad (\beta \geq 0) \; .
\label{eqn:anharmonic-potential}
\end{equation} 
In Fig. \ref{fig:U1overU2}, we plot the ratio $U_{1}/U_{2}$ as a function of anharmonicity $\beta$.  
Clearly, the ratio calculated using Eqs.~\eqref{eqn:def-Uabcd-pxpy} and 
\eqref{eqn:def-U1-U2} deviates from 3 with increasing $\beta$.   In that case ($U_1 < 3 U_2$), 
the original anisotropic model \eqref{eqn:p-band-simple} should be considered.  
%%%%%%%%%%%%%%%%%%%%%%%%%%%%%%%%%%%%%%%%%%%%%%%%%%%%%%%%%%
%%%%%%%%%%%%%%%%%%%%%%%%%%%%%%%%%%%%%%%%%%%%%%%%%%%%%%%%
\subsection{Symmetries}
\label{sec:symmetries}
%%%%%%%%%%%%%%%%%%%%%%%%%%%%%%%%%%%%%%%%%%%%%%%%%%%%%%%%
%%%%%%%%%%%%%%%%%%%%%%%%%%%%%%%%%%%%%%%%%%%%%%%%%%%%%%%%

The different models that we have introduced in the previous section enjoys generically an 
$\text{U}(1)_{\text{c}}\times\text{SU}(N)_{\text{s}}\times \text{U}(1)_{\text{o}}$ continuous symmetry
or an $\text{U}(1)_{\text{c}}\times\text{SU}(N)_{\text{s}}$ symmetry for the $p$-band model . 
On top of these continuous symmetries, the models
display hidden discrete symmetries which are very useful to map out their global zero-temperature
phase diagrams.

%%%%%%%%%%%%%%%%%%%%%%%%%%%%%%%%%%%%%%%%%%%%%%%%%%%%%%%%
\subsubsection{Spin-charge interchange}
\label{sec:spin-charge-interchange}
%%%%%%%%%%%%%%%%%%%%%%%%%%%%%%%%%%%%%%%%%%%%%%%%%%%%%%%%
The first transformation is a direct generalization of the Shiba transformation\cite{Shiba-72,Emery-76} 
for the usual Hubbard model and is defined {\em only} for $N=2$:
\begin{equation}
\begin{split}
& c_{m\uparrow,i} \mapsto \tilde{c}_{m\uparrow,i} \\
& c_{m\downarrow,i} \mapsto (-1)^{i} \tilde{c}^{\dagger}_{m\downarrow,i} 
\quad (m=g,e \text{ or } p_x,p_y)  \; .
\end{split}
\label{eqn:Shiba-tr-SU2-2band}
\end{equation}
It is easy to show that it interchanges spin and charge [see Eq.~\eqref{eqn:second-quant-SUN}]:
\begin{equation}
S^{A}_{m,i} \; \leftrightarrow \; K^{A}_{m,i} \quad (A=x,y,z)  \; , 
\label{eqn:Shiba-S-to-K}
\end{equation}
where $K^{A}_{m,i}$ are defined as
\begin{equation}
\begin{split}
& K^{+}_{m,i} \equiv (-1)^{i} c_{m\uparrow,i}^{\dagger}c_{m\downarrow,i}^{\dagger} , \quad 
K^{-}_{m,i} \equiv (-1)^{i} c_{m\downarrow,i} c_{m\uparrow,i}  , \\
& K_{m,i}^{z} \equiv \frac{1}{2}(n_{m\uparrow,i} + n_{m\downarrow,i} -1 ) 
= \frac{1}{2}(n_{m,i}-1)  \; .
\label{chargepseudospin}
\end{split}
\end{equation}
The latter operator carries charge and is a SU(2) spin-singlet. It generalizes the $\eta$-pairing operator introduced
by Yang for the half-filled spin-$\frac{1}{2}$  Hubbard
model~\cite{yang89} or by Anderson in his study of the BCS superconductivity \cite{anderson58}.

Now let us consider how the transformation \eqref{eqn:Shiba-tr-SU2-2band} affects 
the fermion Hamiltonians $\mathcal{H}_{g\text{-}e}$ [Eq. \eqref{eqn:Gorshkov-Ham}] 
and $\mathcal{H}_{p\text{-band}}$ [Eq. \eqref{eqn:p-band-simple}].  
The first three terms of the alkaline-earth Hamiltonian $\mathcal{H}_{g\text{-}e}$ 
[Eq. \eqref{eqn:Gorshkov-Ham-Hund}] do not change their forms under 
the transformation \eqref{eqn:Shiba-tr-SU2-2band}, while the last two are asymmetric in 
$\bolS_{m,i}$ and $\mathbf{K}_{m,i}$. Hence the $g\text{-}e$ Hamiltonian 
$\mathcal{H}_{g\text{-}e}$ does not preserve its form under 
$\bolS_{m,i} \leftrightarrow \mathbf{K}_{m,i}$.  

On the other hand, the $p$-band Hamiltonian, written in terms of $\bolS_{m,i}$ and $\mathbf{K}_{m,i}$,  
\begin{equation}
\begin{split}
\mathcal{H}_{p\text{-band}} = & 
- t \sum_{i}
( c_{m\alpha,i}^{\dagger}c_{m\alpha,i+1}  + \text{H.c.} ) \\
&+ U_{1}  \sum_{i}
(n_{m\uparrow,i}-1/2)(n_{m\downarrow,i}-1/2)  \\
& - 2 U_{2}\sum_{i} \bolS_{p_x,i}{\cdot} \bolS_{p_y,i}  
+ 2U_{2} \sum_{i} \mathbf{K}_{p_x,i}{\cdot} \mathbf{K}_{p_y,i}  \; ,
\end{split}
\label{pbandSO4}
\end{equation} 
preserves its form 
and the Shiba transformation \eqref{eqn:Shiba-tr-SU2-2band} changes the coupling constants as
\begin{equation}
(U_{1}, \, U_{2}) \rightarrow (-U_{1}, \, -U_{2})  \; .  
\label{eqn:Shiba-tr-SU2-U1U2}
\end{equation}

The expression (\ref{pbandSO4}) reveals the hidden symmetry of the half-filled $p$-band model for $N=2$.
On top of the SU(2) symmetry for the nuclear spins, which is generated by $\sum_{i,m} \bolS_{m,i}$,
the $p$-band Hamiltonian (\ref{pbandSO4}) enjoys a second independent  SU(2) symmetry related to the (charge)
pseudo spin operator (\ref{chargepseudospin}):
$$ \left[ \mathcal{H}_{p\text{-band}} , \sum_{i,m}  \mathbf{K}_{m,i}  \right] = 0. $$
The continuous symmetry group of the  $N=2$ half-filled $p$-band model is therefore:
 SU(2)  $\times$ SU(2) $\sim$ SO(4) for all $U_1,U_2$, i.e., without any fine-tuning.
In this respect, the latter model shares the same continuous symmetry group as the half-filled 
spin-1/2 Hubbard chain \cite{yangZ89,zhang91} but, as we will see later, the physics is strongly different.

%%%%%%%%%%%%%%%%%%%%%%%%%%%%%%%%%%%%%%%%%%%%%%%%%%%%%%%%
\subsubsection{orbital-charge interchange}
\label{sec:orbita-charge-interchange}
%%%%%%%%%%%%%%%%%%%%%%%%%%%%%%%%%%%%%%%%%%%%%%%%%%%%%%%%
For general $N$, we can think of another `Shiba' transformation:
\begin{equation}
\begin{split}
& c_{g\alpha,i} \mapsto \tilde{c}_{g\alpha,i} \\
& c_{e\alpha,i} \mapsto (-1)^{i} \tilde{c}^{\dagger}_{e\alpha,i} 
\quad (\alpha=1,\ldots, N) \; ,
\end{split}
\label{eqn:Shiba-tr-SUN}
\end{equation}
which interchanges the orbital pseudo spin $\mathbf{T}_{i}$ and 
another charge-SU(2) $\boldsymbol{\mathcal{K}}_{i}$.  
Now the charge-SU(2) is generated by the following orbital-singlet operators
\begin{equation}
\begin{split}
& \mathcal{K}_{i}^{+} \equiv  (-1)^{i} c_{g\alpha,i}^{\dagger}c_{e\alpha,i}^{\dagger} \; , \quad 
\mathcal{K}_{i}^{-} \equiv  (-1)^{i} c_{e\alpha,i}c_{g\alpha,i} \\
& \mathcal{K}_{i}^{z} \equiv \frac{1}{2}(n_{g,i}+n_{e,i} -N) 
= \frac{1}{2}( n_{i} -N)  \; .
\end{split}
\end{equation}

The transformation \eqref{eqn:Shiba-tr-SUN} changes the $g\text{-}e$ Hamiltonian \eqref{eqn:Gorshkov-Ham-Hund} 
by flipping the sign of $\left(V - V_{\text{ex}}^{g\text{-}e}/N \right)$ and 
replacing $S^{A}_{e,i}$ with the generators of the conjugate representation.  
Therefore, one sees that only when $J(=V^{g\text{-}e}_{\text{ex}})=0$ 
the $g\text{-}e$ Hamiltonian $\mathcal{H}_{g\text{-}e}$ 
preserves its form after 
\begin{equation}
V \mapsto - V \quad 
(\text{or } J_{z} \leftrightarrow 2U )\; .  
\label{eqn:Gorshkov-VG-flip}
\end{equation}
We will come back to this point later in Sec.~\ref{sec:N4-Gorshkov} in the discussion of the numerical phase diagram of 
the $N=4$ $g\text{-}e$ model.  

The case $N=2$ is special since any SU(2) representations are self-conjugate.  
In fact, when $N=2$, the transformation \eqref{eqn:Shiba-tr-SUN}, supplemented by 
the $\pi$-rotation along the $y$-axis in the SU(2) space ($c_{e\uparrow,i} \mapsto -c_{e\downarrow,i}$, 
$c_{e\downarrow,i} \mapsto c_{e\uparrow,i}$), preserves the form of the Hamiltonian 
after the mapping
\begin{equation}
\begin{split}
& V - \frac{1}{2}V_{\text{ex}}^{g\text{-}e} \rightarrow 
- \left(V - \frac{1}{2}V_{\text{ex}}^{g\text{-}e}\right) \quad \left(
\text{or }  V \rightarrow -V + V_{\text{ex}}^{g\text{-}e} \right) \\
& V_{\text{ex}}^{g\text{-}e} \rightarrow V_{\text{ex}}^{g\text{-}e} \, , \;\;
 U_{m m} \rightarrow  U_{m m}  \; .
\end{split}
\label{eqn:N2-Gorshkov-orbital-charge}
\end{equation}
Due to the orbital anisotropy $\left\{ (T^{x}_{j})^{2} - (T^{y}_{j})^{2} \right\}$ 
in $\mathcal{H}_{p\text{-band}}$ [the last term Eq. \eqref{eqn:p-band}], 
the $p$-band Hamiltonian in general does not preserve its form under 
the orbital-charge interchange \eqref{eqn:Shiba-tr-SUN}.  When $U_2=0$, 
the model is U(1)-orbital symmetric and is invariant (self-dual) under \eqref{eqn:Shiba-tr-SUN}.  
A summary of the effect of the two Shiba transformations on the two models is summarized
in Tables \ref{tab:Gorshkov} and \ref{tab:p-band}.
%%%%%%%%%%%%%%%%%%%% TABLE 1 %%%%%%%%%%%%%%%%%%%%%%%%%%%%%%%%
\begin{table}[H]
\caption{\label{tab:Gorshkov} Two Shiba transformations and $g$-$e$ Hamiltonian 
[Eq. \eqref{eqn:Gorshkov-Ham}].}
\begin{ruledtabular}
\begin{tabular}{lc}
transformation &   mapping  \\
\hline
spin-charge [Eq. \eqref{eqn:Shiba-tr-SU2-2band}] &  
not defined
\\
\hline
orbital-charge [Eq. \eqref{eqn:Shiba-tr-SUN}] & 
\begin{tabular}{lc}
$N=2$: &   $V \rightarrow -V + V_{\text{ex}}^{g\text{-}e}$ \\
$N\geq 3$:  & $V \rightarrow -V$ ($V_{\text{ex}}^{g\text{-}e}=0$)
\end{tabular}  
\\
\end{tabular}
\end{ruledtabular}
\end{table}
%%%%%%%%%%%%%%%%%%%% TABLE 2 %%%%%%%%%%%%%%%%%%%%%%%%%%%%%%%%
\begin{table}[H]
\caption{\label{tab:p-band}Two Shiba transformations and $p$-band Hamiltonian [Eq. \eqref{eqn:p-band-simple}]. 
Orbital-charge interchange exists only when $U_2=0$ and then the Hamiltonian is kept invariant.}
\begin{ruledtabular}
\begin{tabular}{lc}
transformation &   mapping  \\
\hline
spin-charge [Eq. \eqref{eqn:Shiba-tr-SU2-2band}] &  
\begin{tabular}{ll}
$N=2$: &   $U_{1,2} \rightarrow - U_{1,2}$ 
\end{tabular}  
\\
\hline
orbital-charge [Eq. \eqref{eqn:Shiba-tr-SUN}] & 
\begin{tabular}{lr}
$N$ arbitrary: &   invariant (only for $U_2=0$) 
\end{tabular}  
\\
\end{tabular}
\end{ruledtabular}
\end{table}
%%%%%%%%%%%%%%%%%%%%%%%%%%%%%%%%%%%%%%%%%%%%%%%%%%%%%%%%
\subsection{Strong-coupling limits}
\label{sec:strong-coupling}
%%%%%%%%%%%%%%%%%%%%%%%%%%%%%%%%%%%%%%%%%%%%%%%%%%%%%%%%
Useful insight into the global structure of the phase diagram may be obtained by investigating 
the strong-coupling limit where the hopping $t_{(g,e)}$ are very small.  
Then, the starting point is the atomic-limit Hamiltonian \eqref{eqn:atomic-limit-Ham}.  
In the following, we assume that $N=2I +1$ is even since the nuclear
spin $I$ is half-odd-integer for alkaline-earth fermions.  
The dominant phases found in the strong-coupling analysis are summarized in Table~\ref{tab:abbreviation}.
%%%%%%%%%%%%%%%%%%%%%%%%%%%%%%%%%%%%%%%%%%%%%%%%%%%%%%%%
\subsubsection{Positive-$J$}
\label{sec:strong-coupling-positive-J}
%%%%%%%%%%%%%%%%%%%%%%%%%%%%%%%%%%%%%%%%%%%%%%%%%%%%%%%%
First, we assume that $U$ and the chemical potential $\mu_{g}+\mu_{e}$ [see Eq. \eqref{eqn:HG-parameters}] 
are tuned in such a way that the fermion number at each site is $n_{i}=N$.   
Then, the remaining $\bolT$-dependent terms in \eqref{eqn:atomic-limit-Ham} determine 
the optimal orbital and SU($N$) states.   
From Eq.~\eqref{eqn:Gorshkov-Ham-Hund}, we see that for large positive $J(=V_{\text{ex}}^{g\text{-}e})$ 
the orbital pseudo spin $\bolT$ at each site tends to be quenched thereby maximizing the SU($N$) spin as
\begin{equation}
\text{\scriptsize $N/2$} \left\{ 
\yng(2,2,2)
 \right.  \quad (N=\text{even}) \; .
 \label{selfrep}
\end{equation}  
When considering second-order perturbation, it is convenient to view our system as 
a two-leg ladder of SU($N$) fermions [see Fig.~\ref{fig:alkaline-2leg}].  The resulting effective Hamiltonian
reads then as follows
\begin{equation}
\mathcal{H}_{\text{SU(N)}}  = J_{\text{s}} 
\sum_{A=1}^{N^{2}-1} \mathcal{S}_{i}^{A}\mathcal{S}_{i+1}^{A}  + \text{const.}  \; ,
\label{eqn:2nd-order-effective-Ham-Gorshkov}
\end{equation}
where the exchange coupling $J_{\text{s}}$ is $N$-independent
\begin{equation}
J_{\text{s}}  \equiv
\frac{1}{2} \left\{
\frac{(t^{(g)})^{2}}{U+U_{\text{diff}}+J+\frac{J_z}{2}} 
+ \frac{(t^{(e)})^{2}}{U-U_{\text{diff}}+J+\frac{J_z}{2}}
\right\}   \; .
\label{eqn:exch-coupling-Gorshkov}
\end{equation}

In the case of $\mathcal{H}_{\text{$p$-band}}$, $T^{z}$ is no longer conserved and we cannot use 
the same argument as above. However, we found that when $U_{1}>U_{2}(>0)$, the lowest-energy 
state has $T=0$ enabling us to follow exactly the same steps and obtain   
\begin{equation}
\mathcal{H}_{\text{SU(N)}}  = \frac{t^{2}}{U_1 +U_2} 
\sum_{A=1}^{N^{2}-1} \mathcal{S}_{i}^{A}\mathcal{S}_{i+1}^{A}  + \text{const.}  \; .
\label{eqn:2nd-order-effective-Ham-p-band}
\end{equation}

\begin{figure}[htb]
\begin{center}
\includegraphics[scale=0.85]{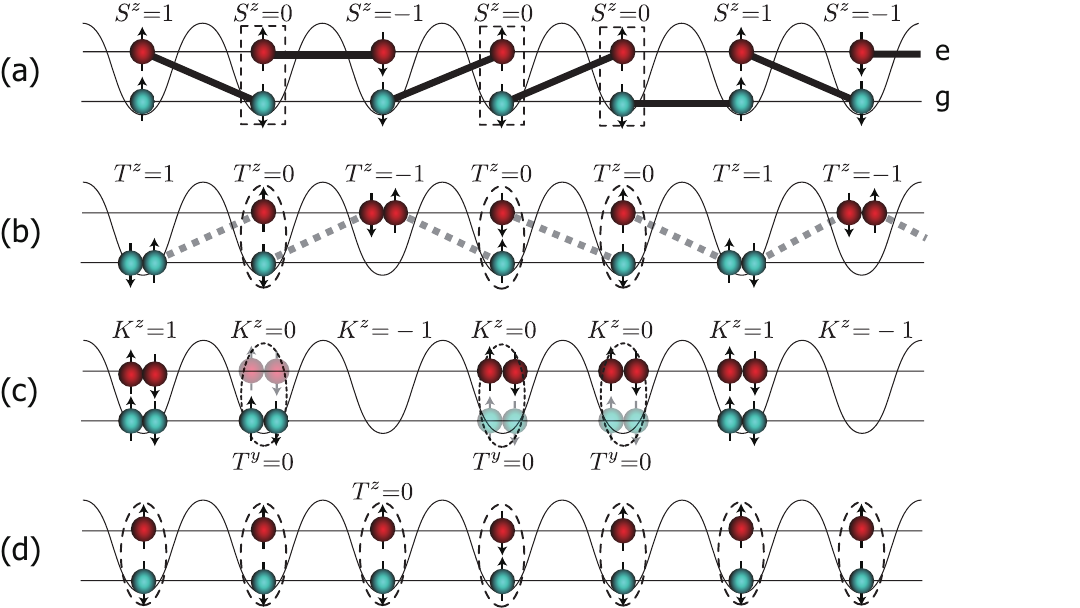}
\caption{(Color online) Four translationally invariant Mott states for $N=2$:  
(a) spin Haldane (SH), (b) orbital Haldane (OH), (c) charge Haldane (CH), 
and (d) rung-singlet (RS) phases (see also Appendix \ref{sec:N2-p-band-continuum}).   
Singlet bonds formed between spins (orbital pseudo spins) are shown by thick solid 
(dashed) lines [singlet bonds are not shown in (c)]. Dashed ovals (rectangles) denote 
spin-singlets (triplets). 
\label{fig:4-MottPhases-SU2}}
\end{center}
\end{figure}

One observes that models \eqref{eqn:2nd-order-effective-Ham-Gorshkov}
and \eqref{eqn:2nd-order-effective-Ham-p-band} take the form of an
SU($N$) spin chain in the self-conjugate representation
(\ref{selfrep}) at each site and is not solvable. The physical
properties of that model are unknown for general $N$.  In the special
$N=2$ case where the model reduces to the SU(2) spin-1 Heisenberg
chain, it is well-known that the Haldane
phase\cite{Haldane-PLA-83,Haldane-PRL-83} is formed by the nuclear
spins. The resulting spin Haldane (SH) phase for $N=2$ is depicted in
Fig.~\ref{fig:4-MottPhases-SU2}(a).  Using the spin-charge interchange
transformation \eqref{eqn:Shiba-tr-SU2-2band}, one concludes, for
$N=2$, the existence of a charge Haldane (CH) phase
\cite{Nonne-L-C-R-B-10} in the $p$-band model for $U_2 <0$ which is
illustrated in Fig.~\ref{fig:4-MottPhases-SU2}(c).  We will come back
to this point later in Sec.~\ref{sec:strong-coupling-negative-J}.

When $N>2$, the situation is unclear and a non-degenerate gapful phase is expected 
from the large-$N$ analysis of Refs.~\onlinecite{Read-S-NP-89,Read-S-90}. 
We will determine the nature of the underlying phase  in the next section. 
%%%%%%%%%%%%%%%%%%%%%%%%%%%%%%%%%%%%%%%%%%%%%%%%%%%%%%%%%%
%%%%%%%%%%%%%%%%%%%%%%%%%%%%%%%%%%%%%%%%%%%%%%%%%%%%%%%%%%%%%%%%%%%%%%%%%%%%%%%%%%%%%%%%%%%%%%%%%%%%%%%%%%%%%%%
\subsubsection{$J=0$}
\label{sec:strong-coupling-J-0}
%%%%%%%%%%%%%%%%%%%%%%%%%%%%%%%%%%%%%%%%%%%%%%%%%%%%%%%%
Another interesting line is the generalized Hund model (\ref{alkaourmodel}) with $J=J_z=0$ 
which becomes equivalent to the U($2N$) Hubbard model. 
In the strong-coupling limit with $U>0$, the lowest-energy states correspond to representations of the
SU($2N$) group which transform in  the antisymmetric self-conjugate representations of SU($2N$), described by a Young diagram with one column of $N$ boxes.  
The model is then equivalent to an SU($2N$)  Heisenberg spin chain 
where the spin operators belong to the antisymmetric self-conjugate representation of SU($2N$). 
The latter model is known for all $N$ to have a dimerized or spin-Peierls (SP) 
twofold-degenerate ground state, where dimers are formed 
between two neighboring 
sites \cite{Affleck-M-88,*Marston-A-89,Affleck-88,OnufrievM99,Assaraf-A-B-C-L-04,Nonne-L-C-R-B-11}.  

\begin{figure}[htb]
\begin{center}
\includegraphics[scale=0.85]{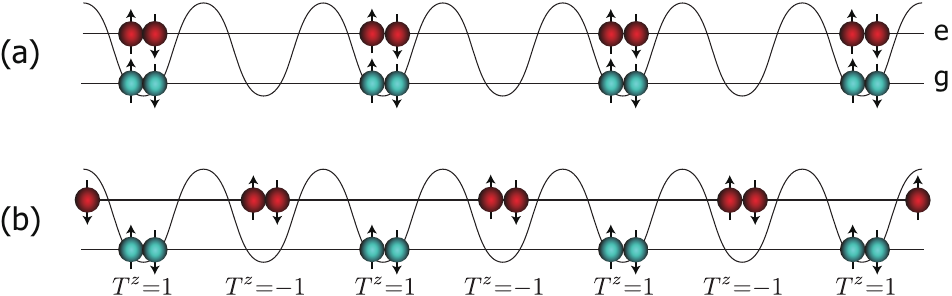}
\caption{(Color online) Two density-wave states for $N=2$. 
In-phase and out-of-phase combinations of two density waves in $g$ and $e$ orbitals respectively form  
(a) CDW and (b) ODW. 
\label{fig:2-DWPhases-SU2}}
\end{center}
\end{figure}

In the attractive case ($U<0$), the lowest-energy states are the empty and the fully occupied state, which is an SU($2N$) singlet. At second order of perturbation theory, 
the effective model reads as follows: \cite{Zhao-U-W-06,*Zhao-U-W-07}
\begin{equation}
	{\cal H}_{\textrm{eff}} = \frac{t^2}{N(2N-1)|U|} \sum_i \left(n_i n_{i+1} - N  n_i\right),
\end{equation}
The first term introduces an effective repulsion interaction between nearest neighbor sites. This leads to a  
two-fold degenerate fully-gapped charge-density wave (CDW) where empty ($n_{i}=0$) 
and fully occupied ($n_{i}=2N$) states alternate. The resulting CDW phase for $N=2$ is depicted in Fig. \ref{fig:2-DWPhases-SU2}(a).

%%%%%%%%%%%%%%%%%%%%%%%%%%%%%%%%%%%%%%%%%%%%%%%%%%%%%%%%%%
%%%%%%%%%%%%%%%%%%%%%%%%%%%%%%%%%%%%%%%%%%%%%%%%%%%%%%%

%%%%%%%%%%%%%%%%%%%%%%%%%%%%%%%%%%%%%%%%%%%%%%%%%%%%%%%%
\subsubsection{Negative-$J$}
\label{sec:strong-coupling-negative-J}
\begin{figure}[htb]
\begin{center}
\includegraphics[scale=0.45]{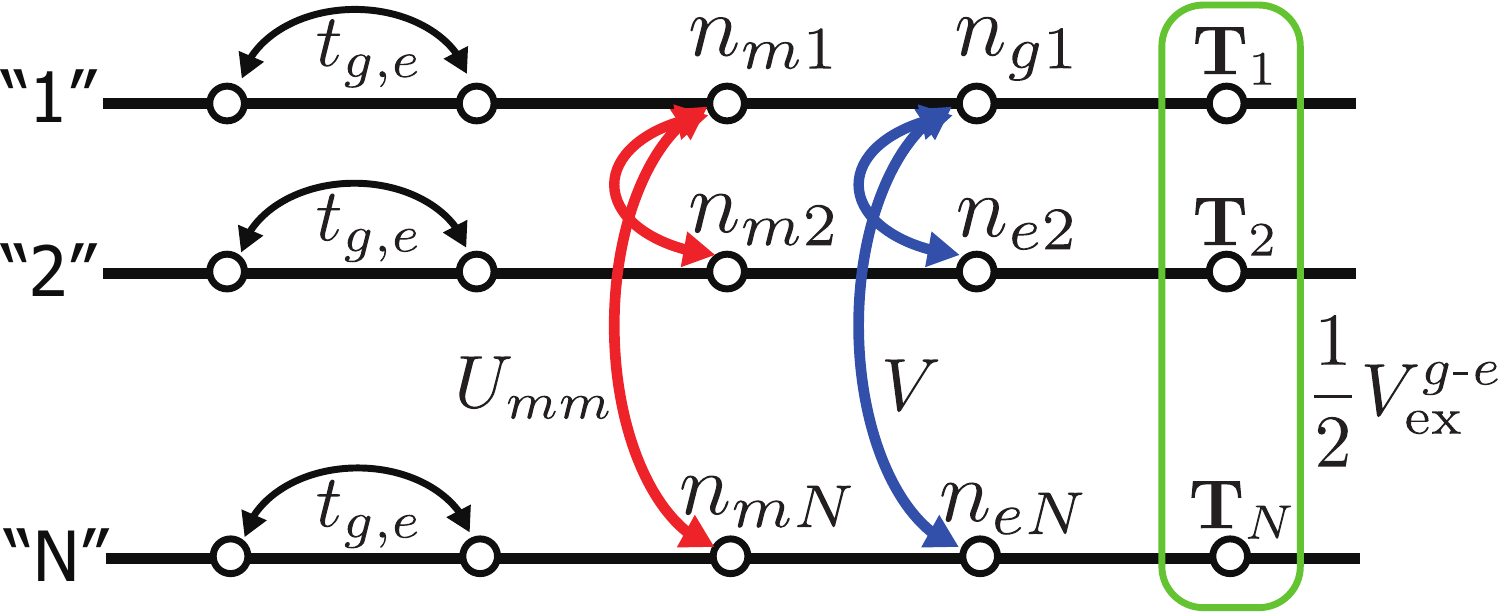}
\caption{(Color online) The $N$-leg ladder representation of the model \eqref{eqn:Gorshkov-Ham}. 
$N$ Hubbard-type chains for ``spinful'' fermions ($g$ and $e$) are coupled to each other by 
$U$ (interchain density-density interaction among like fermions), 
$V$ (that between $g$ and $e$), and 
the inter-chain Hund couplings ($V_{\text{ex}}^{g\text{-}e}$).   
\label{fig:alkaline-Nleg}}
\end{center}
\end{figure}

%%%%%%%%%%%%%%%%%%%%%%%%%%%%%%%%%%%%%%%%%%%%%%%%%%%%%%%%
Now let us discuss the case with $J<0$ (and $U>0$).  
For small enough anisotropies $|J-J_{z}|$, $|\mu_{g}-\mu_{e}|$, the atomic-limit ground states 
are obtained by applying the lowering operators $T^{-}_{i}$ onto the reference state
\begin{equation}
c^{\dagger}_{g1,i}c^{\dagger}_{g2,i}\cdots c^{\dagger}_{gN,i}
|0\rangle   \; .
\end{equation}
To carry out the second-order perturbation, it is convenient to regard the model 
$\mathcal{H}_{g\text{-}e}$ as the $N$ coupled Hubbard-type chains, along which the $g$ and $e$ 
fermions move (see Fig.~\ref{fig:alkaline-Nleg}).  
Since each ``site'' of the chains is occupied by exactly one fermion in the ground states, 
it is clear that the two hopping processes must occur on the same chain.  
Therefore, the calculation is similar to that in the usual single-band Hubbard chain 
(except that we have to symmetrize the $N$ resultant $T=1/2$ chains at the last stage) 
and we finally obtain the pseudo spin $T=N/2$ Hamiltonian
\begin{equation}
\begin{split}
\mathcal{H}_{\text{orb}} = &\sum_{i}
\biggl\{  
\mathcal{J}_{xy} \left(
T^{x}_{i} T^{x}_{i+1} + T^{y}_{i} T^{y}_{i+1} 
\right)
+ \mathcal{J}_{z} T^{z}_{i} T^{z}_{i+1}  \\
& -(J- J_z)  (T_{i}^z)^2 \biggr\}  \\
& + \sum_{i} \left\{ 
N U_{\text{diff}}
- \left(\mu _g -\mu _e\right) 
\right\} T_{i}^{z} 
+ \text{const.}  
\end{split}
\label{eqn:eff-Ham-orbital-Haldane}
\end{equation}
with the following exchange couplings
\begin{subequations}\label{eq:eff_coupling_Hund}
\begin{align}
& \mathcal{J}_{xy} \equiv \frac{4 t_{g}t_{e}}{N\left\{U -J \left( N+\frac{1}{2} \right)\right\}}  \\
& \mathcal{J}_{z} \equiv   
\frac{2 \left\{ t_{g}^{2} + t_{e}^{2}\right\}}{N \left\{U -J \left( N+\frac{1}{2} \right)\right\}} 
\quad (\mathcal{J}_{xy} \leq  \mathcal{J}_{z})   \; .
\end{align}
\end{subequations}
Since the atomic-limit ground state where we have 
started does not depend on $N$, the final effective Hamiltonian \eqref{eqn:eff-Ham-orbital-Haldane} 
is valid for both even-$N$ and odd-$N$.  
When $g$ and $e$ are symmetric (i.e., $U_{\text{diff}}=0$, $\mu_{g}=\mu_{e}$, $t_{g}=t_{e}$), 
$\mathcal{J}_{xy} = \mathcal{J}_{z}$ and 
the above effective Hamiltonian \eqref{eqn:eff-Ham-orbital-Haldane} reduces to 
the usual spin $T=N/2$ Heisenberg model with the single-ion anisotropy, whose phase diagram 
has been studied extensively (see, e.g. Refs.~\onlinecite{Schulz-86,Chen-H-S-03,Tonegawa-O-N-S-N-K-11} 
and references cited therein).    
It is well-known\cite{Haldane-PRL-83,Haldane-PLA-83} that the behavior of the spin-$S(=N/2)$ 
Heisenberg chain differs dramatically depending on the parity of $N$.  
Therefore, we may conclude that, when $N$ is even, 
the gapped ``orbital'' Haldane (OH) phase\cite{Nonne-B-C-L-10} 
appears for large negative $J$ (at least for small anisotropy $J \approx J_{z}$, $t_{g}\approx t_{e}$), 
while, for odd $N$, the same region is occupied by the gapless Tomonaga-Luttinger-liquid phase.   
The non-trivial hidden ordering of orbital degrees of freedom in the OH phase is illustrated for $N=2$  
in Fig.~\ref{fig:4-MottPhases-SU2}(b).  

When we increase $|J-J_{z}|$ ($J<J_{z}$), the OH phase finally gets destabilized and is taken over 
by a gapful SU($N$)-singlet non-degenerate phase.   
This is an orbital-analog of the ``large-$D$ phase'' whose wave function is given essentially by 
a product of $T_{i}^{z} = 0$ states [see Fig.~\ref{fig:4-MottPhases-SU2}(d)].  
In the following, we call it ``rung-singlet (RS)'' as this state reduces in the case of $N=2$ to the well-known 
rung-singlet state in the spin-$\frac{1}{2}$ two-leg ladder.\cite{Dagotto-R-96} 
On the other hand, when $J-J_z$ takes a large positive value 
(as will be seen in Sec.~\ref{sec:N4-gen-Hund-woSU2-J-pm4}), 
the effective Hamiltonian \eqref{eqn:eff-Ham-orbital-Haldane} develops easy-axis anisotropy 
and enters a phase with antiferromagnetic ordering of the orbital pseudo spin $T^{z}$: $-N/2,+N/2,-N/2,+N/2,\cdots$ 
[see Fig.~\ref{fig:2-DWPhases-SU2}(b)].  
This phase will be called `orbital-density wave (ODW)' and is depicted in Fig. \ref{fig:2-DWPhases-SU2}(b)
for $N=2$.
%%%%%%%%%%%%%%%%%%%%%%%%%%%%%%%%%%%%%%%%%%%%%%%%%%%%%%%%%%
%%%%%%%%%%%%%%%%%%%%%%%%%%%%%%%%%%%%%%%%%%%%%%%%%%%%%%%

Due to the strong easy-plane anisotropy in the orbital sector, a different conclusion is drawn  
for the $p$-band model \eqref{eqn:p-band-simple}.  
Now the single-site energy is given as
\begin{equation}
-\mu \, n_{i}
+\frac{1}{4}(U_1 + U_2)  n_{i}^2 
+  \left\{ 2U_2 (T_{i}^x)^2 + (U_1 - U_2)(T_{i}^z)^2\right\} \; .
\end{equation}
Since $V_{\text{ex}}^{g\text{-}e} = J \leftrightarrow U_{2}$, the condition $J<0$ translates to $U_{2}<0$ 
in the $p$-band model.  
Since the condition $U_{2}<0$ in the physical region $U_1 \simeq 3U_2$ implies an attractive interaction 
$U_1 + U_2 <0$, we have to take into account several different values of $n_{i}$.  
We follow the same line of argument as in Sec.~\ref{sec:strong-coupling-J-0} 
to show that at $\mu=-N |U_1+U_2|$, we have two degenerate SU($N$)-singlet states 
$n_{i}=0$ ($T=0$) and $n_{i}=2N$ ($T=0$)  
which feel a repulsive interaction coming from $t^2$-processes.   Therefore, 
$2k_{\text{F}}$-CDW occupies a region around the line $U_1 = 3U_2$ for $N \geq 3$.   

The case $N=2$ is exceptional due to the existence of the spin-charge symmetry 
\eqref{eqn:Shiba-tr-SU2-U1U2}.   
In fact, at $\mu= -4|U_{2}|$, 
the following three spin-singlet states
\begin{equation}
\begin{split}
& c^{\dagger}_{p_x\uparrow,i}c^{\dagger}_{p_x\downarrow,i}
c^{\dagger}_{p_y\uparrow,i}c^{\dagger}_{p_y\downarrow,i}|0\rangle 
\;\; (n_{i}=4) \\
& \frac{1}{\sqrt{2}}\left(  
c^{\dagger}_{p_x\uparrow,i}c^{\dagger}_{p_x\downarrow,i}
+ c^{\dagger}_{p_y\uparrow,i}c^{\dagger}_{p_y\downarrow,i} \right) |0\rangle 
 \; (\equiv |\text{OLD}_y\rangle, n_{i}=2 ) \\
&  |0\rangle \;\; (n_{i}=0)  \; , 
\end{split}
\label{eqn:charge-triplet}
\end{equation}
are degenerate on the U(1)-symmetric line $U_1=3U_2$ and 
form a triplet of charge-SU(2) at each site.  

The effective Hamiltonian for the ground-state manifold spanned by these triplets is readily obtained by applying 
the transformation \eqref{eqn:Shiba-S-to-K} to \eqref{eqn:2nd-order-effective-Ham-p-band}, 
which is nothing but the spin-1 Heisenberg model.  From the known ground state of the effective 
Hamiltonian, one sees that, instead of CDW for $N \geq 3$, CH appears around the line $U_1=3U_2$ 
when $N=2$.   Note that the existence of the Shiba transformation, which guarantees the symmetry 
between spin and charge, is crucial for the appearance of the CH phase in the $N=2$ case.  

%%%%%%%%%%%%%%%%%%%% TABLE 1 %%%%%%%%%%%%%%%%%%%%%%%%%%%%%%%%
\begin{table}
\caption{\label{tab:abbreviation} List of dominant phases and their abbreviations. 
Local SU($N$)/orbital degrees of freedom are shown, too.}
\begin{ruledtabular}
\begin{tabular}{lccc}
Phases &   Abbreviation & SU($N$) & Orbital ($T$) \\
\hline
Spin-Haldane\footnotemark[1] & SH & $S=1$ & Local singlet \\
Orbital-Haldane & 
OH & Local singlet & $N/2$ \\
Charge-Haldane\footnotemark[1] & 
CH & Local singlet & $-$
\\
Orbital large-$D_{x,y}$ & 
OLD$_{x,y}$ & Local singlet & $N/2$\\
Rung-singlet (OLD$_{z}$)\footnotemark[2] & 
RS & Local singlet &  $N/2$
\\
Spin-Peierls  & SP & $-$ & $N/2$ 
\\
Charge-density wave & CDW & Local singlet & Local singlet
\\
Orbital-density wave\footnotemark[3] & ODW & Local singlet & $N/2$
\end{tabular}
\end{ruledtabular}
\footnotetext[1]{Only in $N=2$.}%
\footnotetext[2]{Product of $T^{z}=0$ states (large-$D$ state) of $T=N/2$.}%
\footnotetext[3]{`N\'{e}el-ordered' state of $T=N/2$.}%
\end{table}

%%%%%%%%%%%%%%%%%%%%%%%%%%%%%%%%%%%%%%%%%%%%%%%%%%%%%
\section{SU(\texorpdfstring{$\boldsymbol{N}$}{N}) topological phase}
\label{sec:SUN-topological-phase}
%%%%%%%%%%%%%%%%%%%%%%%%%%%%%%%%%%%%%%%%%%%%%%%%%%%%%
In this section, we investigate the nature of the ground state of the SU($N$) Heisenberg spin
chain \eqref{eqn:2nd-order-effective-Ham-Gorshkov} and its main physical properties.
%%%%%%%%%%%%%%%%%%%%%%%%%%%%%%%%%%%%%%%%%%%%%%%%%%%%%
\subsection{SU(\texorpdfstring{$\boldsymbol{N}$}{N}) valence-bond-solid (VBS) state}
\label{sec:SUN-VBS-state}
%%%%%%%%%%%%%%%%%%%%%%%%%%%%%%%%%%%%%%%%%%%%%%%%%%%%%
In Sec. \ref{sec:models-strong-coupling}, we have seen that for positive $J$ (or positive $U_{2}$), 
we obtain the SU($N$) Heisenberg model \eqref{eqn:2nd-order-effective-Ham-Gorshkov} 
or \eqref{eqn:2nd-order-effective-Ham-p-band} for relatively wide parameter regions.  
This SU($N$) spin chain has the self-conjugate representation (with $N/2$ rows and 2 columns) 
at each site and is not solvable.  Nevertheless, we can obtain\cite{Nonne-M-C-L-T-13}  
a fairly good understanding of the properties of the ground state by constructing 
a series of model ground states, the VBS states~\cite{Affleck-K-L-T-87,Affleck-K-L-T-88},
whose parent Hamiltonian is close to the original ones \eqref{eqn:2nd-order-effective-Ham-Gorshkov} 
and \eqref{eqn:2nd-order-effective-Ham-p-band}.  

\begin{figure}[ht]
\centering
\includegraphics[width=0.9\columnwidth,clip]{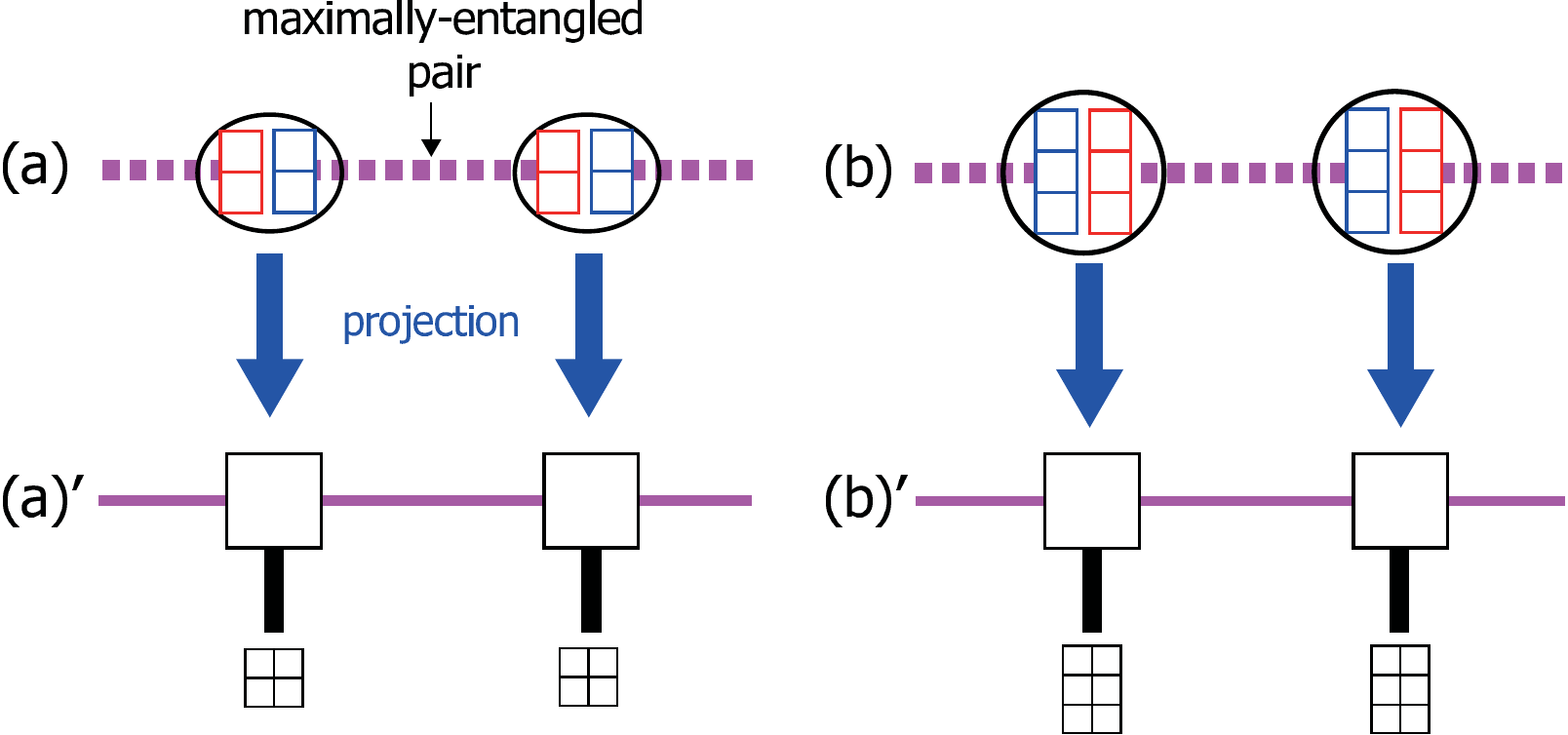}
\caption{(Color online) SU($N$) VBS states are constructed out of a pair of self-conjugate 
representations at each site. Dashed lines denote maximally-entangled pairs. 
(a) SU(4) with 20-dimensional representation and 
(b) SU(6) with 175-dimensional representation. 
(a)' and (b)' are the corresponding matrix-product states. 
\label{fig:VB-construction}}
\end{figure}

We start from a pair of the self-conjugate representations [characterized by a Young diagram  
with {\em one} column and $N/2$ rows; see \eqref{eqn:two-auxiliary-spaces}] 
on each site and create maximally-entangled pairs 
between adjacent sites [see Figs.~\ref{fig:VB-construction}(a) and (b)].  
To obtain the physical wave function, we apply the projection 
[see Figs.~\ref{fig:VB-construction}(a)$^{\prime}$ and \ref{fig:VB-construction}(b)$^{\prime}$]
\begin{equation}
\text{\scriptsize $N/2$} \left\{ 
\yng(1,1,1) \otimes \yng(1,1,1)
 \right.  
\longrightarrow
\yng(2,2,2)  
\label{eqn:two-auxiliary-spaces}
\end{equation}  
onto the tensor-product state obtained above 
and construct the physical Hilbert space [i.e., SU($N$) representation with its Young diagram  
having $N/2$ rows and two columns] at each site.  
This procedure may be most conveniently done by using the matrix-product state (MPS)\cite{Garcia-V-W-C-07} 
\begin{equation}
\sum_{\{m_{i}\}}A_{1}(m_1)A_{2}(m_2)\cdots A_{i}(m_i)\cdots |m_1,m_2,\ldots,m_i
,\ldots\rangle  \; ,
\label{eqn:VBS-state-in-MPS}
\end{equation}
where $m_{i}$ labels the states of the $d$-dimensional local Hilbert space 
at the site-$i$ and $A_{i}(m_{i})$ is $D{\times}D$ 
matrices with $D$ being the bond dimensions.  
The dimensions of the local Hilbert space are $d=20$ [SU(4)], $d=175$ [SU(6)], $d=1764$ [SU(8)], 
and so on.  

Although it is in principle possible to write the MPS for general $N$, the construction rapidly 
becomes cumbersome with increasing $N$. 
Therefore, we focus below only on the $N=4$ case where the ground state is given by the MPS 
with $D=6$ (the dimensions of ${\tiny \yng(1,1)}$).   
The parent Hamiltonian bearing the above VBS state as the exact ground state is not unique 
and, aside from the overall normalization, there are two free (positive) parameters.  
Among them, the one with lowest order in $(S^{A}_{i} S^{A}_{i+1})$ is given by\cite{Nonne-M-C-L-T-13}
\begin{equation}
\begin{split}
&{\cal H}^{(N=4)}_{\text{VBS}}   \\
& = J_{\text{s}} 
\sum_{i}\Big\{ S^{A}_{i} S^{A}_{i+1} + \frac{13}{108}(S^{A}_{i} S^{A}_{i+1})^2 
 + \frac{1}{216}(S^{A}_{i} S^{A}_{i+1})^3 \Big\},
 \end{split}
\label{eqn:SU4-VBSmodel}
\end{equation}
where $S^{A}_{i}$ ($A=1,\ldots,15$) denote the SU($4$) spin operators in the 20-dimensional representation 
[normalized as $\text{Tr}\, (S^{A}S^{B})=16 \delta^{AB}$] 
and $J_{\text{s}}$ is the exchange interaction between SU($N$) ``spins''.\footnote{%
It is evident that one can generalize this strategy to general even-$N$ once we know the Clebsch-Gordan 
decomposition of the two physical spaces on the adjacent sites.}

The ground state is SU(4)-symmetric and featureless {\em in the bulk},  
and has the ``spin-spin'' correlation functions 
\begin{equation}
\langle S^{A}_{j} S^{A}_{j+n} \rangle 
= 
\begin{cases}
\frac{12}{5} \left(-\frac{1}{5} \right)^n   &  n \neq 0 \\
\frac{4}{5} & n=0 
\end{cases}
\end{equation}
that are exponentially decaying 
with a very short correlation length $1/\ln 5\approx 0.6213$.   
In spite of the featureless behavior in the bulk, the system exhibits a certain structure 
near the boundaries.  In fact, if one measures $\VEV{S^{A}_{i}}$ (with $S^{A}_{i}$ being 
any three commuting generators), one can clearly see the structure localized around 
the two edges.  
At each edge, there are six different states distinguished by the value of the set 
of the three generators $\VEV{S^{A}_{i}}$.   
As in the spin-1 Haldane systems where two spin-$\frac{1}{2}$'s 
emerge at the edges,\cite{Affleck-K-L-T-88,Kennedy-90} 
one may regard these six edge states as the emergent SU($N$) `spin' ${\tiny \yng(1,1)}$ 
appearing near the each edge.  
%%%%%%%%%%%%%%%%%%%%%%%%%%%%%%%%%%%%%%%%%%%%%%%%%%%%%%%%%%
%%%%%%%%%%%%%%%%%%%%%%%%%%%%%%%%%%%%%%%%%%%%%%%%%%%%%
%%%%%%%%%%%%%%%%%%%%%%%%%%%%%%%%%%%%%%%%%%%%%%%%%%%%%%%
\subsection{Symmetry-protected topological phases}
%%%%%%%%%%%%%%%%%%%%%%%%%%%%%%%%%%%%%%%%%%%%%%%%%%%%%%%
We observe that the model (\ref{eqn:SU4-VBSmodel}) is not very far from the original pure Heisenberg 
Hamiltonian \eqref{eqn:2nd-order-effective-Ham-Gorshkov} or \eqref{eqn:2nd-order-effective-Ham-p-band} 
obtained by the strong-coupling expansion in Sec.~\ref{sec:strong-coupling}. 
This strongly suggests that the SU(4) topological phase
realizes in the strong-coupling regime of the SU($N$) fermion system $\mathcal{H}_{g\text{-}e}$  
[Eq. \eqref{eqn:Gorshkov-Ham}] or $\mathcal{H}_{p\text{-band}}$ [Eq. \eqref{eqn:p-band}]  
with the emergent edge states that belong to the six-dimensional representation of SU(4).   
In Ref.~\onlinecite{Duivenvoorden-Q-13}, it is predicted using the group-cohomology approach%
\cite{Chen-G-W-11,Fidkowski-K-11,Schuch-G-C-11}, that there are $N$ topologically distinct  
phases (including one trivial phase) protected by PSU($N$) $=$ SU($N$)/$\mathbb{Z}_N$ symmetry, 
which are characterized by the number of boxes $n_{\text{y}}$ (mod $N$) contained 
in the Young diagram corresponding to the emergent edge ``spin'' at the (right) edge.  
Since the six-dimensional representation ${\tiny \yng(1,1)}$ appears at the edge 
of the $N=4$ VBS state \eqref{eqn:VBS-state-in-MPS}, 
one expects that the ground state of the Heisenberg Hamiltonian 
\eqref{eqn:2nd-order-effective-Ham-Gorshkov} [or \eqref{eqn:2nd-order-effective-Ham-p-band}]  
as well as that of the $N=4$ VBS Hamiltonian \eqref{eqn:SU4-VBSmodel} 
belongs to the $n_{\text{y}}=2$ member (we call it 
{\em class-2} hereafter) of the four topological classes. 

Nevertheless, the observation of the edge-state degeneracy alone may lead to erroneous answers. 
A firmer evidence may be provided by the entanglement spectrum,\cite{Li-H-08}  
which is essentially the logarithm of the eigenvalues of the reduced density matrix.   
For instance, by tracing the entanglement spectrum, we can distinguish between different topological 
phases.\cite{Pollmann-T-B-O-10,Zang-J-W-Z-10,Zheng-Z-X-L-11,Pollmann-B-T-O-12}  
On general grounds, one may expect that any representations compatible with 
the group-cohomology classification\cite{Chen-G-W-11,Schuch-G-C-11} 
can appear in the entanglement spectrum.\footnote{%
For instance, according to the group-cohomology scheme\cite{Chen-G-W-11,Schuch-G-C-11}, 
the topological phases protected by $\text{PSU(2)}\simeq \text{SO(3)}$ 
are classified by the parity of integer-spin $S$.  In the topological Haldane phase corresponding to 
odd-$S$, the entanglement spectrum consists of even-fold degenerate levels reflecting 
the even-dimensional edge states emerging at the edge.}  
Quite recently, the entanglement spectrum for the model \eqref{eqn:exch-coupling-Gorshkov} 
has been calculated\cite{Tanimoto-T-14}  
by using the infinite-time evolving block decimation (iTEBD)\cite{Vidal-iTEBD-07,Orus-V-08} method.  
It has been found that the spectrum indeed consists of several different levels whose degeneracies 
are all compatible with the dimensions of the SU(4) irreducible representations allowed for 
the edge states of the class-2 topological phase.  
Specifically, the lowest-lying entanglement levels consist of ${\tiny \yng(1,1)}$ (6-dimensional), 
${\tiny \yng(3,2,1)}$ (64-dimensional), ${\tiny \yng(3,3)}$ (50-dimensional), etc. 
Moreover, the continuity between the ground state of the model \eqref{eqn:2nd-order-effective-Ham-Gorshkov} 
and that of \eqref{eqn:SU4-VBSmodel} has been demonstrated\cite{Tanimoto-T-14} 
by tracing the entanglement spectrum 
along the path ($0 \leq \lambda \leq 1$): 
\begin{equation}
\begin{split}
&{\cal H}(\lambda) 
= J_{\text{s}}  \sum_{i} S^{A}_{i} S^{A}_{i+1}  \\
&  \quad + \lambda J_{\text{s}}  \sum_{i} \Big\{ \frac{13}{108}(S^{A}_{i} S^{A}_{i+1})^2 
 + \frac{1}{216}(S^{A}_{i} S^{A}_{i+1})^3 \Big\}  \; .
 \end{split}
\label{eqn:SU4-VBSmodel-lambda}
\end{equation}   
At $\lambda=0$, ${\cal H}(\lambda)$ reduces to the effective Hamiltonian $\mathcal{H}_{\text{SU(N)}}$ 
[Eq.~\eqref{eqn:2nd-order-effective-Ham-Gorshkov} or \eqref{eqn:2nd-order-effective-Ham-p-band}] 
and ${\cal H}(1)$ is the VBS Hamiltonian \eqref{eqn:SU4-VBSmodel} whose entanglement spectrum 
consists only of the sixfold-degenerate level.  When we move from $\lambda=0$ to 1, 
the entanglement levels other than the lowest one gradually go up and finally disappear from 
the spectrum at $\lambda=1$ while preserving the structure of the spectrum.   

It is interesting to consider the protecting symmetries other than PSU($N$).  
The result from group cohomology\cite{Chen-G-L-W-13} 
$H^{2}(\text{PSU($N$)},\text{U(1)})=H^{2}(\mathbb{Z}_{N}{\times}\mathbb{Z}_{N},\text{U(1)})=\mathbb{Z}_{N}$ 
suggests that  $\mathbb{Z}_{N}{\times}\mathbb{Z}_{N}$ will do the job.   
Since it has been recently demonstrated that the even-fold degenerate structure in the entanglement 
spectrum signals the topological Haldane phase,~\cite{Pollmann-T-B-O-10,Pollmann-B-T-O-12} 
one may ask whether there is a relation between our class-2 topological phase and 
the Haldane phase.  
However, as we will show in the following, the even-fold degeneracy found in the entanglement spectrum 
of our SU(4) state comes from the protecting $\mathbb{Z}_{4}{\times}\mathbb{Z}_{4}$-symmetry 
that is a subgroup of PSU(4).

The first $\mathbb{Z}_{4}$-generator $Q$ is defined in terms of the two commuting SU(4) generators 
(Cartan generators) as
\begin{equation}
\begin{split}
& Q \equiv \be^{i\frac{3\pi}{4}} \exp\left( i \frac{2\pi}{4}G_{Q} \right)  , \;\; Q^{4}=1 \\
& G_{Q} \equiv  2H_{1} + H_{2}  \; .
\end{split}
\end{equation}
On the other hand, the second $\mathbb{Z}_{4}$ is generated by
\begin{equation}
\begin{split}
& P \equiv \be^{i\frac{3\pi}{4}} \exp\left( i \frac{2\pi}{4}G_{P} \right)  , \;\; P^{4}=1 \\
& G_{P} \equiv  - \frac{1}{2} \sum_{\alpha} E_{\alpha}  + \frac{i}{2} \left( \sum_{i=1}^{3}E_{\alpha_{i}}  
- E_{\alpha_{1}+\alpha_{2}+\alpha_{3}} \right)   \\
& \phantom{G_{P} \equiv}
- \frac{i}{2} \left( \sum_{i=1}^{3}E_{-\alpha_{i}}- E_{-\alpha_{1}-\alpha_{2}-\alpha_{3}} \right)   \; .
\end{split}
\end{equation}
In the above equations, we have used the Cartan-Weyl basis $\{ H_{a},E_{\alpha}\}$ 
that satisfies
\begin{equation}
\begin{split}
&  [H_{a},H_{b}]=0 \, , \;\; 
[H_{a},E_{\alpha}] = (\alpha)_{a}E_{\alpha} \, , \\
& [E_{\alpha},E_{-\alpha}] = \sum_{a=1}^{3} (\alpha)_{a}H_{a} \, ,  \; 
\text{Tr}\, (H_{a}H_{b})=16 \delta_{ab} ,
\\
&(a,b=1,2,3) 
\end{split}
\end{equation}
with $\alpha$  being the roots of SU(4) normalized as $|\alpha|=\sqrt{2}$ which are generated 
by the simple roots $\alpha_{i}$ ($i=1,2,3$).  
The summation $\sum_{\alpha}$ is taken over all 12 roots $\alpha$ of SU(4).  
Here we do not give the explicit expressions of the generators which depend on 
a particular choice of the basis, since giving the commutation relations suffices to define 
$\mathbb{Z}_{4}{\times}\mathbb{Z}_{4}$.  
In the actual calculations, one may use, e.g.,  
the generators and the weights given in Sec.~13.1 of Ref.~\onlinecite{Georgi-book-99} 
with due modification of the normalization.~\footnote{%
In order to obtain the generators in the 20-dimensional representation ($\mathbf{20}$), 
one may start, e.g., from $\mathbf{4}\otimes\mathbf{4}\simeq \mathbf{6}\oplus\mathbf{10}$ and then 
use $\mathbf{6}\otimes \mathbf{6} \simeq \mathbf{1}\oplus\mathbf{15}\oplus\mathbf{20}$. 
}%%% 

It is important to note that the two $\mathbb{Z}_{4}$s defined above commute with each other 
(i.e., $[Q,P]=0$) {\em only} when the number of boxes in the Young diagram is an integer multiple of 4.  
To put it another way, the two operators $Q$ and $P$ constructed here 
generate $\mathbb{Z}_{4}{\times}\mathbb{Z}_{4}$ only for PSU(4) as the two $\pi$-rotations 
along the $x$ and $z$ axes generate $\mathbb{Z}_{2}{\times}\mathbb{Z}_{2}$ only when the spin 
quantum number is integer.  

Now let us consider the relation between the PSU($4$) topological classes and 
the above $\mathbb{Z}_{4}{\times}\mathbb{Z}_{4}$ symmetry.  To this end, we recall the fundamental 
property of MPS. 
If a given MPS generated by the matrices $\{A(m)\}$ is invariant under 
the $\mathbb{Z}_{4}{\times}\mathbb{Z}_{4}$ symmetry introduced above, 
there exists a set of unitary matrices $U_{Q}$ and $U_{P}$ satisfying~\cite{Garcia-W-S-V-C-08}
\begin{equation}
\begin{split}
&   A(m) \xrightarrow{Q}
   e^{i\theta_{Q}}{U_{Q}}^{\dagger} A(m) U_{Q}  \\
& A(m) \xrightarrow{P}
   e^{i\theta_{P}}{U_{P}}^{\dagger} A(m) U_{P}  \; .
\end{split}
\label{unitary_matrix}
\end{equation}
Then, the property $QP=PQ$ mentioned above implies\footnote{%
We use an argument similar to that in Refs.~\onlinecite{Garcia-W-S-V-C-08,Pollmann-T-B-O-10}.}
that they obey the following non-trivial relation\cite{Duivenvoorden-Q-ZnxZn-13}:
\begin{equation}
U_{Q}U_{P} = \omega^{n_{\text{y}}} U_{P}U_{Q} \;\; 
(\omega \equiv \be^{i \frac{2\pi}{N}})  
\label{eqn:commutation-projective}
\end{equation}  
with the same $n_{\text{y}}(=0,1,2,3)$ as above.  
Reflecting the entanglement structure, $U_{P}$ and $U_{Q}$ are both block-diagonal.  
By taking the determinant of both sides, one immediately sees that 
the degree of degeneracy $D_{\xi}$ of each entanglement level $\xi$ 
(i.e., the size of each block) satisfies $\omega^{D_{\xi} n_{\text{y}}}=1$.  
In our SU(4) case, 
$D_{\xi}=4n$ ($n$: positive integer) for class-1 ($n_{\text{y}}=1$) and class-3 ($n_{\text{y}}=3$), while 
$D_{\xi}=2n$ for class-2 ($n_{\text{y}}=2$).
The relation \eqref{eqn:commutation-projective} implies that the crucial information on 
the PSU(4) topological phase is encoded in the exchange property of the {\em projective} 
representations $U_Q$ and $U_P$ of $\mathbb{Z}_{4}{\times}\mathbb{Z}_{4}$.   
This is the key to the construction of non-local string order parameters of our PSU($N$) topological phases.  
%%%%%%%%%%%%%%%%%%%%%%%%%%%%%%%%%%%%%%%%%%%%%%%%%%%%
\subsection{Non local order parameters}
\label{sec:non-local-OP}
%%%%%%%%%%%%%%%%%%%%%%%%%%%%%%%%%%%%%%%%%%%%%%%%%%%% 
By definition, local order parameters are not able to capture the SU($N$) SPT phases. 
Nevertheless, elaborate choice \cite{Haegeman-G-C-S-12,Pollmann-T-12,Hasebe-T-13} 
of {\em non-local} order parameters could detect hidden topological orders in those phases.  
We adapt the method\cite{Duivenvoorden-Q-ZnxZn-13} of constructing non-local order parameters 
in generic $(\mathbb{Z}_{N}{\times}\mathbb{Z}_{N})$-invariant systems 
to our SU(4) system.  
As in the usual spin systems\cite{Kennedy-T-92-PRB,Kennedy-T-92-CMP}, 
one can construct the following sets of order parameters in terms of SU(4) generators
\begin{subequations}
\begin{align}
\begin{split}
& \mathcal{O}_{1}(m,n) \\
&\equiv \lim_{|i-j|\nearrow \infty} 
\Biggl\langle \left\{\hat{X}_{P}(i)\right\}^{m} \left\{
\prod_{i\leq k <j} \hat{Q}(k)^{n}
\right\} \left\{ \hat{X}_{P}^{\dagger}(j) \right\}^{m}  \Biggr\rangle
\label{eqn:def-stringOP-1b} 
\end{split}
\\
\begin{split}
& \mathcal{O}_{2}(m,n) \\
& \equiv \lim_{|i-j|\nearrow \infty} 
\Biggl\langle
\left\{  \hat{X}_{Q}(i) \right\}^{m}  
\left\{ \prod_{i < k \leq j} \hat{P}(k)^{n}
\right\} \left\{\hat{X}_{Q}^{\dagger}(j)  \right\}^{m} \Biggr\rangle  \\
& \qquad (0 \leq m,n < N) 
 \; . 
 \end{split}
\label{eqn:def-stringOP-2b}
\end{align}
\end{subequations}
The subscripts 1 and 2 refer to the string order parameters corresponding to the two commuting 
$\mathbb{Z}_{N}$'s.  
The operators $\hat{X}_{Q}$ and $\hat{X}_{P}$ appearing in the above can be expressed by the SU(4) 
generators as
\begin{equation}
\begin{split}
& \hat{X}_{Q} = \frac{1}{2}( E_{-\alpha_1}+E_{-\alpha_2}+E_{-\alpha_3} + E_{\alpha_1+\alpha_2+\alpha_3})   \\
& \hat{X}_{P} = \frac{1}{\sqrt{2}}(H_{1} - i H_{3} )   
\end{split}
\end{equation}
and obey the following relations ($\omega=\be^{i\frac{2\pi}{4}}$)
\begin{equation}
\begin{split}
& \hat{Q}^{\dagger}\hat{X}_{Q} \hat{Q} = \omega \hat{X}_{Q} \;\; , \quad 
\hat{P}^{\dagger}\hat{X}_{Q} \hat{P} = \hat{X}_{Q} \\
& \hat{Q}^{\dagger}\hat{X}_{P} \hat{Q} = \hat{X}_{P} \;\; , \quad 
\hat{P}^{\dagger}\hat{X}_{P} \hat{P} = \omega^{-1}\hat{X}_{P}  
\end{split}
\end{equation}
for {\em any} irreducible representations of SU(4).  

It is known\cite{Pollmann-T-12,Hasebe-T-13} that the boundary terms of $\mathcal{O}_{1,2}(m,n)$ 
carry crucial information about the projective representation under which the physical edge states 
transform and hence give a physical way of characterizing the topological phases.  
By carefully analyzing the phase factors appearing in the boundary terms, one sees that 
the three sets of non-local string order parameters 
$\{\mathcal{O}_{1,2}(1,3),\mathcal{O}_{1,2}(2,1),\mathcal{O}_{1,2}(1,1)\}$ can distinguish 
among the four distinct phases (one trivial and three topological) protected by PSU(4) symmetry 
(see Table \ref{tab:3-string-OP}).\cite{Tanimoto-T-14}   
In fact, one can check\cite{Tanimoto-T-14} numerically that $\mathcal{O}_{1,2}(2,1)$ remains finite 
all along the interpolating path 
$\mathcal{H}(\lambda)$, while all the others are zero (at the solvable point $\lambda=1$, 
$\mathcal{O}_{1,2}(2,1)=1$).      
%%%%%%%%%%%%%%%%%%%% TABLE 1 %%%%%%%%%%%%%%%%%%%%%%%%%%%%%%%%
\begin{table}[H]
\caption{\label{tab:3-string-OP} Three sets of string order parameters characterizing the four distinct phases 
protected by PSU(4). The entry `finite' means that the corresponding $\mathcal{O}_{1,2}$ in principle can 
take non-zero values.}
\begin{ruledtabular}
\begin{tabular}{lccc}
Phases &   $\mathcal{O}_{1,2}(1,3)$ & $\mathcal{O}_{1,2}(2,1)$ & $\mathcal{O}_{1,2}(1,1)$ \\
\hline
Trivial  ($n_{\text{y}}=4n$) &  
0 & 0 & 0 
\\
Class-1 ($n_{\text{y}}=4n+1$) & Finite & 0 & 0
\\
Class-2 ($n_{\text{y}}=4n+2$) & 0 & Finite & 0
\\
Class-3 ($n_{\text{y}}=4n+3$) & 0 & 0 & Finite 
\end{tabular}
\end{ruledtabular}
\end{table}
%%%%%%%%%%%%%%%%%%%%%%%%%%%%%%%%%%%%%%%%%%%%%%%%%%%%
%%%%%%%%%%%%%%%%%%%%%%%%%%%%%%%%%%%%%%%%%%%%%%%%%%%%
\section{The weak-coupling approach}
\label{sec:weak-coupling}
%%%%%%%%%%%%%%%%%%%%%%%%%%%%%%%%%%%%%%%%%%%%%%%%%%%%
In this section, we map out the zero-temperature phase diagram of the different lattice models \eqref{eqn:Gorshkov-Ham},
\eqref{alkaourmodel} and \eqref{eqn:p-band-simple} 
related to the physics of the 1D  two-orbital SU($N$) cold fermions by means of a low-energy approach.
In particular, we will investigate the fate of the different topological Mott-insulating phases, revealed in the strong-coupling
approach, in the regime where the hopping term is not small.
%%%%%%%%%%%%%%%%%%%%%%%%%%%%%%%%%%%%%%%%%%%%%%%%%%%%
\subsection{Continuum description}
\label{sec:continuum_description}
%%%%%%%%%%%%%%%%%%%%%%%%%%%%%%%%%%%%%%%%%%%%%%%%%%%%
The starting point of the analysis is the continuum description of the lattice fermionic operators
$c_{m\alpha,\,i}$ in terms of $2N$ left-right moving Dirac fermions 
($m=g,e$ or $m=p_x, p_y$, $\alpha=1,\ldots,N$): \cite{Gogolin-N-T-book,Giamarchi-book}
\begin{equation}
c_{m \alpha,\,i} \rightarrow \sqrt{a_0} \left(L_{m \alpha}(x)
e^{-i k_\text{F} x} + R_{m \alpha}(x) e^{i k_\text{F} x} \right),
\label{contlimitDirac}
\end{equation}
where $x= i a_0$ ($a_0$ being the lattice spacing). 
Here we assume $t_{g}=t_{e}$ and $\mu^{(g)}=\mu^{(e)}$, and hence 
$k^{(g)}_\text{F} =k^{(e)}_\text{F} =k_\text{F} = \pi/(2 a_0)$ for half-filling.  
The non-interacting Hamiltonian is equivalent to that of $2N$ left-right moving Dirac fermions:
\begin{equation}
  {\cal H}_0=-i v_\text{F} \left(R_{m \alpha} ^\dag \partial_x R_{m \alpha} ^{\phantom \dag} - 
  L_{m \alpha}^\dag \partial_x L_{m \alpha}^{\phantom \dag}\right) ,
\label{HamcontDirac}  
\end{equation}  where $v_\text{F} =  2t a_0$ is the Fermi velocity. The non-interacting model \eqref{HamcontDirac} 
enjoys an U(2$N$)$|_\text{L}$ $\otimes$ U(2$N$)$|_\text{R}$ continuous symmetry which results from
its invariance under independent unitary transformations on the $2N$ left and right Dirac fermions. 
It is then very helpful to express the Hamiltonian \eqref{HamcontDirac} directly in terms 
of the currents generated by these continuous symmetries.
To this end, we introduce the U(1)$_{\text{c}}$ charge current  and
the SU(2$N$)$_1$ current which underlie the conformal field theory (CFT) of massless $2N$ 
Dirac fermions: \cite{Affleck-NP86,Affleck-88}
\begin{equation}
\begin{split}
& J_{\text{c} \text{L}} = :L_{n\alpha}^\dagger L_{n\alpha}: \quad\textrm{U(1)$_\text{c}$ charge current} \\
&  J_\text{L}^{A} = L_{m\alpha}^\dagger {\cal T}^{A}_{m,\alpha;n,\beta} L_{n\beta} \quad 
	  \text{SU($2N$)}_1 \text{ currents}, 
	\end{split}
	\label{U(2N)currents}
\end{equation}
with $m,n = g,e$ (or $m,n = p_x, p_y$ for the $p$-band model), $\alpha,\beta = 1, \ldots, N$, and we have similar definitions for the right currents. In Eq. (\ref{U(2N)currents}),  the symbol $::$ denotes  the normal ordering with respect to the Fermi sea, and ${\cal T}^{A}$ ($A =1, \ldots, 4 N^2 -1$) stand for the generators of SU($2N$) in the fundamental
representation normalized such that: $\text{Tr}({\cal T}^A {\cal T}^B)=\delta^{A\,B}/2$.
 The non-interacting model (\ref{HamcontDirac}) can then be written in terms of these currents
 (the so-called Sugawara construction of the corresponding CFT\cite{DiFrancesco-M-S-book}): 
\begin{equation}
\begin{split}
{\cal H}_0 =& \frac{\pi v_\text{F}}{2N}
 \left[ : J^2_{c\text{R}} : + : J^2_{c\text{L}} :
\right]   \\
&+ \frac{2\pi v_\text{F}}{2N + 1} \left[ : J^A_{\text{R}} J^A_{\text{R}}: + : J^A_{\text{L}} J^A_{\text{L}}: 
\right] .
\end{split}
\label{contfreehambis}
\end{equation}

The non-interacting part is thus described by an U(1)$_{\text{c}}$ $\times$ SU(2$N$)$_1$ CFT.
 Since the lattice model has a lower SU($N$) symmetry originating from the nuclear spin degrees of freedom, 
 it might be useful to consider the following conformal embedding \cite{DiFrancesco-M-S-book}, which is also relevant to multichannel Kondo problems \cite{Affleck-L-91}: 
U(1)$_{\text{c}}$ $\times$ SU(2$N$)$_1$  $\supset$ U(1)$_{\text{c}}$ $\times$ SU($N$)$_2$ $\times$ SU($2$)$_N$.
In this respect,  let us define the following currents which 
generate the  SU($N$)$_2$ $\times$ SU($2$)$_N$ CFT:
\begin{equation}
\begin{split}
& J_\text{L}^a = L_{n\alpha}^\dagger (T^a)_{\alpha,\beta} L_{n\beta} \quad \text{SU($N$)}_2 
\text{ (nuclear) spin currents}\\
&  j_{\text{L}}^i = L_{m\alpha}^\dagger (\sigma^i/2)_{m,n} L_{n\alpha} \quad
\text{SU(2)}_N \text{ orbital currents}  \\
&  \mathcal{J}_{\text{L}}^{a,i} = L_{m\alpha}^\dagger T^{a,i}_{m,\alpha;n,\beta} L_{n\beta} \quad
\text{remaining SU($2N$)$_1$ currents},
\end{split}
\label{defalkacurrents}
\end{equation}
where $T^a$ ($a=1,\ldots,N^{2}-1$) and $\sigma^i$ ($i=x,y,z$) respectively are the SU($N$) generators and the Pauli matrices.  
The $4N^2-1$ SU(2$N$) generators can be expressed in a unifying manner 
as a direct product between 
the SU($N$) and the SU(2) generators:
\begin{equation}
\begin{split}
& T^{a,0} = \frac{1}{\sqrt{2}}T^a\otimes I_2  \\
& T^{0,i} = \frac{1}{2\sqrt{N}}I_N\otimes \sigma^i  \\
& T^{a,i} = \frac{1}{\sqrt{2}}T^a\otimes \sigma^i ,
\end{split}
\end{equation}
where all the above generators are normalized in such a way that: $\text{Tr}(T^X T^Y)=\delta^{X\,Y}/2$ ($X,Y=(a,i)$).
The current $j_{\text{L}}^i$, being the sum of $N$ SU(2)$_1$ currents, 
the CFT corresponding to spin-1/2 degrees of freedom
\cite{Gogolin-N-T-book}, becomes an SU(2)$_N$ current, that accounts for
the critical properties of the orbital degrees of freedom.
Similarly, $J_\text{L}^a$ is a sum of two level-1 SU($N$) currents and 
the low-energy properties of the nuclear spin degrees are governed by an SU($N$)$_2$ CFT
which is generated by the  $J_{\text{L}}^a$ ($a = 1, \ldots, N^2 -1$) current.

At half-filling, we need to introduce, on top of these currents, additional operators which carry the U(1) charge
to describe various umklapp operators in the continuum limit:
\begin{equation}
\begin{split}
& A^{\alpha \beta +}_{mn \text{L}} =
\frac{-i}{2} \left(  L^{\dagger}_{m \alpha} L^{\dagger}_{n \beta} 
- L^{\dagger}_{m \beta}  L^{\dagger}_{n \alpha}  \right)  \\
& S^{\alpha \beta +}_{\text{L}} =  \frac{1}{2} \left(  L^{\dagger}_{g \alpha} L^{\dagger}_{e \beta} 
+ L^{\dagger}_{g \beta}  L^{\dagger}_{e \alpha}  \right) , 
\end{split}
\label{umklappop}
\end{equation}
with $m,n = g,e$ (or $m,n = p_x, p_y$ for the $p$-band model), 
and $\alpha,\beta = 1, \ldots, N$.  We introduce a similar 
set of operators for the right fields as well.

With all these definitions at hand, we are able to derive the continuum limit of two-orbital SU($N$) models 
of Sec. II. We will neglect all the velocity anisotropies for the sake of simplicity. Performing the continuum limit, we get 
 the following interacting Hamiltonian density:
\begin{equation}
\begin{split}
& \mathcal{H}_{\text{int}} \\
&= g_1 J_\text{L}^a J_\text{R}^a +\frac{g_2}{2}\left(\mathcal{J}_\text{L}^{a,+} \mathcal{J}_\text{R}^{a,-} 
+\textrm{H.c.}\right) + g_3 \mathcal{J}_\text{L}^{a,3} \mathcal{J}_\text{R}^{a,3}  \\
  & + \frac{g_4}{2}\left(j_{\text{L}}^+j_{\text{R}}^-+\textrm{H.c.}\right) 
  +g_5 j_{\text{L}}^zj_{\text{R}}^z + g_6 J_{\text{c L}} J_{\text{c R}}  \\
  & + g_7 \left(S^{\alpha \beta +}_{\text{L}}  S^{\alpha \beta -}_{R} +\textrm{H.c.}\right) 
  + \frac{g_8}{2} \sum_{m=g,e} \left(A^{\alpha \beta +}_{mm \text{L}}  
  A^{\alpha \beta -}_{mm \text{R}} +\textrm{H.c.}\right) \\
 &+ g_9  \left(A^{\alpha \beta +}_{g e \text{L}}  A^{\alpha \beta -}_{g e \text{R}} +\textrm{H.c.}\right) . 
\end{split}
  \label{lowenergyham}
\end{equation}

Although the different lattice models, having the same
continuous symmetry, share the same continuum Hamiltonian \eqref{lowenergyham} 
in common, the sets of initial coupling constants are different.  
For the generalized Hund model \eqref{alkaourmodel}, 
we find the following identification for the coupling constants:
\begin{equation}
\begin{split}
& g_1 = -  \left( U + J + \frac{J_z}{2}\right)a_0 \\
& g_2 =   \left(-2U+J_z\right)a_0 \\
& g_3 =  \left(-2U+2J-J_z\right)a_0 \\
& g_4 =  \left(-\frac{2U}{N}+2J+\frac{J_z}{N}\right)a_0 \\
& g_5 =  \left( -\frac{2U}{N} +\frac{2J}{N}+\frac{2N-1}{N}J_z \right) a_0 \\
& g_6 =  \left(\frac{U(2N-1)}{2N} - \frac{J}{2N} - \frac{J_z}{4N}\right)a_0  \\
& g_7 =  - \left(-U + J  + \frac{J_z}{2}\right)a_0  \\
& g_8 =   \left(U  + \frac{J_z}{2}\right)a_0 \\
& g_9 =  \left(U  + J - \frac{J_z}{2}\right) a_0  , 
\end{split}
\label{couplings}
\end{equation}
while, for the $g\text{-}e$ model with fine-tuning $U_{gg} = U_{ee} = U_{mm}$, 
we use Eq. \eqref{eqn:Gorshkov-to-Hund} to obtain:
\begin{equation}
\begin{split}
& g_1 = - a_0 \left( U_{mm} + V_{\text{ex}}^{g\text{-}e} \right) \\
& g_2 =  -2 a_0 V \\
& g_3 =  2 a_0 \left(V_{\text{ex}}^{g\text{-}e}  - U_{mm}  \right) \\
& g_4 = 2 a_0 \left(V_{\text{ex}}^{g\text{-}e}  - \frac{V}{N}\right)\\
& g_5 = 2 a_0 \left( \frac{(N-1)}{N} U_{mm}  + \frac{1}{N} V_{\text{ex}}^{g\text{-}e}  - V\right)\\
& g_6 =  \frac{a_0}{2N} \left( - V_{\text{ex}}^{g\text{-}e}  +  (N-1) U_{mm}  + N V \right)\\
& g_7 =  a_0 \left(V - V_{\text{ex}}^{g\text{-}e}  \right)\\
& g_8 =  a_0 U_{mm}  \\
& g_9 =  a_0 \left(V + V_{\text{ex}}^{g\text{-}e}  \right) . 
\end{split}
\label{couplingsGorsh}
\end{equation}

Since the effective Hamiltonian \eqref{lowenergyham} enjoys an $\text{U}(1)_{\text{c}}\times\text{SU}(N)_{\text{s}}\times \text{U}(1)_{\text{o}}$ continuous symmetry, it governs also the low-energy properties the $p$-band 
model \eqref{eqn:p-band-simple} with an harmonic confinement potential where $U_1 = 3 U_2$ and
also along the line $U_1 =U_2$ as discussed in Sec.~\ref{sec:p-band-definition}. In absence of the U(1)$_{o}$ orbital symmetry,  
model \eqref{lowenergyham} will be more complicated with 12 independent coupling constants 
and we will not investigate this case here.

%%%%%%%%%%%%%%%%%%%%%%%%%%%%%%%%%%%%%%%%%%%%%%%%%%%%
\subsection{RG analysis}
\label{sec:RG_analysis}
%%%%%%%%%%%%%%%%%%%%%%%%%%%%%%%%%%%%%%%%%%%%%%%%%%%%
The interacting part \eqref{lowenergyham} consists of marginal current-current interactions.  
The one-loop RG calculation enables one to deduce the infrared (IR) properties of that model and thus 
the nature of the phase diagram of the SU($N$) two-orbital models. 
After very cumbersome calculations, we find the following one-loop RG equations:
%%%%%%% RGE %%%%%%%%%%%%%%%%%%%%%%%%%%%%
\begin{equation}
\begin{split}
\dot{g_1} =& \frac{N}{4\pi} g^2_1 + \frac{N}{8\pi} g^2_2 + \frac{N}{16\pi} g^2_3
  + \frac{N+2}{4\pi} g^2_7 + \frac{N-2}{4\pi} \left( 2 g^2_8 + g^2_9  \right)  \\
\dot{g_2} =&  \frac{N}{2\pi} g_1 g_2 + 
    \frac{N^2 -4}{4\pi N} g_2 g_3 +   \frac{1}{2\pi} (g_2 g_5+g_3 g_4) + \frac{N}{\pi} g_7 g_8  \\
&+  \frac{N-2}{\pi} g_8 g_9  \\
\dot{g_3} =&  \frac{N}{2\pi} g_1 g_3 +  
    \frac{N^2 -4}{4\pi N} g^2_2 +    \frac{1}{\pi} g_2 g_4 
     + \frac{N}{\pi} g_7 g_9 +  \frac{N-2}{\pi} g^2_8  \\
\dot{g_4} =&  \frac{1}{2\pi} g_4 g_5 + 
    \frac{N^2 - 1}{2\pi N^2} g_2 g_3 +    \frac{2(N-1)}{\pi N} g_8 g_9  \\
 \dot{g_5} =& \frac{N^2 -1}{2\pi N^2} g^2_2 + \frac{1}{2\pi} g^2_4 + \frac{2(N-1)}{\pi N} g^2_8   \\
\dot{g_6} =& \frac{N +1}{4\pi N} g^2_7 +  \frac{N - 1}{2\pi N} g^2_8 + \frac{N-1}{4\pi N} g^2_9  \\
\dot{g_7} =&  \frac{(N + 2)(N - 1)}{2\pi N} g_1 g_7 +  \frac{2}{\pi} g_6 g_7 
+    \frac{N-1}{4\pi} (2 g_2 g_8 + g_3 g_9) \\
\dot{g_8} =&  \frac{N + 1 }{4\pi} g_2 g_7 +  \frac{2}{\pi} g_6 g_8  
+  \frac{1}{2\pi} \left(g_4 g_9 + g_5 g_8 \right) \\
& +  \frac{(N-2)(N+1)}{4\pi N} \left(2 g_1 g_8 + g_2 g_9 + g_3 g_8 \right) \\
\dot{g_9} =&  \frac{N + 1 }{4\pi} g_3 g_7 +  \frac{1}{\pi} (g_4 g_8 + 2 g_6 g_9 ) \\
&+  \frac{(N-2)(N+1)}{2\pi N} \left( g_1 g_9 + g_2 g_8 \right),
\end{split}
\label{RG}
\end{equation}
where ${\dot g}_i = \partial g_i/ \partial l (i =1,\ldots, 9)$ with $l$ being the RG time. 
First, we note that the RG flow of these equations is drastically different for $N=2$ and $N>2$ as
we observe, from Eqs. (\ref{RG}), that some terms vanish in the special $N=2$ case. 
In the latter case, the RG analysis has been done in detail already in 
Refs. \onlinecite{Nonne-B-C-L-10,Nonne-B-C-L-11}, where the phase diagram
of the generalized Hund and $g\text{-}e$ cold fermions have been mapped out.
We thus assume $N>2$ hereafter and, for completeness, we will also determine the 
phase diagram of the half-filled $p$-band model \eqref{eqn:p-band-simple} for $N=2$ 
(see Appendix \ref{sec:N2-p-band-continuum}).

The next step is to solve the RG equations \eqref{RG} numerically using the Runge-Kutta procedure.
For the initial conditions  (\ref{couplings}, \ref{couplingsGorsh})  corresponding to the different lattice models of Sec.~\ref{sec:models-strong-coupling}, 
the numerical analysis reveals the existence of the two very different regimes that we will  now investigate 
carefully below.
%%%%%%%%%%%%%%%%%%%%%%%%%%%%%%%%%%%%%%%%%%%%%%%%%%%%
\subsubsection{Phases with dynamical symmetry enlargement}
\label{sec:Phases_with_dynamical_symmetry_enlargement}
%%%%%%%%%%%%%%%%%%%%%%%%%%%%%%%%%%%%%%%%%%%%%%%%%%%%
One striking feature of 1D interacting Dirac fermions is that when the interaction is marginally 
relevant, a dynamical symmetry enlargement (DSE)\cite{Lin-B-F-98,Boulat-A-L-09,Konik-S-L-02} 
emerges very often in the far IR.  
Such DSE corresponds to the situation where the Hamiltonian is attracted
under an RG flow to a manifold possessing a symmetry higher than that of 
the original field theory. Most of DSEs have been discussed within the one-loop RG 
approach. Among those examples is the emergence of SO(8) symmetry in the low-energy description
of the half-filled two-leg Hubbard model \cite{Lin-B-F-98,Chen-C-L-C-M-04} 
and the SU(4) half-filled Hubbard chain model. \cite{Assaraf-A-B-C-L-04}

It is convenient to introduce the following rescaling of the coupling constants to identify the possible
DSEs compatible with the one-loop RG Eqs. (\ref{RG}): 
\begin{equation}
\begin{split}
& f_{1,7,8,9} = \frac{N}{\pi} g_{1,7,8,9}  \, , \;\;   f_{2,3} = \frac{N}{2 \pi} g_{2,3} \\
&  f_{4,5} = \frac{N^2}{2 \pi} g_{4,5}  \, , \;\;  f_{6} = \frac{2 N^2}{\pi} g_{6} . 
\end{split}
 \label{rescaling}
 \end{equation}
One then observes that along a special direction of the flow (dubbed `ray'\footnote{%
The simplest example of such rays is the separatrices in the Kosterlitz-Thouless RG flow.}) where $f_i = f$,
all the nine one-loop RG equations (\ref{RG}) reduces to a single equation:
\begin{equation}
\dot{f} = \frac{2N-1}{N} f^2 \; . 
\end{equation}
This signals the emergence of an SO($4N$) symmetry
which is the maximal continuous symmetry enjoyed by $2N$ Dirac fermions,
i.e., $4N$ Majorana (real) fermions. To see this, one notes that 
along this special ray, model (\ref{lowenergyham}) 
reduces to the SO($4N$)  Gross-Neveu (GN) model: \cite{Gross-N-74} 
\begin{equation}
\begin{split}
{\cal H}_{\text{GN}} =& -i v_\text{F}\left(R_{m \alpha} ^\dag \partial_x R_{m \alpha} ^{\phantom \dag} - 
  L_{m \alpha}^\dag \partial_x L_{m \alpha}^{\phantom \dag}\right) \\
  &+ \frac{\pi f}{2N} \left( L_{m \alpha} ^{\dag}  R_{m \alpha} - \text{H.c.} \right)^2,
  \end{split}
\label{GN}
\end{equation}
where the SO($4N$)  symmetry 
stems from the decomposition of Dirac fermions into Majorana fermions: $L_{m \alpha} = 
\xi_{m \alpha} + i \chi_{m \alpha}$.
The  GN model (\ref{GN}) is a massive integrable field theory when $f>0$ 
whose mass spectrum is known exactly \cite{Zamolodchikov-Z-79,Karowski-T-81}.  

The numerical integration of RG Eqs.~(\ref{RG}) revealed that for some set of initial conditions, 
the coupling constants flow along the highly-symmetric ray where $f_i = f >0$ in the far IR 
(see Sec.~\ref{sec:RG-phase-diag}).
The model is then equivalent to the SO($4N$)  GN model  and a non-perturbative spectral
gap is generated. The development of this strong-coupling regime in the SO($4N$) GN model
signals the formation of a SP phase for all $N \ge 2$ with 
the order parameter: 
\begin{equation}
 {\cal O}_{\rm SP}  = 
 i \left( L_{m \alpha}^\dagger R_{m \alpha} - H.c.\right),
\label{SP}
\end{equation}
which is the continuum limit of the SP operator on a lattice
\begin{equation}
{\cal O}_{\rm SP}(i) = (-1)^i\sum_{m \alpha}c^\dagger_{m \alpha,i+1}c_{m \alpha,i} \; .
\label{SP-lattice}
\end{equation}
Since the interacting part of the GN model (\ref{GN}) can be written directly in terms of 
${\cal O}_{\rm SP}$:
${\cal H}^{\rm int}_{\text{GN}} = - \pi f {\cal O}_{\rm SP}^2 /(2N)$, we may conclude 
that $ \langle {\cal O}_{\rm SP} \rangle \ne 0$ in
the ground state  for $f>0$, i.e., the emergence of a dimerized phase.
The latter is two-fold degenerate and breaks spontaneously the one-step
translation symmetry: 
\begin{equation}
T_{a_0} : \quad 
L_{m \alpha} \to -i L_{m \alpha}\, ,  \;\; R_{m \alpha} \to i R_{m \alpha}  ,
\end{equation}
since ${\cal O}_{\text{SP}} \to - {\cal O}_{\text{SP}}$ under  $T_{a_0}$.    
It turns out that the SU$(2N)$ line ($J=J_{z}=0$) with $U >0$ of the generalized Hund model \eqref{alkaourmodel} 
is described by the $f_i = f >0$  manifold with an SO($4N$)  DSE.  This is in full agreement with the fact 
that the repulsive SU($2N$) Hubbard model  
for $N \ge 2$ displays a SP phase at half filling.\cite{Nonne-L-C-R-B-11}

On top of this phase, we can define other DSE phases with global SO($4N$) symmetry.
These phases are described by RG  trajectories along the rays 
$f_i = \epsilon_i f $ ($\epsilon_i = \pm 1$) in the long-distance limit. 
The physical properties of these phases are related to 
those of the  SO($4N$) GN model up to some duality symmetries on the  Dirac fermions. \cite{Boulat-A-L-09}
These duality symmetries can be determined using the symmetries of the RG Eqs. (\ref{RG}):
\begin{subequations}
\begin{align}
 & \Omega_1: f_{7,8,9} \rightarrow  - f_{7,8,9} 
 \label{eq:Omega_1} \\
 & \Omega_2 : f_{2,4,8}   \rightarrow - f_{2,4,8} 
 \label{eq:Omega_2} \\
 & \Omega_3 (=   \Omega_1  \Omega_2): f_{2,4,7,9}   \rightarrow - f_{2,4,7,9} ,
 \label{eq:Omega_3}
\end{align}
\end{subequations} 
which are indeed symmetries of Eqs. (\ref{RG}) in the general $N$ case.
Using the definitions \eqref{defalkacurrents}, \eqref{umklappop}, and \eqref{lowenergyham}, 
one can represent these duality symmetries simply in terms of the Dirac fermions:
\begin{equation}
\begin{split}
& \Omega_1: 
L_{m \alpha}
\rightarrow  i L_{m \alpha}  \\
 & \Omega_2 :
 L_{m \alpha}  \rightarrow  \left(-1\right)^{m} i L_{m \alpha} \\
& \Omega_3:
 L_{m \alpha}  \rightarrow   \left(-1\right)^{m+1}  L_{m \alpha} ,
 \end{split}
 \label{dualitiesDirac}
 \end{equation}
while the right fermions remain invariant. These transformations are automorphisms of the different
current algebra in Eq.  (\ref{defalkacurrents}). \cite{Boulat-A-L-09}

Starting from the gapful SP phase found above, one can deduce the three other insulating
phases by exploiting the duality symmetries \eqref{dualitiesDirac}:
\begin{equation}
\begin{split}
& {\cal O}_{\rm SP}  \xrightarrow{\Omega_1} {\cal O}_{\rm CDW} \equiv 
L_{m \alpha}^\dagger R_{m \alpha}  +  \text{H.c.}  \\
& {\cal O}_{\rm SP}   \xrightarrow{\Omega_2} {\cal O}_{\rm ODW}  \equiv
\sum_m  \left(- 1\right)^{m} L_{m \alpha}^\dagger R_{m \alpha}  +  \text{H.c.} \\
&  {\cal O}_{\rm SP}  \xrightarrow{\Omega_3} {\cal O}_{\rm SP_{\pi}}  \equiv
   \sum_m  \left(-1 \right)^{m} i  \left(  L_{m \alpha}^\dagger R_{m \alpha}  -  \text{H.c.} \right) \; .
\end{split}
 \label{dualityorderparameters}
 \end{equation}
 Using \eqref{contlimitDirac}, one can identify the lattice order parameters corresponding 
 to these operators as:
\begin{equation}
\begin{split}
& {\cal O}_{\rm CDW}(i) = (-1)^i n_i
 \\
& {\cal O}_{\rm ODW}(i) =
 (-1)^i\sum_{m}   \left(- 1\right)^{m}  c^\dagger_{m \alpha,i}c_{m \alpha,i} 
 \\
&  {\cal O}_{\rm SP_{\pi}}(i) =  (-1)^i\sum_{m}   \left(- 1\right)^{m} c^\dagger_{m \alpha,i+1}c_{m \alpha,i} ,
\end{split}
 \label{latticedualityorderparameters}
\end{equation}
which describe respectively a CDW, an orbital-density wave (ODW), 
and an alternating SP phase (SP$_{\pi}$).  For instance, by using $\Omega_{1}$, 
one can immediately conclude that 
on the SU$(2N)$ line ($J=J_{z}=0$) with $U <0$, the generalized Hund model is in a CDW phase 
$ \langle {\cal O}_{\rm CDW} \rangle \ne 0$ exhibiting the SO($4N$) DSE.   
This is fully consistent with the known result that the {\em attractive} SU($2N$) Hubbard model 
for $N \ge 2$ displays a CDW phase at half filling.\cite{Zhao-U-W-06,Zhao-U-W-07,Nonne-L-C-R-B-11}

In summary, in the first regime of the RG flow characterized by DSE, 
we found four possible Mott-insulating phases which 
are two-fold degenerate and spontaneously break the one-site translation symmetry. 
The RG approach developed here tells that each of these four phases is characterized by one of 
the four SO($4N$)-symmetric DSE rays related to each other by the duality symmetries $\Omega_{1,2,3}$.  
%%%%%%%%%%%%%%%%%%%%%%%%%%%%%%%%%%%%%%%%%%%%%%%%%%%%%
\subsubsection{Non-degenerate Mott insulating phases}
\label{sec:Non-degenerate_Mott_insulating_phases}
%%%%%%%%%%%%%%%%%%%%%%%%%%%%%%%%%%%%%%%%%%%%%%%%%%%%%
In the second regime, the RG flow displays no symmetry enlargement, 
and we can no longer use any duality symmetry
to relate the underlying insulating phases to a single phase ({\em e.g.} the SP phase in the above). 
Indeed, in stark contrast, 
the numerical solution of the one-loop RG equations \eqref{RG} for $N > 2$ 
reveals that the coupling constant $g_1$ in the low-energy effective
Hamiltonian (\ref{lowenergyham}) reaches the strong-coupling regime
before the other coupling constants such as $g_{2,4,5,8}$.  
Since the operator corresponding to $g_1$ depends only on the nuclear spin
degrees of freedom,  one expects  a separation of the energy scales in this second region of the RG flow.
Neglecting all the other couplings for the moment, the resulting perturbation corresponds
to an SU($N$)$_2$ CFT perturbed by a marginally relevant current-current interaction $g_1 >0$.
This model is an integrable massive field theory  \cite{Ahn-B-L-90,Babichenko-04} and a
 spin gap $\Delta_{\text{s}}$ thus opens  for the SU($N$) (nuclear) spin sector in this regime.
The next task is to integrate out these (nuclear) spin degrees of freedom to derive
an effective Hamiltonian for the remaining degrees of freedom 
in the low-energy limit $E \ll \Delta_{\text{s}}$ from which the physical
properties of  the second regime of the RG approach will be determined. 

%%%%%%%%%%%%%%%%%%%%%%%%%%%%%%%%%%%%%%%%%%%%%%
\vspace{0.5cm}
\paragraph{SU(2)$_\text{o}$ symmetric case.}
\label{sec:SU2o_symmetry}
%%%%%%%%%%%%%%%%%%%%%%%%%%%%%%%%%%%%%%%%%%%%%%

Let us first consider the SU(2)$_{\text{o}}$ symmetric case to derive the low-energy limit 
$E \ll \Delta_\text{s}$. In this case,
the model (\ref{lowenergyham}) simplifies as:
\begin{equation}
\begin{split}
\mathcal{H}^{\text{SU(2)}_{\text{o}}}_{\text{int}} =&
g_1 J_\text{L}^a J_\text{R}^a  +
g_2  \mathcal{J}_\text{L}^{a,i} \mathcal{J}_\text{R}^{a,i} 
 + g_4 \, \mathbf{j}_\text{L} \cdot  \mathbf{j}_\text{R} 
\\
& + g_6 J_{\text{c L}} J_{\text{c R}} 
    + g_7 \left(S^{\alpha \beta +}_{\text{L}}  S^{\alpha \beta -}_{\text{R}} +\textrm{H.c.}\right) 
 \\
& + \frac{g_8}{2} \left[  A^{\alpha \beta +}_{mn \text{L}}  A^{\alpha \beta -}_{mn \text{R}} 
  +\textrm{H.c.} \right] ,
\end{split}
  \label{lowenergyhamsu2}
\end{equation}
since $g_2=g_3, g_4= g_5$ and $g_8 = g_9$ as a consequence of the SU(2)$_\text{o}$-symmetry.
At this point, we need to express all operators appearing in Eq. (\ref{lowenergyhamsu2}) in 
the U(1)$_\text{c} \times$ SU(2)$_N \times$ SU($N$)$_2$ basis.	
To this end, we will use the so-called non-Abelian bosonization \cite{Knizhnik-Z-84,Affleck-NP86}:
\begin{equation}
\begin{split}
& L^{\dagger}_{m \alpha} R_{n \beta} \simeq  \exp \left( i \sqrt{2 \pi /N} \Phi_\text{c} \right) g_{nm} G_{\beta \alpha} ,
 \\
& R^{\dagger}_{m \alpha} L_{n \beta} \simeq  \exp \left( - i \sqrt{2 \pi /N} \Phi_\text{c} \right) g^{\dagger}_{mn} 
G^{\dagger}_{\alpha \beta} ,
\end{split}
\label{nonabelboso}
\end{equation}
where the charge field  $\Phi_{\text{c}} $ is a compactified bosonic field with radius 
$R_\text{c} =  \sqrt{N/2\pi} $: $ \Phi_{\text{c}}  \sim \Phi_{\text{c}} +   \sqrt{2 \pi N}$.
This field describes the low-energy properties of the 
charge degrees of freedom.  In Eq. \eqref{nonabelboso}, $g$ (respectively $G$)
is the SU(2)$_N$ (respectively  SU($N$)$_2$) primary field with spin-1/2
(respectively which transforms in the fundamental representation of SU($N$)).
The scaling dimensions of these fields are given as 
\begin{equation}
\Delta_{g}= \frac{3}{N+2} \, , \;\; \Delta_{G} = \frac{N^{2}-1}{N(N+2)} 
\end{equation}
(see Appendix \ref{sec:CFT-data}) 
so that Eq. (\ref{nonabelboso}) is satisfied at the level of the scaling dimension:
$1 = 1/2N + 3/(N+2) + (N^2-1)/N(N+2)$. 

By the correspondence (\ref{nonabelboso}), 
the different operators of the  low-energy effective Hamiltonian \eqref{lowenergyhamsu2} can
then be expressed in terms of the U(1)$_\text{c} \times$ SU(2)$_N \times$ SU($N$)$_2$ basis.
Let us first find the decomposition of $\mathcal{J}_\text{L}^{a,i} \mathcal{J}_\text{R}^{a,i}$ 
of Eq. (\ref{lowenergyhamsu2}).
Using the SU($N$) identity 
\begin{equation}
 \sum_a T^a_{\alpha \beta} T^a_{\gamma \rho} = 
\frac{1}{2} \left(\delta_{\alpha \rho} \delta_{\beta \gamma}
- \frac{1}{N}\; \delta_{\alpha \beta} \delta_{\gamma \rho} \right),
\label{SUNident}
\end{equation}
and ${\vec \sigma}_{m n} \cdot {\vec \sigma}_{p q} = 
2  \left(\delta_{m q} \delta_{n p}
- \frac{1}{2}\; \delta_{m n} \delta_{p q} \right)$, we obtain:
\begin{equation}
\begin{split}
\mathcal{J}_\text{L}^{a,i} \mathcal{J}_\text{R}^{a,i} = &
- \frac{1}{2} L^{\dagger}_{l \alpha} R_{l \alpha} R^{\dagger}_{m \beta} L_{m \beta}
+ \frac{1}{4} L^{\dagger}_{l \alpha} R_{m \alpha} R^{\dagger}_{m \beta} L_{l \beta}
\\
 &+ \frac{1}{2N} L^{\dagger}_{l \alpha} R_{l \beta} R^{\dagger}_{m \beta} L_{m\alpha}
 -  \frac{1}{4N} L^{\dagger}_{l \alpha} R_{m \beta} R^{\dagger}_{m \beta} L_{l \alpha} .
 \end{split}
 \label{g2term}
\end{equation}
Using Eq. (\ref{nonabelboso}), we get:
\begin{equation}
\begin{split}
\mathcal{J}_\text{L}^{a,i} \mathcal{J}_\text{R}^{a,i} = &
-  \frac{1}{2} \left[ {\rm Tr} \left( g \right) {\rm Tr} \left( g^{\dagger} \right) 
-   \frac{1}{2} g_{mn}  g^{\dagger}_{mn}   \right]  \\
& \times \left[
{\rm Tr} \left( G^{\dagger} \right) {\rm Tr} \left( G \right) 
- G_{\beta \alpha}  G^{\dagger}_{\beta \alpha} /N \right] . 
\end{split}
\label{g2termbis}
\end{equation}

Now we use the expression of the trace of the SU(2)$_N$ primary field 
 which transforms in the spin-1 representation that we have derived in Appendix \ref{sec:CFT-data}
 [Eq. \eqref{adjointSU2App}] and a similar one for  the SU($N$)$_2$ primary field in the adjoint
 representation of SU($N$):
 \begin{equation}
{\rm Tr} \left(  \Phi^{\text{SU($N$)}_2}_{\text{adj}} \right) =  {\rm Tr} \left( G^{\dagger} \right) {\rm Tr} \left( G \right) 
-  \frac{1}{N} G_{\beta \alpha}  G^{\dagger}_{\beta \alpha} ,
\label{adjointSUN}
\end{equation}
 so that Eq. (\ref{g2term}) simplifies as follows:
\begin{equation}
\mathcal{J}_\text{L}^{a,i} \mathcal{J}_\text{R}^{a,i} \sim  
-  {\rm Tr} \left(  \Phi^{\text{SU(2)}_N}_{j=1} \right) 
{\rm Tr} \left(  \Phi^{\text{SU($N$)}_2}_{\text{adj}} \right) .
\label{g2decomp}
\end{equation}
The expression of the operator $S^{\alpha \beta +}_{\text{L}}  S^{\alpha \beta -}_{\text{R}}$ in 
Eq. (\ref{lowenergyhamsu2}) in the U(1)$_\text{c} \times$ SU(2)$_N \times$ SU($N$)$_2$ 
basis can be obtained by observing that  $S^{\alpha \beta +}_{\text{L}} $ is symmetric 
with respect to the exchange $ \alpha \leftrightarrow  \beta$ and a singlet under the SU(2) orbital. 
The decomposition will then involve the SU($N$)$_2$ primary field
in the symmetric representation of SU($N$) with dimension $N(N+1)/2$:
\begin{equation}
S^{\alpha \beta +}_{\text{L}}  S^{\alpha \beta -}_{\text{R}} \sim \exp \left(i \sqrt{8 \pi /N} \Phi_\text{c} \right) 
{\rm Tr} \left(  \Phi^{\text{SU($N$)}_2}_{\text{s}} \right) .
\label{g7decomp}
\end{equation}
Finally, the last operator in Eq. (\ref{lowenergyhamsu2}) is symmetric under the SU(2) orbital
symmetry and antisymmetric with respect to the exchange $ \alpha \leftrightarrow  \beta$ of SU($N$).
Therefore, it will involve the spin 1 operator $\Phi^{\text{SU(2)}_N}_{j=1}$ and SU($N$)$_2$ primary field
in the antisymmetric representation of SU($N$) with dimension $N(N-1)/2$:
\begin{eqnarray}
 A^{\alpha \beta +}_{mn L}  A^{\alpha \beta -}_{mn R} 
 \sim 
 e^{ i \sqrt{8 \pi /N} \Phi_\text{c} } {\rm Tr} \left(  \Phi^{\text{SU(2)}_N}_{j=1} \right) 
{\rm Tr} \left(  \Phi^{\text{SU($N$)}_2}_{\text{a}} \right).
\label{g8decomp}
\end{eqnarray}

In the low-energy limit $E \ll \Delta_\text{s}$, we can average the SU($N$) degrees of freedom
in the decompositions \eqref{g2decomp}, \eqref{g7decomp},  and \eqref{g8decomp} to get 
the effective interacting Hamiltonian which controls the physics in the second region of the RG analysis:
 \begin{equation}
 \begin{split}
\mathcal{H}^{\text{SU(2)}_\text{o}}_{\text{eff}} =&
    \lambda_2  \text{Tr} \left(  \Phi^{\text{SU(2)}_N}_{j=1} \right) 
    + g_4 \, \mathbf{j}_{\text{L}} \cdot  \mathbf{j}_{\text{R}}  \\
    &+  \frac{2N g_6}{\pi} \;  \partial_x  \Phi_{\text{c L}}  \partial_x  \Phi_{\text{c R}}
     + \lambda_7 \cos \left(\sqrt{8 \pi /N} \Phi_\text{c} \right)   \\
  &
  + \lambda_8  {\rm Tr} \left(  \Phi^{\text{SU(2)}_N}_{j=1} \right)  \cos \left(\sqrt{8 \pi /N} \Phi_\text{c} \right) ,
\end{split}
  \label{effhamregion3}
\end{equation}
where we have used the bosonized description of the chiral charge currents:
$J_{\text{c L,R}} =  \sqrt{2N/ \pi}  \; \partial_x  \Phi_{\text{c L,R}}$.
In Eq. (\ref{effhamregion3}), the coefficients are phenomenological since
they involve the form factors of the SU($N$) operators in the  integrable model
with SU($N$)$_2$ current-current interaction which are not known to the best of our knowledge: 
$  \lambda_2 \simeq - 2 g_2 \left\langle {\rm Tr} \left(  \Phi^{\text{SU($N$)}_2}_{\rm adj} \right)\right\rangle$, 
$ \lambda_{7,8}  \simeq g_{7,8} 
\left\langle \text{Tr} \left(  \Phi^{\text{SU($N$)}_2}_{\rm S,A} \right) + \text{H.c.} \right\rangle$.
We assume, in the following, that the expectation values of the SU($N$)$_2$ operators are positive.
We can safely neglect the last term ($\lambda_{8}$) in Eq. (\ref{effhamregion3}) which is less relevant 
than the perturbations with $ \lambda_2$ and $ \lambda_7$ to obtain the following residual interaction
for the charge and the orbital sectors:
\begin{equation}
\begin{split}
  \mathcal{H}^{\text{\text{SU(2)}}_\text{o}}_{\rm eff}  =&
    \lambda_2  {\rm Tr} \left(  \Phi^{\text{SU(2)}_N}_{\rm j=1} \right) 
    + g_4 \, \mathbf{j}_{\text{L}} \cdot  \mathbf{j}_{\text{R}}  \\
    &+ \lambda_7 \cos \left(\sqrt{8 \pi K_\text{c} /N} \Phi_\text{c} \right),
\end{split}
\label{effhamregion3fin}
\end{equation}
where the charge Luttinger parameter $K_\text{c}$ satisfies 
\begin{equation}
K_\text{c} =  \frac{1}{\sqrt{1 + 2 N g_6/\pi v_\text{F}}} < 1,
\label{Luttingerpara}
\end{equation}
since $g_6 >0$ from the numerical solution of the RG flow in the second region.

Therefore, for the energy scale lower than the gap $\Delta_{\text{s}}$ in the nuclear-spin sector, 
the effective Hamiltonian for the charge degrees of freedom is the well-known 
sine-Gordon model at $\beta_{\text{c}}^2  =  8 \pi K_\text{c} /N$.  The model is known to develop 
a charge gap $\Delta_\text{c}$  for all $N$ satisfying $K_\text{c} < N$,  
which is always the case as far as the weak-coupling expression  \eqref{Luttingerpara} is valid.  
The development of the strong-coupling regime of the sine-Gordon 
model is accompanied by the pinning of the charged field on either of the two minima:
\begin{equation}
\langle \Phi_{\text{c}} \rangle = \sqrt{\frac{N\pi}{8 K_\text{c}}} + p \sqrt{N \pi/2K_\text{c}} \quad 
(p=0,1) \; , 
\label{pinningcharge}
\end{equation}
since $\lambda_7  > 0$ in the second region of the RG flow. 

For energy smaller than the charge gap $\Delta_\text{c}$, the effective interaction (\ref{effhamregion3fin})
governing the fate of the orbital degrees of freedom simplifies as follows:
 \begin{eqnarray}
   \mathcal{H}^{\text{SU(2)}_\text{o}}_{\text{eff}} =
    \lambda_2  {\rm Tr} \left(  \Phi^{\text{SU(2)}_N}_{\rm j=1} \right) 
    + g_4 \, \mathbf{j}_{\text{L}} \cdot  \mathbf{j}_{\text{R}},
  \label{effhamregion3orbital}
\end{eqnarray}
which is nothing but the low-energy
theory of the spin-$N/2$ SU(2) Heisenberg chain derived by Affleck and Haldane
in Ref. \onlinecite{Affleck-H-87}.  
This is quite natural in view of the strong-coupling effective Hamiltonian \eqref{eqn:eff-Ham-orbital-Haldane} 
obtained in Sec. \ref{sec:strong-coupling}.   

The nature of the ground state of this Hamiltonian can be inferred
from a simple semiclassical approach. The operator with the coupling constant $\lambda_{2}$ 
in Eq. \eqref{effhamregion3orbital} has the scaling dimension $4/(N+2)$ and 
is strongly relevant. By using Eq. \eqref{adjointSU2App},  
the minimization of that operator in the second regime of the RG flow with $\lambda_2 >0$ (since $g_2 <0$)
gives the condition $\text{Tr} \,g =0$, $g$ being an SU(2) matrix. We have thus
$g = i \boldsymbol{\sigma} \cdot \boln$, with $\boln$ being an unit vector. 
From Eq. (\ref{nonabelboso}), one may expect that the `dimerization' operator 
for the orbital pseudo spin 
  $\mathbf{T}_i = c^{\dagger}_{ m \alpha,\,i} \boldsymbol{\sigma}_{m n} c_{ n \alpha,\,i}/2$ 
would be related, when $E \ll \Delta_{\text{c}}$, to $g$ as
 \begin{equation}
 (-1)^{i} \bolT_{i+1} {\cdot} \mathbf{T}_i \sim  \text{Tr}\, g .
 \label{orbitaldim}
 \end{equation}
Therefore, the ground state is not dimerized when $\lambda_2 >0$. The nature of the phase can be determined by 
 exploiting  the result of  Affleck and Haldane in Ref. \onlinecite{Affleck-H-87} that 
 model (\ref{effhamregion3orbital}) with  $g = i \boldsymbol{\sigma} \cdot \boln$
 is the non-linear sigma model with the topological angle $\theta = \pi N$. 
 Since $N$ is even in our cold fermion problem, the topological term is trivial 
and the resulting model is then equivalent to the non-linear sigma model 
which is a massive field theory in $(1+1)$-dimensions. \cite{Zamolodchikov-Z-79} 
As is well-known, the latter model describes
the physics of integer-spin Heisenberg chain in the large-spin limit \cite{Haldane-PLA-83}.

To summarize, in the $\text{SU(2)}_\text{o}$ symmetric case, the second region 
of the RG flow describes the emergence of 
a non-degenerate gapful phase with no  CDW or SP ordering.  Such phase is 
an Haldane phase for the orbital pseudo spin $\mathbf{T}$, i.e., the OH phase that
we found in the strong-coupling investigation for all even $N$ (see Sec.~\ref{sec:strong-coupling}).
The resulting OH phase
exhibits an hidden ordering which is revealed by a non-local string order
parameter. On top of this hidden ordering, the OH phase  has edge state
with pseudo spin $T_{\text{edge}}= N/4$.   
According to Ref. \onlinecite{Pollmann-B-T-O-12}, this is a SPT phase
when $N/2$ is odd.

%%%%%%%%%%%%%%%%%%%%%%%%%%%%%%%%%%%%%%%%%%%%%%
\vspace{0.5cm}
\paragraph{U(1)$_\text{o}$ symmetric case.}
\label{sec:U1o_symmetry}
%%%%%%%%%%%%%%%%%%%%%%%%%%%%%%%%%%%%%%%%%%%%%%

We now investigate the nature of the RG flow in the second regime in the generic
case $J \neq J_{z}$ with an U(1)$_\text{o}$ symmetry. 
For energy $E \ll \Delta_\text{c}$, the interacting part \eqref{effhamregion3orbital} of the effective Hamiltonian 
for the orbital sector now takes the following anisotropic form:
\begin{equation}
\begin{split}
 \mathcal{H}^{\text{U(1)}_\text{o}}_{\rm eff}  =&
\lambda_{2\|} \left(  \Phi^{1}_{1, 1} + \Phi^{1}_{-1, -1} \right) +
\lambda_{2\perp} \Phi^{1}_{0, 0} \\ 
 &+  \frac{g_{4\perp}}{2}\left(j_{\text{L}}^+j_{\text{R}}^-+\textrm{H.c.}\right) +g_{4\|} j_{\text{L}}^zj_{\text{R}}^z, 
 \end{split}
 \label{HamU1}
\end{equation}
where the   SU(2)$_N$  primary operators with spin $j = 0,\ldots, N/2$  are denoted by 
$\Phi^{j}_{m, \bar m}$  ($-j \le m, \bar m \le j$) with scaling dimension $d_j = 2 j (j+1)/(N+2)$ 
(see Appendix \ref{sec:CFT-data}).

The low-energy properties of model (\ref{HamU1}) can then be determined by introducing 
${\mathbb{Z}}_N$ parafermion degrees of freedom and relating the fields of the SU(2)$_N$ CFT to those of 
the U(1)$_\text{o}$ CFT.  Such a mapping is realized by the conformal embedding:
${\mathbb{Z}}_N$ $\sim$  SU(2)$_N$ / U(1)$_\text{o}$, 
which defines the series of the $\mathbb{Z}_N$ parafermionic CFTs 
with central charge $c= 2 (N -1)/(N+2)$. \cite{Zamolodchikov-F-JETP-85,Gepner-Q-87}  
These CFTs describe the critical properties of 
two-dimensional ${\mathbb{Z}}_N$ generalizations of the Ising model,\cite{Zamolodchikov-F-JETP-85}  
where the lattice spin $\sigma_r$ takes values:
$e^{i 2 \pi m /N }, m=0, \ldots, N-1$ and the corresponding generalized Ising lattice Hamiltonian 
is ${\mathbb{Z}}_N$ invariant. In the scaling limit, the conformal fields $\sigma_k$ 
with scaling dimensions $\Delta_k = k(N-k)/N(N+2)$ $(k=1, \ldots, N -1)$ 
describe the long-distance correlations of $\sigma^{k}_r$ at the critical point. \cite{Zamolodchikov-F-JETP-85}
In the context of cold atoms,
the  ${\mathbb{Z}}_N$ CFT is also very useful to map out  the zero-temperature phase diagram
of general 1D higher-spin cold fermions. \cite{Nonne-L-C-R-B-11,Lecheminant-B-A-05,Lecheminant-A-B-08} 

The orbital SU(2)$_N$ currents can be directly expressed in terms of 
the first parafermionic current $\Psi_{1 L,R}$ with scaling dimension $1 - 1/N$ and a 
bosonic field $\Phi_{o}$ which accounts for orbital  fluctuations: \cite{Zamolodchikov-F-JETP-85}
\begin{eqnarray}
 j_{\text{L,R}}^{\dagger} 
&\simeq& \frac{\sqrt{N}}{2\pi}
:\exp \left(\pm i \sqrt{8 \pi/N} \; \Phi_{\text{o L,R}} \right): \Psi_{1 \text{L,R}}
\nonumber \\
 j_{\text{L,R}}^{z} 
&\simeq& \sqrt{\frac{N}{2\pi}} \partial_x \Phi_{\text{o L,R}} ,
\label{parasu2Ncurrent}
\end{eqnarray}
where the orbital bosonic field  $\Phi_{\text{o}} = \Phi_{\text{o L}} + \Phi_{\text{o R}}$ 
is a compactified bosonic field with radius 
$R_\text{o} =  \sqrt{N/2\pi} $: $ \Phi_{\text{o}}  \sim \Phi_{\text{o}} +   \sqrt{2 \pi N}$. 
Under the ${\mathbb{Z}}_N$ symmetry, the parafermionic currents $\Psi_{1 \text{L,R}}$ transform as 
\cite{Zamolodchikov-F-JETP-85}
\begin{eqnarray}
\Psi_{1 \text{L,R}} &\rightarrow& e^{i 2 \pi k /N} \Psi_{1 \text{L,R}} , 
\label{chargepara}
\end{eqnarray}
with $k=0,\ldots, N-1$.
Using Eq. (\ref{parasu2Ncurrent}), we identify the ${\mathbb{Z}}_N$ symmetry of the parafermions directly
 on the Dirac fermions through:
\begin{equation}
 L_{g\alpha} \to  e^{- i \pi k/N} L_{g\alpha}, \;  \;  \;  L_{e\alpha} \to  e^{i \pi k/N} L_{e\alpha},
\label{ZN}
\end{equation}
with  a similar transformation for the right-moving Dirac fermions. It is easy to check that
the low-energy description (\ref{lowenergyham}) is invariant under this transformation, 
and thus ${\mathbb{Z}}_N$-symmetric.  
Using the definition (\ref{contlimitDirac}), one can deduce a lattice representation of this ${\mathbb{Z}}_N$
in terms of the original fermions $c_{m \alpha,i}$:
\begin{equation}
 c_{g\alpha} \to  e^{- i \pi k/N} c_{g\alpha}, \;  \;  \;  c_{e \alpha} \to  e^{i \pi k/N} c_{e \alpha} ,
\label{ZNlattice}
\end{equation}
which is indeed a symmetry of all lattice models introduced in Sec. \ref{sec:models-strong-coupling}.

As described in the Appendix, the  SU(2)$_N$  primary operators 
can be related to that of the ${\mathbb{Z}}_N$  CFT.
Using the results (\ref{identiprimZnApp}) and (\ref{spin1primaryparaApp}) of Appendix \ref{sec:CFT-data} and 
Eq. (\ref{parasu2Ncurrent}),
the low-energy effective Hamiltonian (\ref{HamU1}) can then be expressed
in terms of ${\mathbb{Z}}_N$  primary fields as follows:
\begin{equation}
\begin{split}
\mathcal{H}^{\text{U(1)}_\text{o}}_{\rm eff}  = &
\lambda_{2\|} \left\{ \mu_2 \exp \left(-i \sqrt{8 \pi/N} \; \Phi_{\text{o}} \right)
 + \textrm{H.c.} \right\} -
\lambda_{2\perp}  \epsilon_1  \\
 &+ \frac{g_{4\perp} N}{2 \pi}  \left\{  \Psi_{\text{1 L}}  \Psi^{\dagger}_{\text{1 R}}
 \exp \left(i \sqrt{8 \pi/N} \; \Phi_{\text{o}} \right) + \textrm{H.c.} \right\}   \\
 &+  \frac{N g_{4\|}}{2 \pi} \partial_x \Phi_{\text{oL}} \partial_x \Phi_{\text{oR}},
 \end{split}
 \label{effectiveHamU1para}
\end{equation}
where $\epsilon_1$ (respectively $\mu_2$)   
is the thermal (respectively second disorder) operator of the ${\mathbb{Z}}_N$  CFT  with scaling dimension 
$4/(N+2)$ (respectively $(N-2)/N(N+2)$).
In our convention, $\langle \epsilon_1 \rangle > 0 $ in a phase where the  ${\mathbb{Z}}_N$-symmetry 
is broken so that
the disorder parameters do not condense $\langle \mu_k\rangle=0$ ($k=1,\ldots, N-1$), 
as they are dual to the order fields $\sigma_k$.  
Since the second disorder and the thermal operators themselves are known to be ${\mathbb{Z}}_N$-invariant,
the model (\ref{effectiveHamU1para}) is invariant under the ${\mathbb{Z}}_N$-symmetry as it should be.

The low-energy effective field theory (\ref{effectiveHamU1para}) appears in such different contexts
as the field theory approach to the Haldane's conjecture \cite{Cabra-P-R-98} and the half-filled 1D general spin-$S$ cold fermions.\cite{Nonne-L-C-R-B-11} 
It was shown\cite{Nonne-L-C-R-B-11} that the phase diagram of the latter model strongly depends on the parity of $N$.
The numerical solution of the RG flow shows that the operator with the coupling constant $\lambda_{2\perp}$ dominates
the strong-coupling regime. Such perturbation describes an integrable deformation  
of the ${\mathbb{Z}}_N$ CFT\cite{Fateev-91} which is always a massive field theory 
for all sign of $\lambda_{2\perp}$; 
when $\lambda_{2\perp} >0$ (i.e. $g_3 < 0$), we have $\langle \epsilon_1 \rangle > 0 $ 
and the mass is generated from the spontaneous ${\mathbb{Z}}_N$-symmetry breaking and 
all the order fields of the ${\mathbb{Z}}_N$ CFT condense: $\langle \sigma_k \rangle \ne 0$, while the disorder one
$\langle \mu_k \rangle = 0$ for all $k=1,\ldots, N-1$. 

One can immediately see that the nature of the underlying
phase can be captured neither by the SP nor by the density-order parameters \eqref{SP-lattice} and 
\eqref{dualityorderparameters} since they are all invariant under the ${\mathbb{Z}}_N$ symmetry (\ref{ZN}).
In fact, by using the identifications (\ref{identiprimZnApp}), it is straightforward to check 
that these order parameters
involve the first disorder operator $\mu_1$ and therefore cannot sustain a long-range ordering in
the ${\mathbb{Z}}_N$-broken phase.
In this respect, the first regime, in which we have DSE, corresponds 
to a region where the ${\mathbb{Z}}_N$-symmetry is not broken spontaneously.

Since all the  parafermionic operators in \eqref{effectiveHamU1para} average to zero in 
the ${\mathbb{Z}}_N$ broken phase, one has to consider higher orders in perturbation theory
to derive an effective theory for the orbital bosonic field $\Phi_\text{o}$.
When $N$ is even, one needs the $N/2$-th order
of perturbation theory to cancel out the operator $\mu_2$ in Eq. (\ref{effectiveHamU1para}).
The resulting low-energy Hamiltonian then reads as follows: 
\begin{equation}
\begin{split}
\mathcal{H}^{\rm even}_{\text{o}} =& \frac{v_\text{o}}{2} \left\{ 
\frac{1}{K_\text{o}} \left(\partial_x \Phi_\text{o}\right)^{2}
+ K_\text{o} \left(\partial_x \Theta_\text{o}\right)^{2} \right\}  \\
&+ g_\text{o}  \cos \left(\sqrt{2 \pi N} \; \Phi_{\text{o}} \right) ,
\end{split}
\label{hoNeven}
\end{equation}
where $v_\text{o}$ and $K_\text{o}$ are the velocity and the Luttinger parameters 
for the orbital boson $\Phi_\text{o}$:
\begin{equation}
K_\text{o} =  \frac{1}{\sqrt{1 + N g_{4\|}/(2\pi v_\text{F}})}.
\label{Luttingerparaorbital}
\end{equation}
A naive estimate of the coupling
constant $g_\text{o}$ in higher orders of perturbation theory reads as: $g_\text{o} \sim - (-\lambda_{2\|})^{N/2}$.

The resulting low-energy Hamiltonian  (\ref{hoNeven}) which governs the physical properties of the orbital
sector takes the form of the sine-Gordon model at $\beta_{\text{o}}^2  =  2 \pi N K_\text{o}$ . The latter turns out to be the
effective field theory of  a spin-$S=N/2$ Heisenberg chain with a single-ion anisotropy  as shown 
by Schulz in Ref. \onlinecite{Schulz-86}. From the integrability of the quantum sine-Gordon model, we expect that 
a gap for orbital degrees of freedom opens when $K_\text{o} < 4/N$. As usual, it is very difficult to extract
the precise value of the Luttinger parameter $K_\text{o}$ from a perturbative RG analysis.  
Along the SU(2)$_\text{o}$ line, 
the exact value $K_\text{o}$ is known by the SU(2)-symmetry, 
i.e. $K_\text{o} = 1/N$, since the $\beta_{\text{o}}^2  =  2 \pi$ sine-Gordon model  is known to
display a hidden SU(2) symmetry. \cite{Affleck-chiral-86} 
In the vicinity of that line, we thus expect that there is a region where $K_\text{o} < 4/N$ and a Mott-insulating
phase emerges. In that situation, the orbital bosonic field is pinned into the following configurations:
\begin{alignat}{2}
\langle \Phi_{\text{o}} \rangle &= \sqrt{\frac{\pi}{2N}} + p \sqrt{\frac{2\pi}{N}}\; ,& & \qquad {\rm if}   
\;  g_\text{o} > 0 \notag \\
\langle \Phi_{\text{o}} \rangle &= p \sqrt{\frac{2\pi}{N}} \; ,&  &\qquad  {\rm if}   \;  g_\text{o} < 0 ,
\label{pinningNeven}
\end{alignat}
where $p=0,\ldots, N-1$.
This semiclassical analysis naively gives rise to a ground-state degeneracy. However, there is a gauge-redundancy
in the continuum description. On top of the ${\mathbb{Z}}_N$ symmetry (\ref{ZN}) of the parafermions CFT,
there is an independent discrete symmetry, ${\tilde {\mathbb{Z}}}_N$, such that the parafermionic currents
transform as follows: \cite{Zamolodchikov-F-JETP-85}
\begin{eqnarray}
\Psi_{1 \text{L,R}} &\rightarrow& e^{ \pm i 2 \pi m /N} \Psi_{1 \text{L,R}} , 
\label{chargeparadual}
\end{eqnarray}
with $m=0,\ldots, N-1$.
The two ${\mathbb{Z}}_N$ symmetries are related by a Kramers-Wannier duality 
transformation. \cite{Zamolodchikov-F-JETP-85}
The thermal operator $\epsilon_1$ is a singlet under the ${\tilde {\mathbb{Z}}}_N$ while
the disorder operator $\mu_2$ transforms as: $\mu_{2} \rightarrow e^{i 4 \pi m /N} \mu_{2}$. 
\cite{Zamolodchikov-F-JETP-85}
The combination of the ${\tilde {\mathbb{Z}}}_N$ (\ref{chargeparadual}) and the identification on the orbital bosonic
field:
\begin{equation}
\Phi_\text{o} \sim \Phi_\text{o} - m \sqrt{\frac{2 \pi}{N}} + p \sqrt{\frac{N \pi}{2}}, 
\; m=0, \ldots, N -1,
\label{tildeznorbitalbose}
\end{equation}
becomes a symmetry of model (\ref{effectiveHamU1para}), as it can be easily seen. 
In fact, this symmetry is a gauge redundancy since it corresponds to the identity in terms of the Dirac fermions.
Using the redundancy (\ref{tildeznorbitalbose}), we thus conclude that the gapful phase of the quantum 
sine-Gordon model (\ref{hoNeven}) is non-degenerate with ground state:
\begin{eqnarray}
\langle \Phi_{\text{o}} \rangle &=& \sqrt{\frac{\pi}{2N}}, \; {\rm if}   \;  g_\text{o} > 0 \nonumber \\
\langle \Phi_{\text{o}} \rangle &=& 0, \;  \;  \;  \;  \;  \;  \;   \; \; {\rm if}   \;  g_\text{o} < 0 .
\label{pinningGSNeven}
\end{eqnarray}

The lowest massive excitations are
the soliton and the antisoliton of the quantum sine-Gordon model; they carry the orbital pseudo spin:
 \begin{equation}
T^z = \pm  \sqrt{N/ 2\pi} \int dx \;  \partial_x \Phi_\text{o} =  \pm 1,
\label{chargekinkHaldane}
\end{equation}
and correspond to massive spin-1 magnon excitations. 

At this point, it is worth observing that the duality symmetry $\Omega_2$ of Eq. (\ref{dualitiesDirac}) plays a subtle
role in the even $N$ case. Indeed, the change of sign of the coupling constants $g_{2,4}$ can be 
implemented by the shift:
$\Phi_{\text{o}}  \rightarrow \Phi_{\text{o}}  +  \sqrt{N/ 8\pi}$ so that
the cosine term of Eq. (\ref{hoNeven}) transforms as
\begin{equation}
 \cos \left(\sqrt{2 \pi N} \; \Phi_{\text{o}} \right) \to \left(-1 \right)^{N/2} \cos \left(\sqrt{2 \pi N} \; \Phi_{\text{o}} \right).
\label{dualcosNeven}
\end{equation}
The latter result calls for a separate analysis depending on the parity of $N/2$.

%%%%%%%%%
\noindent%
\emph{\underline{$N/2$ odd case}.}

When $N/2$ is odd, the cosine term of Eq.~(\ref{hoNeven}) is odd under the $\Omega_2$ duality transformation
and there is thus two distinct fully gapped phases depending on the 
sign of $g_\text{o}$. 
The numerical solution of the RG equations shows that  $g_2 < 0$, i.e. $\lambda_{2\|} > 0$, 
in the vicinity of the SU(2)$_\text{o}$ line. We thus expect that $g_\text{o} > 0$ in this region and the ground state
of the sine-Gordon model (\ref{Luttingerparaorbital}) with $K_\text{o} < 4/N$  is described by the pinning:
$\langle \Phi_{\text{o}} \rangle =  \sqrt{\pi/2N} $ [first line of Eq. \eqref{pinningGSNeven}]. 
The corresponding Mott-insulating phase is the 
continuation of the OH phase that we have found along the SU(2)$_\text{o}$ line. This phase can be described by 
a string-order ordermeter which takes the form:
\begin{equation}
\begin{split}
& \lim_{|i-j| \rightarrow \infty}
 \left\langle
{T}^{z}_i
e^{ i \pi \sum_{k=i+1}^{j-1}{T}^{z}_k}
{T}^{z}_j 
\right\rangle  \simeq   \\
& \lim_{|x-y| \rightarrow \infty}
 \left\langle{\sin \left( \sqrt{N \pi/2} \; \Phi_\text{o} \left( x \right)  \right) 
 \sin \left( \sqrt{N \pi/2} \; \Phi_\text{o} \left( y \right)  \right)} \right\rangle  \ne 0 .
 \end{split}
\label{stringorbitalNodd}
\end{equation}
This result is in full agreement with the known properties of the Haldane phase when the orbital 
pseudo spin $T=N/2$ is odd.  

According to Eq.~\eqref{dualcosNeven}, the duality symmetry $\Omega_2$ changes the sign of 
the cosine operator in the sine-Gordon model \eqref{hoNeven} 
when $N/2$ is odd.  Therefore, there exists yet another Mott-insulating phase obtained by 
the duality $\Omega_{2}$ when  $K_\text{o} < 4/N$ which is characterized by the pinning:
$\langle \Phi_{\text{o}} \rangle =  0$ [the second of Eq. \eqref{pinningGSNeven}]. 
In this phase, the string-order parameter \eqref{stringorbitalNodd} 
vanishes, i.e., we have a new fully gapped non-degenerate phase which is different from the OH phase.
A simple non-zero string order parameter in this phase,  that we can estimate within our low-energy approach, reads as follows
\begin{equation}
\begin{split}
& \lim_{|i-j| \rightarrow \infty}
\left\langle{\cos \left(\pi \sum_{k<i} T^{z}_k \right)
\cos \left(\pi \sum_{k<j} T^{z}_k \right)}\right\rangle    \\
& \simeq \lim_{|x-y| \rightarrow \infty}
 \left\langle{\cos \left( \sqrt{N \pi/2} \; \Phi_\text{o} \left( x \right)  \right) 
 \cos \left( \sqrt{N \pi/2} \; \Phi_\text{o} \left( y \right)  \right)} \right\rangle   \\
 &\ne 0 .
\end{split}
\label{HaldanestringN4}
\end{equation}
The latter phase is expected to be the RS phase (i.e., the orbital-analogue of the large-$D$ phase 
with $T^z =0$) that we have  already identified in the strong-coupling analysis of Sec.~\ref{sec:strong-coupling}.

\noindent%
\emph{\underline{$N/2$ even case}.}
 
When $N/2$ is even, the cosine term of Eq. \eqref{hoNeven}  is now even under the $\Omega_2$ duality transformation
and there is thus  a single fully gapped phase. In this phase, we have $g_\text{o} <0$ and the orbital bosonic
field is pinned when  $K_\text{o} < 4/N$  into configurations: $\langle \Phi_{\text{o}} \rangle =  0$.
The phase is thus characterized by the long-range ordering of the string-order parameter \eqref{HaldanestringN4} 
while the standard one \eqref{stringorbitalNodd} vanishes. In this respect, the physics is very similar
to the properties of the even-spin Haldane phase. The authors of Ref.~\onlinecite{Pollmann-B-T-O-12}
have conjectured that there is an adiabatic continuity between the Haldane and large-$D$ phases
in the even-spin case. Such continuity has been shown numerically in the spin-2 XXZ Heisenberg chain with a single-ion
anisotropy by finding a path where the two phases are connected without any 
phase transition.  \cite{Tonegawa-O-N-S-N-K-11}
The Haldane phase for integer spin is thus equivalent to a topologically trivial insulating phase in this 
case.  In our context, the two non-degenerate Mott insulating OH and RS (the orbital large-$D$) phases belong 
to the same topologically trivial phase when $N/2$ is even, 
while they exhibit very different topological properties for odd $N/2$.

\noindent% 
\emph{\underline{Orbital Luttinger liquid phase}.}

 Regardless of the parity of $N/2$, 
there is a room to have, on top of the Mott-insulating phases, an algebraic (metallic) one since 
the Luttinger parameter $K_\text{o}$ can be large in the second region of the RG flow.  
 When $K_\text{o} > 4/N$, the interaction of the sine-Gordon model (\ref{hoNeven}) becomes irrelevant and
 a critical Luttinger-liquid phase emerges having one gapless mode in the orbital sector.
 At low energies $E \ll \Delta_\text{c}$, the staggered part of the orbital-pseudo spin $\mathbf{T}_i$ simplifies
 as follows using the identifications (\ref{identiprimZnApp}):
 \begin{eqnarray}
T^{+}_{\pi} &\sim&  \sigma_1  e^{ i  \sqrt{2 \pi/N} \; 
 \Theta_{\text{o}} } \left(  \left\langle    e^{ i \sqrt{2 \pi K_\text{c} /N} \Phi_\text{c} }  \right\rangle   
 \left \langle  \text{Tr}\, G  \right \rangle  \;+ \text{c.c.} \right)
 \nonumber \\
 T^{z}_{\pi} &\sim& \left\langle    e^{ i \sqrt{2 \pi K_\text{c}  /N} \Phi_\text{c} }  \right\rangle    \langle  {\rm Tr} G  \rangle 
 \left(  \mu_1  e^{- i  \sqrt{2 \pi/N}  \Phi_{\text{o}} } -  \mu^{\dagger}_1  e^{ i  \sqrt{2 \pi/N}\Phi_{\text{o}} }  \right) 
   \nonumber \\
 &+& \text{H.c.}
 \label{staggeredorbitalspin}
\end{eqnarray}
Since the $\mathbb{Z}_N$-symmetry is broken in the second region of the RG flow, 
we have $\langle \sigma_1 \rangle \ne 0$
and $\langle \mu_1 \rangle = 0$  so that the $z$-component of 
$\mathbf{T}_{\pi}$ is thus short-range while the transverse ones are gapless:
$T^{+}_{\pi} \sim  e^{ i  \sqrt{2 \pi/N} \Theta_{\text{o}} }$.
Taking into account the uniform part of the $z$-component  of the orbital-pseudo spin $\mathbf{T}_i$, i.e. 
the SU(2)$_N$ current $j^z_{\text{L}} + j^z_{\text{R}} $,
we get the following leading asymptotics for the equal-time orbital pseudo spin correlations:
 \begin{eqnarray}
\langle   T^{+} \left(x\right)   T^{-} \left(0\right) \rangle
&\sim&  \left(-1\right)^{x/a_0} x^{- 1/NK_\text{o}} \nonumber \\
\langle   T^{z} \left(x\right)  T^{z} \left(0\right) \rangle &\sim&
-\frac{NK_\text{o}}{4\pi^2 x^2}.
\label{correlBCSNeven}
\end{eqnarray}
The leading instability is thus the transverse orbital correlation when $K_\text{o} > 4/N$, i.e.,
the formation of a critical orbital-XY phase, i.e., an orbital Luttinger-liquid phase.

%%%%%%%%%%%%%%%%%%%%%%%%%%%%%%%%%%%%%%%%%%%%%%%%%%%%%%%%%%
\begin{figure}[!ht]
\centering
\includegraphics[width=0.99\columnwidth,clip]{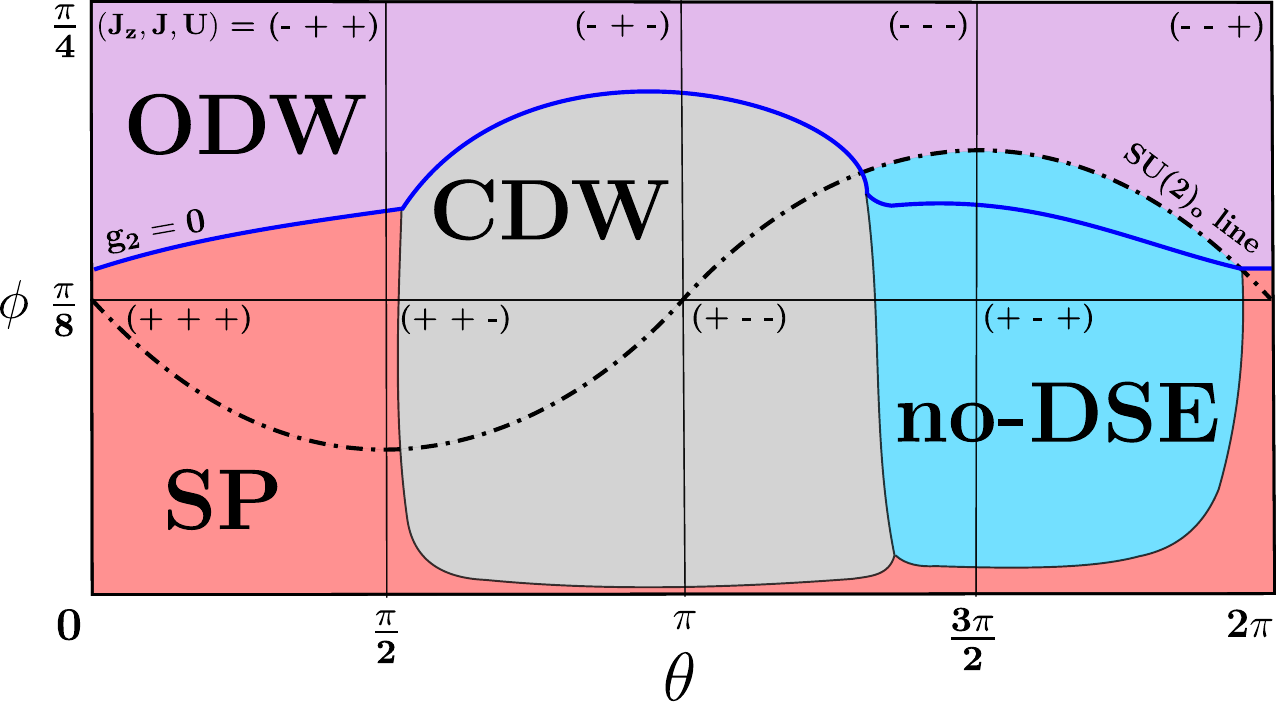}
\caption{(Color online) General phase diagrams for the $N=6$ generalized Hund model (\ref{alkaourmodel}) obtained by solving numerically the one-loop RG equations (\ref{RG}) with initial conditions (\ref{couplings}). 
T coupling constants $(J_z, J, U)$ are parametrized by $(\theta, \phi)$ as Eqs.~\eqref{eq:sphere_variable} 
and the meaning of the extra bold lines is discussed in the text. 
The signs of $J_z$, $J$ and $U$ in each quadrant are indicated. 
In the region shown as ``no-DSE'', RGE flow does not exhibit dynamical symmetry enlargement.  
For other abbreviations, see TABLE \ref{tab:abbreviation}.
\label{fig:RG_general_phase_diagram}}  
\end{figure}
%%%%%%%%%%%%%%%%%%%%%%%%%%%%%%%%%%%%%%%%%%%%%%%%%%%%%
%%%%%%%%%%%%%%%%%%%%%%%%%%%%%%%%%%%%%%%%%%%%%%%%%%%%
\subsection{Phase diagrams}
\label{sec:RG-phase-diag}
%%%%%%%%%%%%%%%%%%%%%%%%%%%%%%%%%%%%%%%%%%%%%%%%%%%%
We have determined the possible phases of general 1D two-orbital SU($N$) models in the weak-coupling regime
by means of the one-loop RG analysis combined with CFT techniques.
We now exploit all these results to map out the zero-temperature phase diagram of the generalized Hund model \eqref{alkaourmodel} and the $g$-$e$ model \eqref{eqn:Gorshkov-Ham} defined in Sec.~\ref{sec:models-strong-coupling}. 
The phase diagram of the $N=2$ $p$-band model \eqref{eqn:p-band-simple} is presented in Appendix~\ref{sec:N2-p-band-continuum}, together with the study of its low-energy limit. 
The correspondence between the parameters used in the phase diagrams and 
the physical interactions is summarized in TABLE \ref{tab:three-models}.  

Before solving numerically the one-loop RG analysis, one immediately observes that our global approach
of the phases in the weak-coupling regime does not give any SPT phases
when $N>2$ in stark contrast to the strong coupling result of Sec. \ref{sec:strong-coupling}. It might suggest that
there is no adiabatic continuity between weak and strong coupling regimes and necessarily
a quantum phase transition occurs in some intermediate regime which is not reachable by the one-loop RG
analysis. In this respect, a two-loop analysis might be helpful but it is well beyond the scope of this work.
The possible occurence of a quantum phase transition
will be investigated in Sec. \ref{sec:DMRG} by means of DMRG calculations to study the extension of the SU(4) SPT
phase.

The sets of first-order differential equations obtained with the one-loop RG analysis, $\{ {\dot g}_i \} = \{ \partial g_i/ \partial l\}$, $l$ being the RG time, can be solved numerically with Runge-Kutta methods.
The initial conditions $g_{i,0}$ depend on the lattice model and we loop on values of the couplings taken in $[-0.1;0.1]$ to scan the zero-temperature phase diagrams in the weak coupling regime. 
For each run, the couplings $g_i$ flow to the strong coupling regime as the RG time increases. 
The procedure is stopped at $l_{\text{max}}$ when one of the couplings, which turns out 
to be $g_1$ (see Sec.~\ref{sec:Non-degenerate_Mott_insulating_phases}), reaches 
an arbitrary large value $g_{\text{max}}$.  
Typically, we choose $g_{\text{max}} \ge 10^{10}$ so that the directions taken by the RG flow in the far IR appear clearly. 
For simplicity, we consider renormalized ratios $g_i(l_{\text{max}})/g_1(l_{\text{max}})$.
For instance, when the procedure stops in the SP phase, all the couplings have reached 
a value $g_i(l_{\text{max}})/g_1(l_{\text{max}}) \sim +1$, as a signature of the SO(4$N$) maximal DSE. 

As discussed in Sec.~\ref{sec:RG_analysis}, we distinguish in the weak coupling limit two types of regimes: phases with DSE and non-degenerate Mott insulating phases. 
On the one hand, the first ones can be readily identified by looking at the ratios 
$g_i(l_{\text{max}})/g_1(l_{\text{max}})$ that are either $+1$ in the SP phase or $\pm 1$ in the phases obtained by applying the duality symmetries Eqs.~(\ref{eq:Omega_1}-\ref{eq:Omega_3}). 
On the other hand, couplings $g_{2,4,5,8}$ flow very slowly to the strong coupling regime in the non-degenerate phases. Determining the exact nature of the phase is thus more approximative in that case. In particular, as detailed in Sec.~\ref{sec:U1o_symmetry}, the sign of $g_2$ allows to distinguish between OH and the RS phase 
only in the $N/2$ odd case. Next, we therefore show results for $N=6$. \footnote{Actually, for $N>2$ the position of the different regions obtained by solving numerically the RG equations is almost not affected by $N$. However, the nature of the phases differs. For instance, for $N$ odd, the `No DSE' region in Fig.~\ref{fig:RG_general_phase_diagram} turns out to be critical.}
In order to have an overview of the phases that appear, we first compute the general phase diagram of the generalized Hund model (\ref{alkaourmodel}) for all $J_z$, $J$ and $U$, see Fig.~\ref{fig:RG_general_phase_diagram}. 
We solve the RG equations (\ref{RG}) using the initial conditions (\ref{couplings}) and introduce sphere variables:
\begin{eqnarray}
    U   &=& R \cdot \sin 4\phi \cdot\cos\theta  \nonumber \\
    J   &=& R \cdot \sin 4\phi \cdot\sin\theta  \nonumber \\
    J_z &=& R \cdot \cos 4\phi  ,
    \label{eq:sphere_variable} 
\end{eqnarray}
where $R=0.1$.
Eight quadrants are required to get all the possible combinations of signs for $U$, $J$ and $J_z$ ($\theta \in [0,2\pi]$ and $\phi \in [0,\frac{\pi}{4}]$).
We directly identify three phases with DSE (SP, CDW and ODW) while the SP$_{\pi}$ phase obtained by applying the duality $\Omega_3$ (\ref{eq:Omega_3}) is not realized. \footnote{Our global RG approach, based on duality symmetries, give all possible DSE phases compatible with the global symmetry group of the low-energy Hamiltonian. Some phases might not be realized in concrete lattice model with the same continuous symmetry. The SP$_{\pi}$ phase  is one example and
a more general lattice model is necessary to stabilize such a phase.}
The SU(2)$_o$ symmetry ($J=J_z$) corresponds to $\theta = \arcsin (\cot 4\phi)$ 
and is showed with bold dashed lines in Fig.~\ref{fig:RG_general_phase_diagram}.
In the `no-DSE' region, the sign of $g_2$ changes on the blue line and the nature of the phases obtained is discussed next, in special cuts of the phase diagram. The one-loop RG analysis does not allow to confirm 
if the SU(2)$_{\text{o}}$ line is exactly at the ODW/`No DSE' transition but the latter is clearly 
in its vicinity as seen in Fig. \ref{fig:RG_general_phase_diagram}.

%%%%%%%%%%%%%%%%%%%%%%%%%%%%%%%%%%%%%%%%%%%%%%%%%%%%
\subsubsection{Generalized Hund model}
%%%%%%%%%%%%%%%%%%%%%%%%%%%%%%%%%%%%%%%%%%%%%%%%%%%%
%%%%%%%%%%%%%%%%%%%%%%%%%%%%%%%%%%%%%%%%%%%%%%%%%%%%%%%%%%
\begin{figure}[!ht]
\centering
\includegraphics[width=0.5\columnwidth,clip]{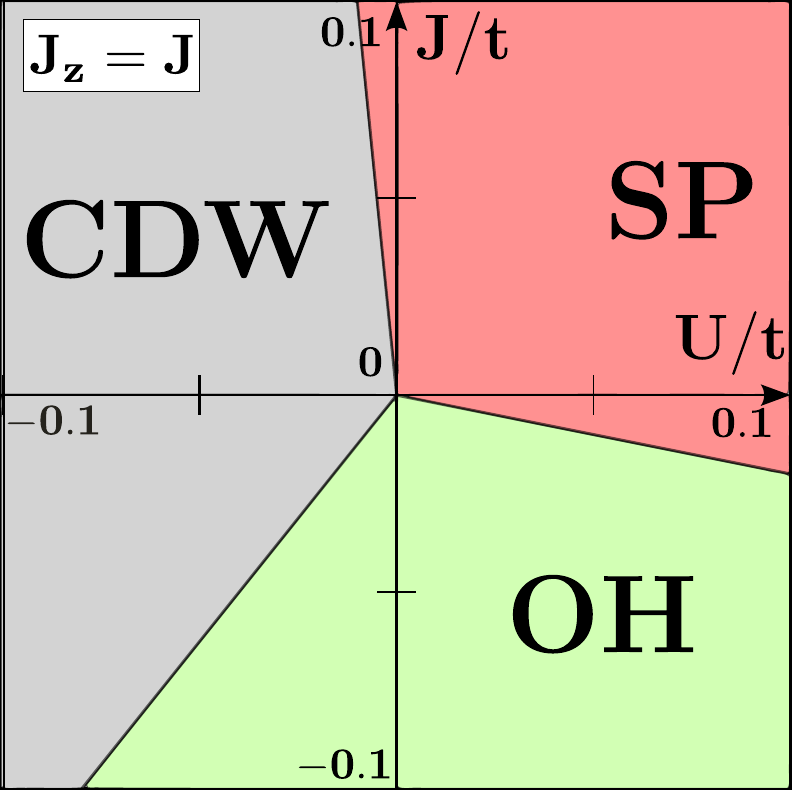}
\caption{(Color online) Phase diagram for the $N=6$ generalized Hund model 
(\ref{alkaourmodel}) with SU(2)$_o$ symmetry obtained by solving numerically 
the one-loop RG equations (\ref{RG}) with initial conditions (\ref{couplings}).
\label{fig:RG_phasediag_alka_N6_SU2o}}
\end{figure}
%%%%%%%%%%%%%%%%%%%%%%%%%%%%%%%%%%%%%%%%%%%%%%%%%%%%%
Let us continue with the generalized Hund model (\ref{alkaourmodel}) and take a 
closer look at special cuts in the general phase diagram Fig.~\ref{fig:RG_general_phase_diagram}.
%%%%%%%%%%%%%%%%%%%%%%%%%%%%%%%%%%%%%%%%%%%%%%%%%%%%%%%%%%
\begin{figure}[!ht]
\centering
\includegraphics[width=0.5\columnwidth,clip]{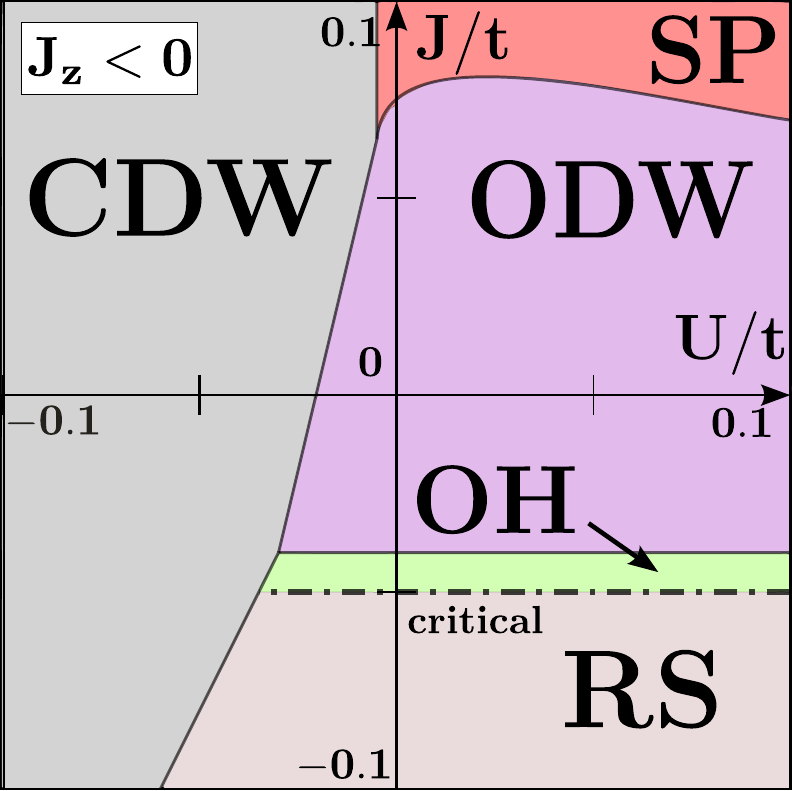}\vspace{0.2cm}
%\begin{tabular}{cc}
\includegraphics[width=0.49\columnwidth,clip]{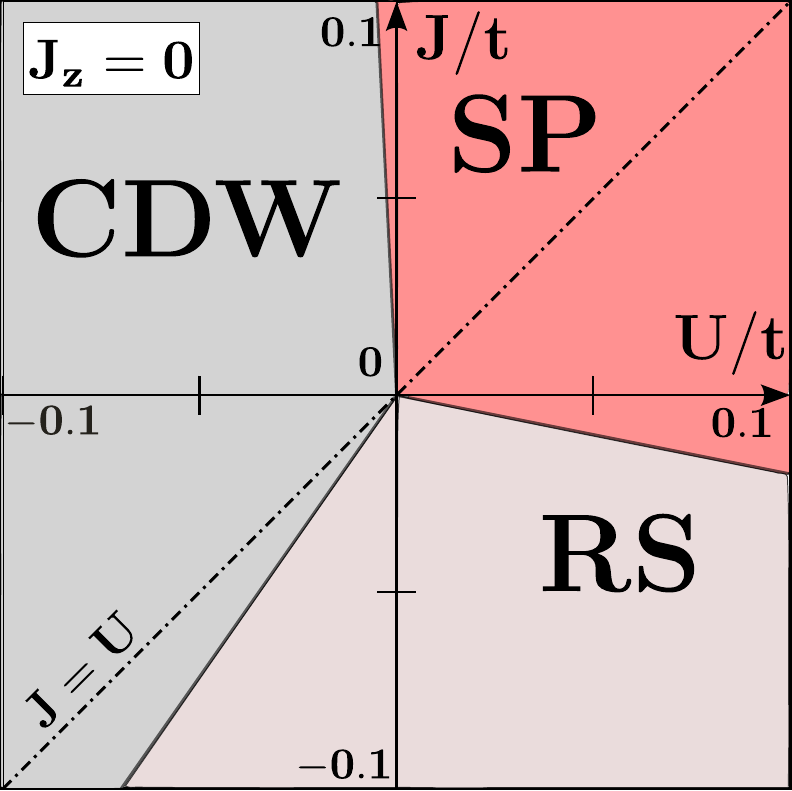}% &
\includegraphics[width=0.49\columnwidth,clip]{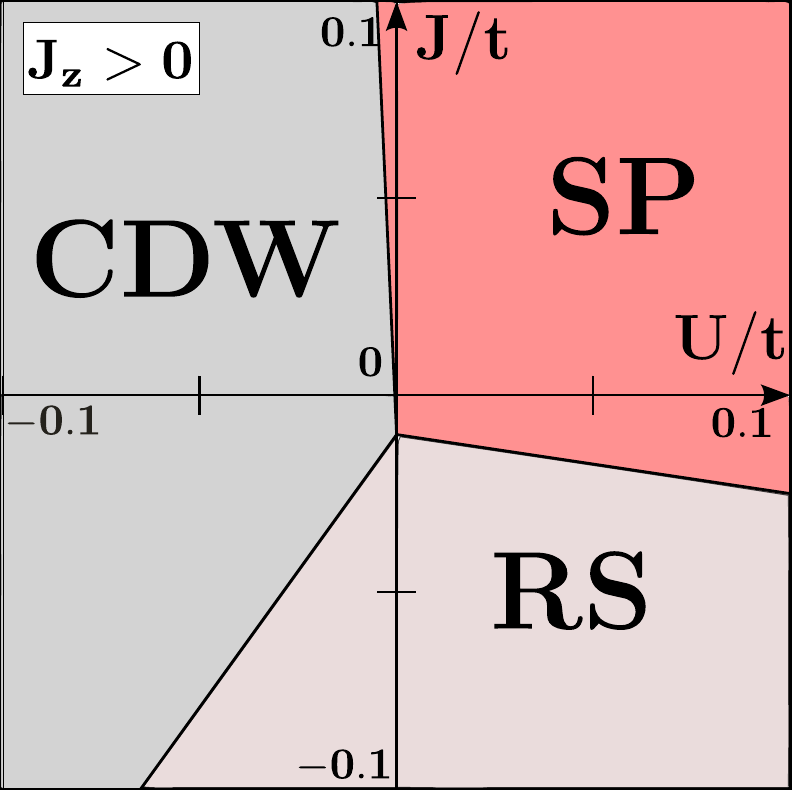}
%\end{tabular}
\caption{(Color online) Phase diagram for the $N=6$ generalized Hund model 
(\ref{alkaourmodel}) obtained by solving numerically the one-loop RG equations 
(\ref{RG}) with initial conditions (\ref{couplings}).
From top to bottom, and from right to left: $J_z/t=-0.03$, $J_z/t=0$ and $J_z/t=0.03$. 
\label{fig:RG_phasediag_alka_N6_Jz}}
\end{figure}
%%%%%%%%%%%%%%%%%%%%%%%%%%%%%%%%%%%%%%%%%%%%%%%%%%%%%
%%%%%%%%%%%%%%%%%%%%%%%%%%%%%%%%%%%%%%%%%%%%%%
\vspace{0.5cm}
\paragraph{SU(2)$_\text{o}$ symmetric case.}
%%%%%%%%%%%%%%%%%%%%%%%%%%%%%%%%%%%%%%%%%%%%%%
We first consider the case of SU(2) orbital symmetry ($J=J_z$, along bold dashed lines in Fig.~\ref{fig:RG_general_phase_diagram}). 
We focus on $N=6$, although the position of the phases is almost not sensitive to the value of $N$ in this case.
In Fig.~\ref{fig:RG_phasediag_alka_N6_SU2o}, we identify three regions: the SP phase, the degenerate CDW phase obtained by applying the duality symmetry $\Omega_1$ (\ref{eq:Omega_1}) and a region that displays no DSE with $|g_{2,4,5,8}(l_{\text{max}})| \ll g_{\text{max}}$.
The latter was identified in Sec.~\ref{sec:SU2o_symmetry} as the non-degenerate OH phase for even $N$. 
It is a SPT phase for $N/2$ odd. 
Besides, on the particular SU(2$N$) line $J = 0$, for $U>0$ (respectively $U<0$) we recover the SP (respectively CDW) phase expected for the repulsive (respectively attractive) SU(2$N$) Hubbard model at half-filling.
%%%%%%%%%%%%%%%%%%%%%%%%%%%%%%%%%%%%%%%%%%%%%%
\vspace{0.5cm}
\paragraph{U(1)$_\text{o}$ symmetric case.}
%%%%%%%%%%%%%%%%%%%%%%%%%%%%%%%%%%%%%%%%%%%%%%
We now turn to the phase diagrams of the generic case of U(1) orbital symmetry ($J \ne J_z$) at $N=6$.
We chose arbitrary cuts of the general phase diagram Fig.~\ref{fig:RG_general_phase_diagram} at constant $J_z$: $J_z =-0.03$, $J_z = 0 $ and $J_z = 0.03$ (see Fig.~\ref{fig:RG_phasediag_alka_N6_Jz}). 
As discussed the \emph{$N/2$ odd case} of Sec.~\ref{sec:U1o_symmetry}, the sign of $g_2$ allows us to determine if the non-degenerate Mott insulating phase (blue `no-DSE' region in Fig.~\ref{fig:RG_general_phase_diagram}) is either OH or RS. We find that the change of sign takes place at $J_z^* < 0$. 
The one-loop RG analysis does not allow us to determine the value of the Luttinger parameter $K_\text{o}$ except in the vicinity of the SU(2)$_\text{o}$ symmetric line where $K_\text{o}$ is fixed by symmetry. We cannot thus conclude that the phases, obtained by varying $J_z$, are indeed fully gapped from this analysis. However, the DMRG calculations 
in this regime of parameters strongly support that  $K_\text{o}$ is small enough to get gapful phases.
In Fig.~\ref{fig:RG_phasediag_alka_N6_Jz}, for $J_z=0$ and $J_z>0$, we find thus that the non-degenerate Mott insulating phase is the RS phase, while for $J_z<0$, a transition takes place between RS and OH. 
At the transition, the line $g_2=0$ (bold dashed line in Fig.~\ref{fig:RG_phasediag_alka_N6_Jz}, top panel) corresponds to the Luttinger critical line in which the cosine term of Eq.~(\ref{hoNeven}) is canceled. 
Interestingly, the phase diagram for $J_z<0$ obtained in the weak coupling regime is in agreement with the prediction from the strong coupling regime, i.e., an OH region followed by a RS region as $|J-J_z|$ increases (see Sec. II D 2).

%%%%%%%%%%%%%%%%%%%%%%%%%%%%%%%%%%%%%%%%%%%%%%%%%%%%
\subsubsection{$g$-$e$ model}
%%%%%%%%%%%%%%%%%%%%%%%%%%%%%%%%%%%%%%%%%%%%%%%%%%%%
%%%%%%%%%%%%%%%%%%%%%%%%%%%%%%%%%%%%%%%%%%%%%%%%%%%%%%%%%%
\begin{figure}[!ht]
\centering
\begin{tabular}{cc}
\includegraphics[width=0.5\columnwidth,clip]{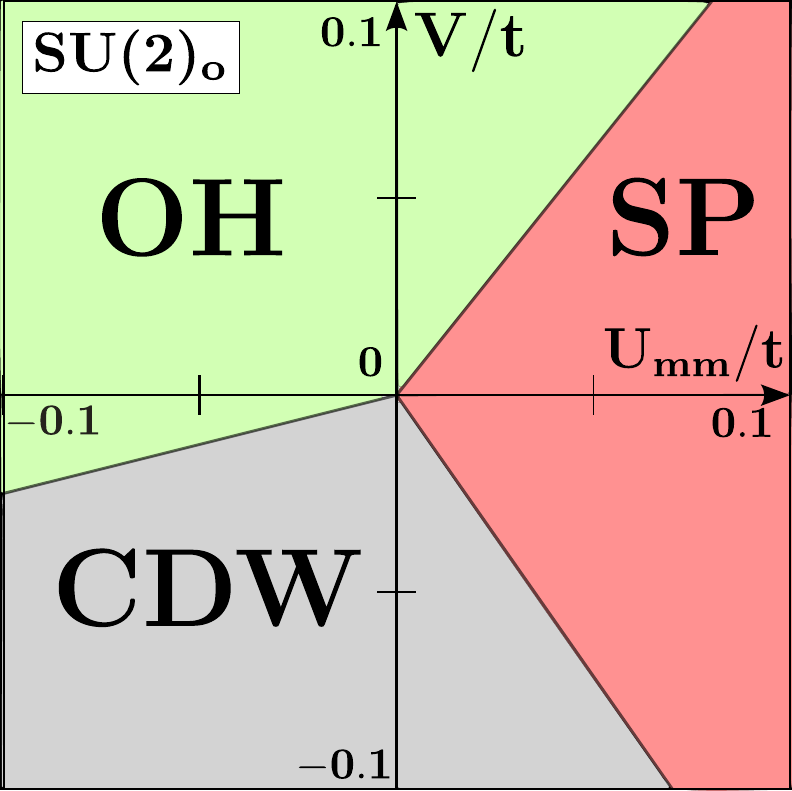} &
\includegraphics[width=0.5\columnwidth,clip]{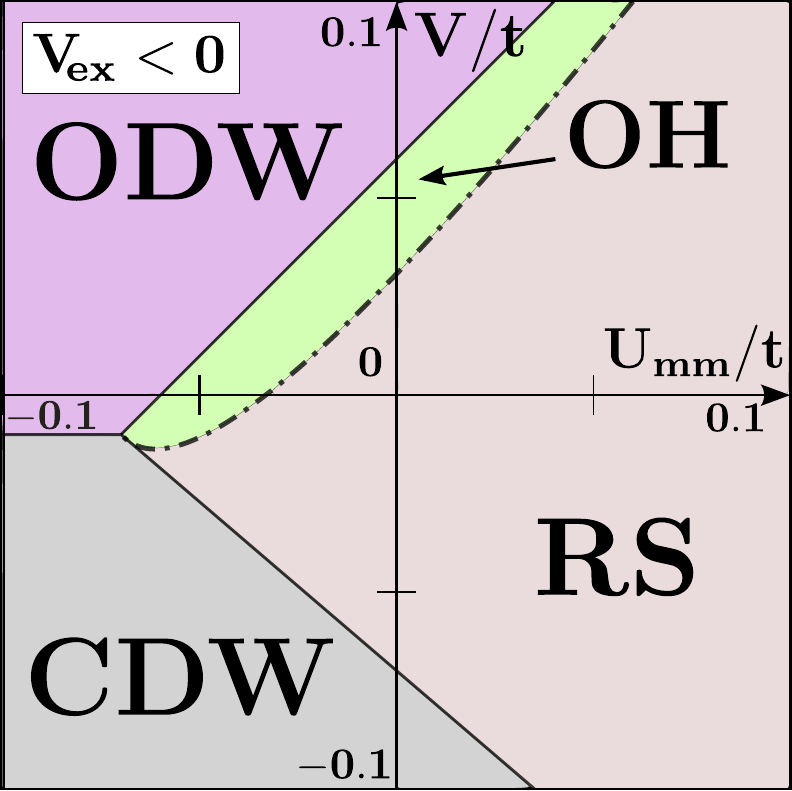} \\
\includegraphics[width=0.5\columnwidth,clip]{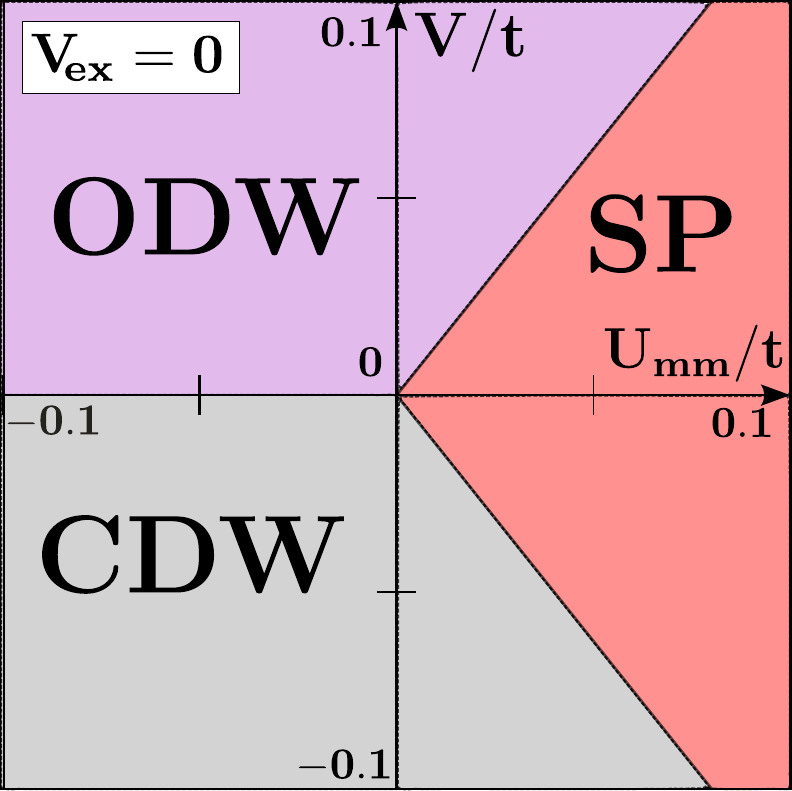} &
\includegraphics[width=0.5\columnwidth,clip]{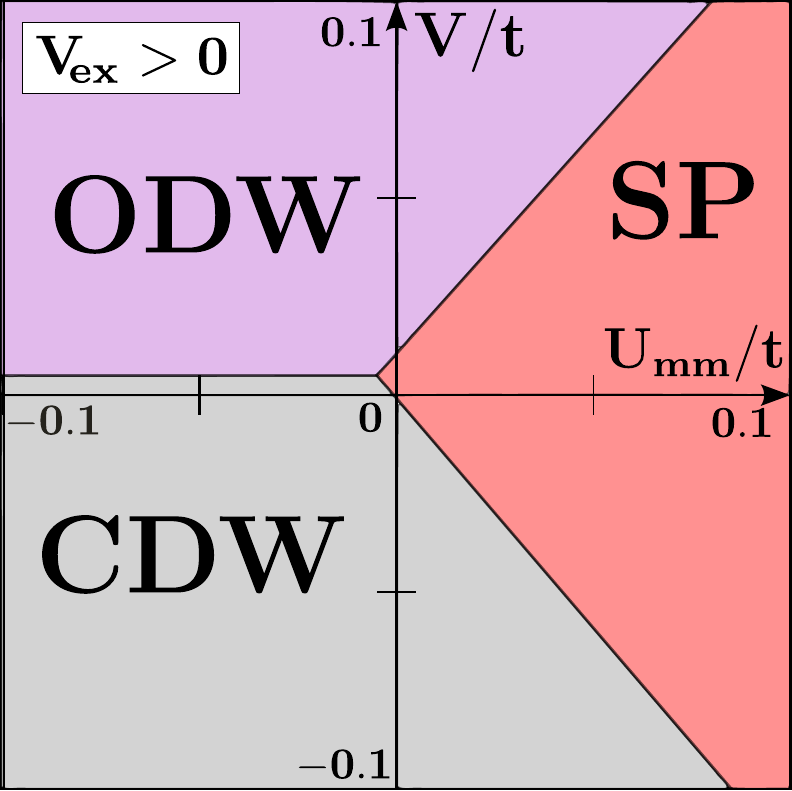}
\end{tabular}
\caption{(Color online) Phase diagram for the $N=6$ $g$-$e$ model 
(\ref{eqn:Gorshkov-Ham}) obtained by solving numerically the one-loop RG 
equations (\ref{RG}) with initial conditions (\ref{couplingsGorsh}).
From top to bottom, and from right to left: SU(2)$_o$ symmetry, 
$V_{\text{ex}}^{g\text{-}e}/t=-0.06$, $V_{\text{ex}}^{g\text{-}e}/t=0$ and 
$V_{\text{ex}}^{g\text{-}e}/t=0.02$.  
\label{fig:RG_phasediag_Gorshkov_N6_Vex}}
\end{figure}
%%%%%%%%%%%%%%%%%%%%%%%%%%%%%%%%%%%%%%%%%%%%%%%%%%%%%
For completeness, we also present the phase diagrams of the  $g$-$e$ model (\ref{eqn:Gorshkov-Ham}) with coupling constants $U_{gg} = U_{ee} = U_{mm}, V$ and $V_{\text{ex}}^{g\text{-}e}$. The mapping to the couplings $J$, $J_z$ and $U$ is defined in Eqs.~(\ref{eqn:Gorshkov-to-Hund}), in particular, $V_{\text{ex}}^{g\text{-}e}=J$. 
As explained in Sec.~\ref{sec:continuum_description}, the  $g$-$e$ model shares the same continuum Hamiltonian with the generalized Hund model. Only the initial conditions differ and we solve the set of equation (\ref{RG}) 
starting from (\ref{couplingsGorsh}). 
In Fig.~\ref{fig:RG_phasediag_Gorshkov_N6_Vex}, we show the phase diagrams 
for the SU(2)$_\text{o}$-symmetric (i.e. $V=U_{mm}-V_{\text{ex}}^{g\text{-}e}$) cases,  
$V_{\text{ex}}^{g\text{-}e}/t=-0.06$, $V_{\text{ex}}^{g\text{-}e}/t=0$ and $V_{\text{ex}}^{g\text{-}e}/t=0.02$. 
In the presence of the orbital SU(2)$_\text{o}$-symmetry, 
we recover the SP, CDW and OH phases from Fig.~\ref{fig:RG_phasediag_alka_N6_SU2o}.
For $V_{\text{ex}}^{g\text{-}e}> 0$ and $V_{\text{ex}}^{g\text{-}e}= 0$, the phase diagram exhibits only regions with DSE (SP, CDW and ODW), in agreement with the phase found for $J \ge 0$ in the preceding section.
Their positions are little affected by the value of $V_{\text{ex}}^{g\text{-}e}$.
Finally, for $V_{\text{ex}}^{g\text{-}e} < 0$, as for $J <0$, we have a non-degenerate Mott insulating region in which the sign of $g_2$ changes. We thus identify the OH and the RS regions.

%%%%%%%%%%%%%%%%%%%%%%%%%%%%%%%%%%%%%%%%%%%%%%%%%%%%
\section{DMRG calculations}  
\label{sec:DMRG}
%%%%%%%%%%%%%%%%%%%%%%%%%%%%%%%%%%%%%%%%%%%%%%%%%%%%

We now turn to numerical simulations using DMRG algorithm in order to
determine some of the phase diagrams that were discussed in the
previous sections (Sections \ref{sec:strong-coupling} and \ref{sec:RG-phase-diag}), 
namely the $g$-$e$ model with orbital SU(2)$_{\text{o}}$ symmetry (\ref{eqn:Gorshkov-Ham}), 
the generalized Hund model  with or without SU(2)$_\text{o}$ symmetry  \eqref{alkaourmodel}, 
and the $p$-band model \eqref{eqn:p-band-simple}.  
As already mentioned in Sec.~\ref{sec:Gorshkov-Hamiltonian}, for concreteness we assume 
that the two orbitals behave in a similar manner, i.e. we restrict ourselves to the case 
\begin{equation}
t_g = t_e=t, \; \; U_{g g} = U_{ee} =U_{mm}, \;\;  \mu_{g}=\mu_{e}
\end{equation}
of the $g$-$e$ model \eqref{eqn:Gorshkov-Ham} or the generalized Hund one 
\eqref{alkaourmodel}. 
The parametrization used in the three models ($g$-$e$ model, generalized Hund model, 
and $p$-band model) considered here is summarized in TABLE~\ref{tab:three-models}.   
Also the definitions of the abbreviations used in the phase diagrams are given in TABLE.~\ref{tab:abbreviation}.

This numerical investigation is especially needed (i) to check our
weak-coupling predictions (Sec.~\ref{sec:RG-phase-diag}) and (ii) to go beyond this regime and make
connection with strong-coupling results (Sec.~\ref{sec:strong-coupling}). 
Moreover, it allows us to get
precise numerical estimates of the locations of the phases and the transitions among them,
which is of fundamental importance to decide whether they could be
accessed experimentally. Typically, we used open boundary
conditions, keeping between 2000 and 4000 states depending on the
model and the parameters in question in order to keep a discarded weight below
$10^{-6}$.  Note also that for the sake of the efficiency of the simulations, 
for all models with $N=4$ and for the $p$-band with $N=2$ too, 
we map the \emph{original} two-orbital SU($N$) models onto the \emph{equivalent} (pseudo)spin-1/2 (where the
pseudo spin corresponds to the orbital) fermionic models on some $N$-leg ladder
(with generalized rung interactions which are tailored to reproduce the original interactions) 
shown in Fig.~\ref{fig:alkaline-Nleg}.  
As a last remark, let us mention that we worked at half-filling and except for the $p$-band model, 
we have implemented the abelian U(1) symmetry corresponding to the conservation of particles in each orbital.

In order to map out the phase diagrams, we worked at fixed length $L=36$ (for $N=4$) or $L=64$ (for $N=2$) 
and measured the local quantities (densities, pseudospin densities, kinetic energies, etc.) 
as well as the presence/absence of edge states.  
One may wonder why we do not use the string order parameters introduced in Sec.~\ref{sec:non-local-OP} 
in determining (a part of) the phase diagram.  In fact, for purely bosonic models, the string order parameters 
combined with, e.g., the Binder-parameter analysis may yield a reasonably good results \cite{Totsuka-N-H-S-95}.  
However, the string order parameters are defined for {\em fixed} SU($N$) `spins' which are meaningful   
only deep inside the Mott phases \cite{Manmana-H-C-F-R-11}.  
When the charge fluctuations are not negligible, entanglement spectrum necessarily contains the contribution 
from the fermionic sector\cite{Hasebe-T-13}, for which the relation between the SPTs 
and the string order parameters mentioned in Sec.~\ref{sec:non-local-OP} is not very clear.  
For this reason, in order for the search in the full parameter space, 
more conventional methods seem robust.  
We refer the interested reader to Refs.~\onlinecite{Nonne-B-C-L-10,Nonne-B-C-L-11} 
which contain more details on our procedure. 
%%%%%%%%%%%%%%%%%%%%%%%%%%%%%%%%%%%%%%%%%%%%%%%%%%%%%%%%
\subsection{\texorpdfstring{$\boldsymbol{N=2}$}{N=2} \texorpdfstring{$\boldsymbol{g}$}{g}-\texorpdfstring{$\boldsymbol{e}$}{e} model}
\label{sec:N2-Gorshkov}
\begin{figure}[ht]
\centering
\includegraphics[scale=0.33]{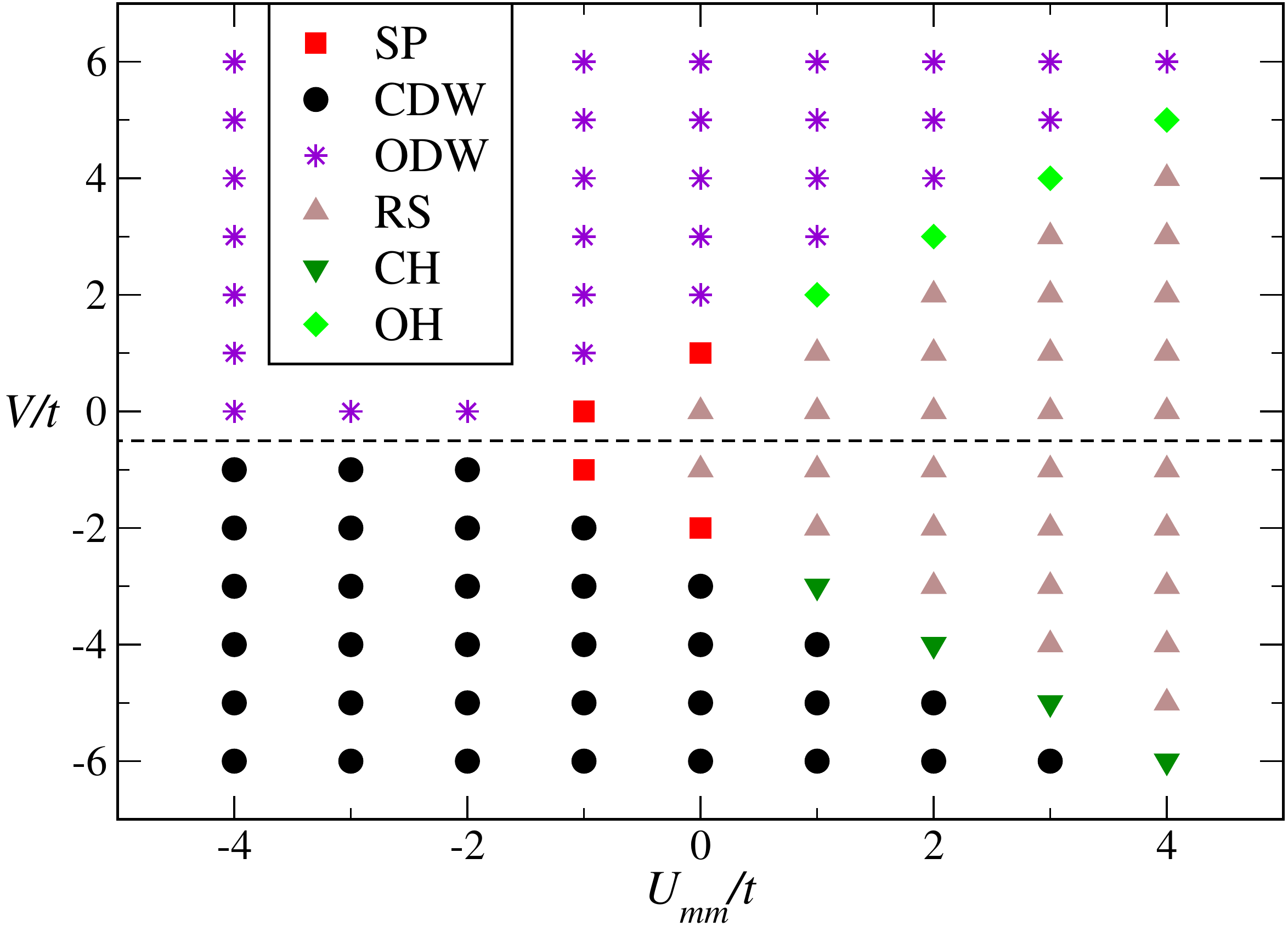}
\includegraphics[scale=0.33]{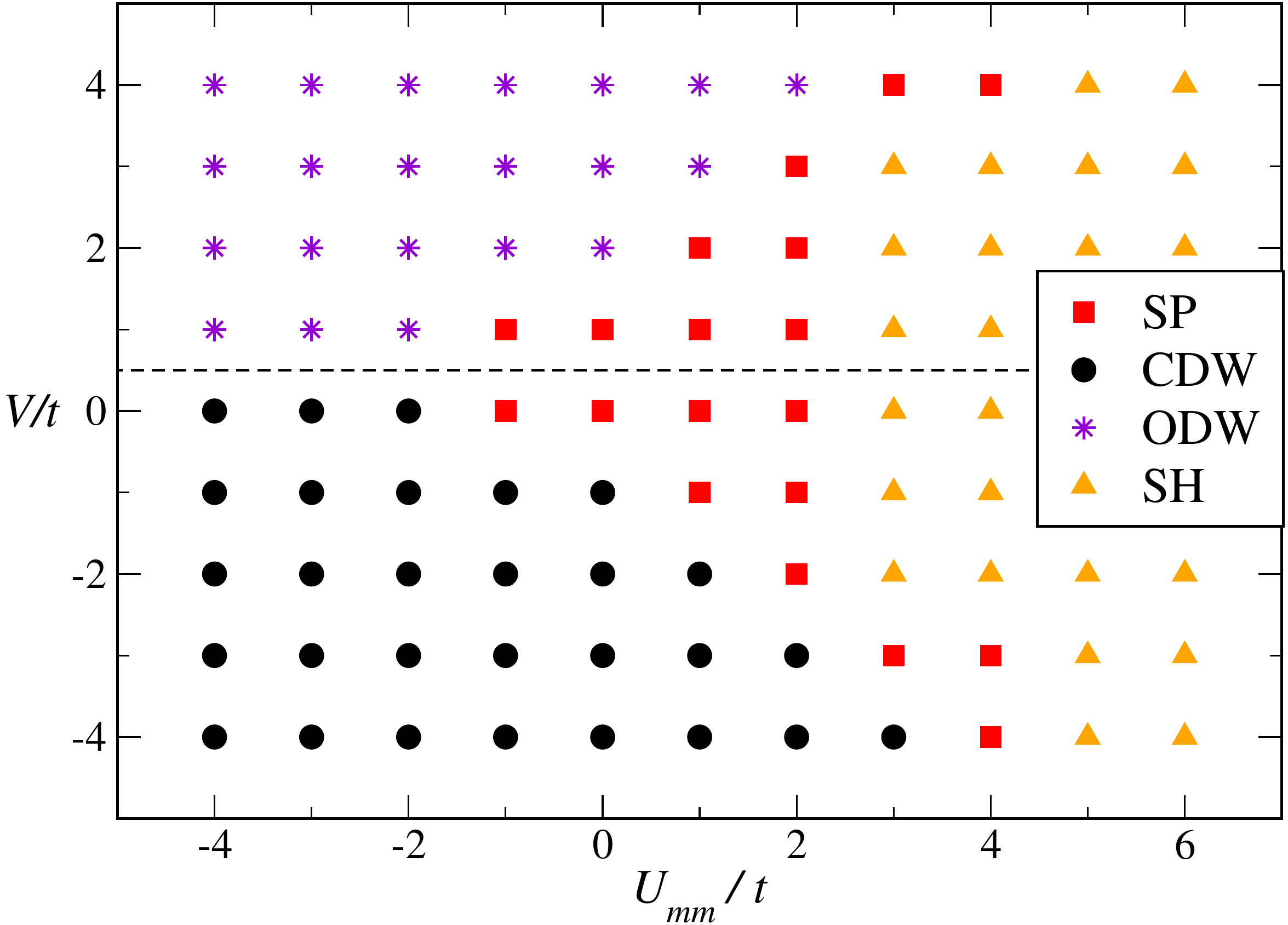}
\caption{(Color online) Phase diagram for $N=2$ $g$-$e$ model (\ref{eqn:Gorshkov-Ham}) 
obtained by DMRG. Top  and bottom panels correspond respectively to $V_{\text{ex}}^{g\text{-}e}/t=-1$   and $V_{\text{ex}}^{g\text{-}e}/t=1$.  
Due to the symmetry Eq.~\eqref{eqn:N2-Gorshkov-orbital-charge} (which exists only for $N=2$), 
CH and OH, as well as CDW and ODW, appear in a symmetrical way with respect to the symmetry axis $V=V_{\text{ex}}^{g\text{-}e}/2$ indicated with a dashed line.
\label{fig:phasediag_Gorshkov_N2_Vex}}
\end{figure}

%%%%%%%%%%%%%%%%%%%%%%%%%%%%%%%%%%%%%%%%%%%%%%%%%%%%%%%%
For completeness, we present, in Figs.~\ref{fig:phasediag_Gorshkov_N2_Vex}, 
some phase diagrams of the $g$-$e$ model (\ref{eqn:Gorshkov-Ham}) 
with $N=2$ which exhibit a large variety of phases: (i) charge density wave (CDW), (ii) orbital density wave (ODW), 
(iii) spin-Peierls (SP), (iv) charge Haldane (CH), (v) orbital Haldane (OH), (vi) spin Haldane (SH), 
and (vii) rung singlet (RS) (see the previous sections 
and TABLE.~\ref{tab:abbreviation} for the definitions). 
These very rich phase diagrams are in rather good agreement with the low-energy predictions, and they were already discussed in Ref.~\onlinecite{Nonne-B-C-L-11}.  
In Figs.~\ref{fig:phasediag_Gorshkov_N2_Vex}, one notes that the phases concerning the charge sector 
(CDW and CH) and those concerning the orbital sector (ODW and OH) appear in a very symmetric manner.  
In fact, this is quite natural since the $N=2$ $g$-$e$ model possesses the symmetry 
discussed in Sec.~\ref{sec:orbita-charge-interchange}:
\begin{equation}
\begin{split}
& V \rightarrow -V + V_{\text{ex}}^{g\text{-}e}  \\
& V_{\text{ex}}^{g\text{-}e} \rightarrow V_{\text{ex}}^{g\text{-}e} \, , \;\;
 U_{mm} \rightarrow  U_{mm}  \; ,
\end{split}
\end{equation}
that swaps a phase related to charge and the corresponding orbital phase. 
%%%%%%%%%%%%%%%%%%%%%%%%%%%%%%%%%%%%%%%%%%%%%%%%%%%%%%%%
\begin{widetext}
\begin{center}
%%%%%%%%%%%%%%%%%%%% TABLE %%%%%%%%%%%%%%%%%%%%%%%%%%%%%%%%
\begin{table}
\caption{\label{tab:three-models} Three models considered in Sec.~\ref{sec:weak-coupling}, \ref{sec:DMRG}   
and their parametrization.  See Figs.~\ref{fig:alkaline-2leg} and \ref{fig:p-band-2leg}  
for the physical process to which each parameter corresponds.   
In the first two models, pair-hopping does not exist.}
\begin{ruledtabular}
\begin{tabular}{lcccccc}
models &  parameters & hopping & intra-orbital & inter-orbital  & Hund & pair-hop. \\
\hline
$g$-$e$ model\footnotemark[1] [eq.\eqref{eqn:Gorshkov-Ham}] 
& $(t,U_{mm},V,V_{\text{ex}}^{g\text{-}e})$ & $t$ & $U_{mm}$ & $V$ & $V_{\text{ex}}^{g\text{-}e}$ & -- \\
generalized Hund model\footnotemark[2] [eq.\eqref{alkaourmodel}] 
& $(t,U,J,J_{z})$ & $t$ & $U+J_z/2$ & $U-J_z/2$ & $J$ &  -- \\
$p$-band model\footnotemark[3] [eq.\eqref{eqn:p-band}] 
& $(t,U_1,U_2)$ & $t$ & $U_1$ & $U_2$ & $U_2$ & $U_2$ 
\end{tabular}
\end{ruledtabular}
\footnotetext[1]{We have set: $t_g = t_e =t$, $U_{gg}=U_{ee}=U_{mm}$.}%
\footnotetext[2]{Equivalent to $g$-$e$ model through eq.\eqref{eqn:Gorshkov-to-Hund}.}%
\footnotetext[3]{$U_1=3 U_2$ for axially-symmetric trap.}%
\end{table}
%%%%%%%%%%%%%%%%%%%%%%%%%%%%%%%%%%%%%%%%%%%%%%%%%%%%%%%%
\end{center}
\end{widetext}
%%%%%%%%%%%%%%%%%%%%%%%%%%%%%%%%%%%%%%%%%%%%%%%%%%%%

%%%%%%%%%%%%%%%%%%%%%%%%%%%%%%%%%%%%%%%%%%%%%%%%%%%%%%%%%%
%%%%%%%%%%%%%%%%%%%%%%%%%%%%%%%%%%%%%%%%%%%%%%%%%%%%%

%%%%%%%%%%%%%%%%%%%%%%%%%%%%%%%%%%%%%%%%%%%%%%%%%%%%%
\subsection{\texorpdfstring{$\boldsymbol{N=2}$}{N=2}  {\em p}-band model}
\label{sec:DMRG_N2pband}
%%%%%%%%%%%%%%%%%%%%%%%%%%%%%%%%%%%%%%%%%%%%%%%%%%%%%
We now map out the phases of the $N=2$ $p$-band model \eqref{eqn:p-band-simple}  
as a function of $(U_1/t,U_2/t)$.  
While the physical realization with a harmonic trap imposes $U_1=3U_2$,  
we think that it is worth investigating the full phase diagram which could be accessible 
using other trapping schemes for instance (see Sec.~\ref{sec:p-band-definition}).  
Note also that Kobayashi {\it et al.} have recently reported in Ref.~\onlinecite{Kobayashi-O-O-Y-M-12} 
the presence of the spin Haldane (SH) phase in the same model at a slightly different ratio $U_2/U_1$. 

\begin{figure}[ht]
\centering
\includegraphics[width=\columnwidth,clip]{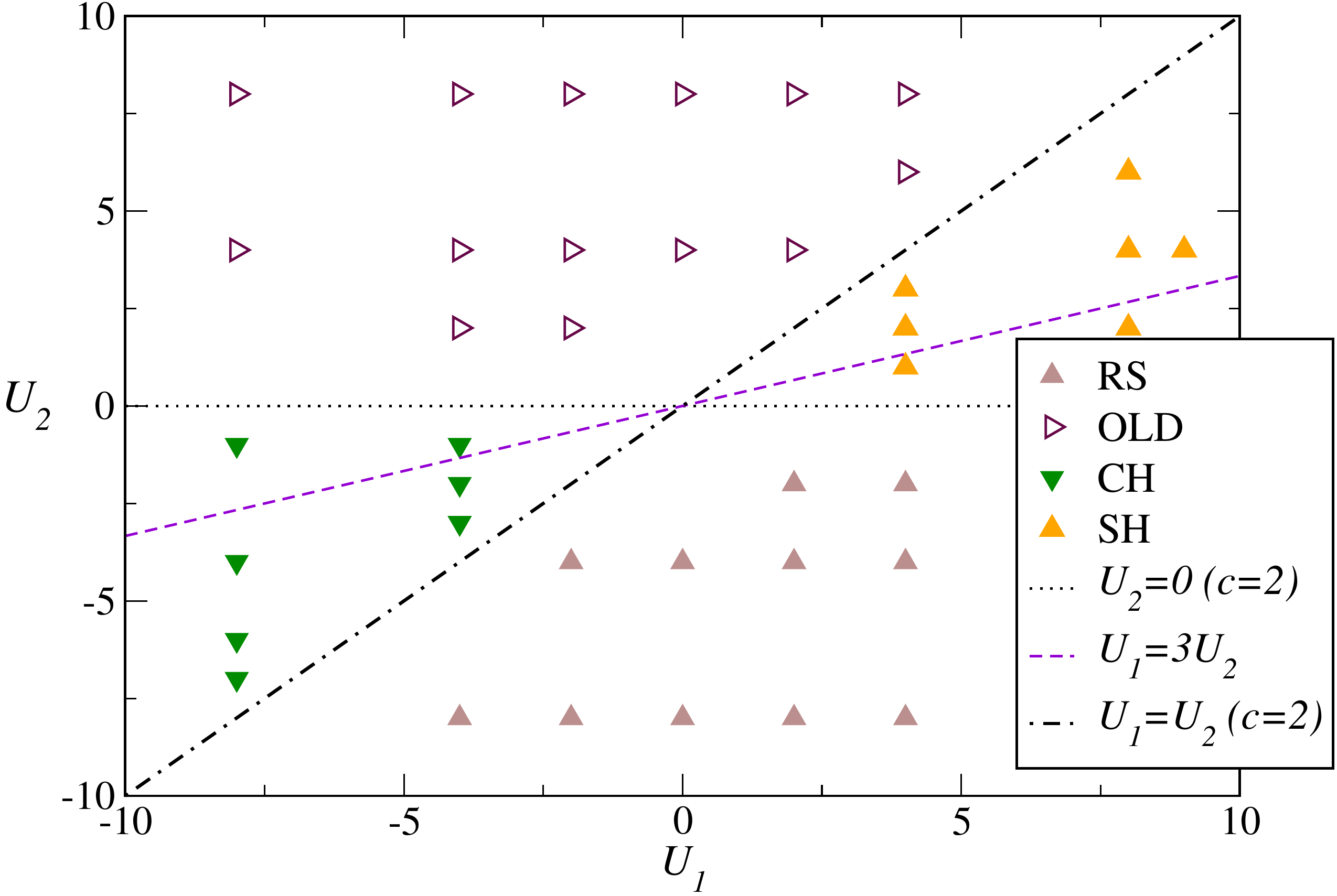}
\caption{(Color online) Phase diagram for $N=2$ $p$-band model \eqref{eqn:p-band-simple} 
obtained by DMRG. Note the mapping \eqref{eqn:Shiba-tr-SU2-U1U2} 
$(U_1,U_2) \rightarrow (-U_1,-U_2)$ which interchanges spin and charge. The line $U_1=3U_2$ corresponds to the axially symmetric trapping scheme. The two other lines denote the transitions and are compatible with the expected $c=2$ Luttinger liquid behavior.
\label{fig:phasediag_pband_N2}}
\end{figure}

The phase diagram (Fig.~\ref{fig:phasediag_Gorshkov_N2_Vex}) obtained 
exhibits a remarkable symmetry with respect to the origin.  In fact, as has been discussed in 
Sec.~\ref{sec:spin-charge-interchange}, the $p$-band model possesses the symmetry under 
the Shiba transformation \eqref{eqn:Shiba-tr-SU2-2band} under which spin and charge are interchanged 
by the mapping: $(U_1,U_2) \mapsto (-U_1,-U_2)$.   
Consequently, the SH and CH phases appear in a symmetric manner 
in Fig.~\ref{fig:phasediag_pband_N2}.  
The remaining areas of the phase diagram are filled respectively with the trivial RS phase 
(with $T^{z}=0$) and 
its symmetry partner, the orbital large-$D$ (OLD$_{x,}$) one. 
We have not investigated in details the transition between these phases, but their locations are in excellent agreement with the weak coupling predictions (i.e. $U_2=0$ and $U_1=U_2$). Moreover, using block entanglement entropy scaling at the transition, one can obtain an estimate of the central charge~\cite{Calabrese-C-04,Capponi-L-M-13}, estimated to
 be 1.8 (on $L=64$ chain with $U_1=U_2=-8t$ for instance, data not shown), 
rather close to the expected $c=2$ behavior discussed in Appendix D.

%%%%%%%%%%%%%%%%%%%%%%%%%%%%%%%%%%%%%%%%%%%%%%%%%%%%%%%%%%
%%%%%%%%%%%%%%%%%%%%%%%%%%%%%%%%%%%%%%%%%%%%%%%%%%%%%
%%%%%%%%%%%%%%%%%%%%%%%%%%%%%%%%%%%%%%%%%%%%%%%%%%%%%
\subsection{\texorpdfstring{$\boldsymbol{N=4}$}{N=4}  \texorpdfstring{$\boldsymbol{g}$}{g}-\texorpdfstring{$\boldsymbol{e}$}{e} model}
\label{sec:N4-Gorshkov}
\begin{figure}[!ht]
\centering
\includegraphics[width=\columnwidth,clip]{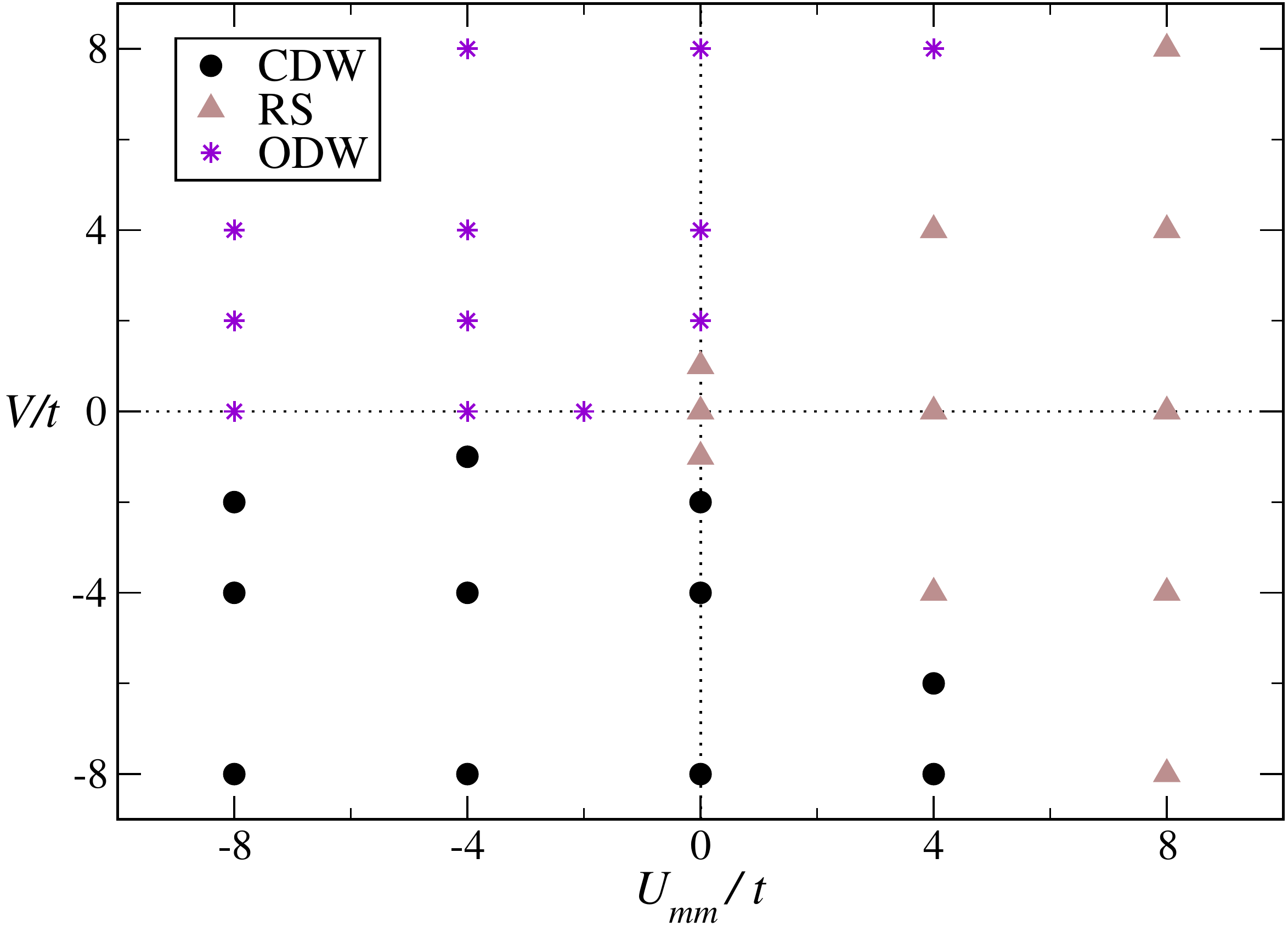}
\includegraphics[width=\columnwidth,clip]{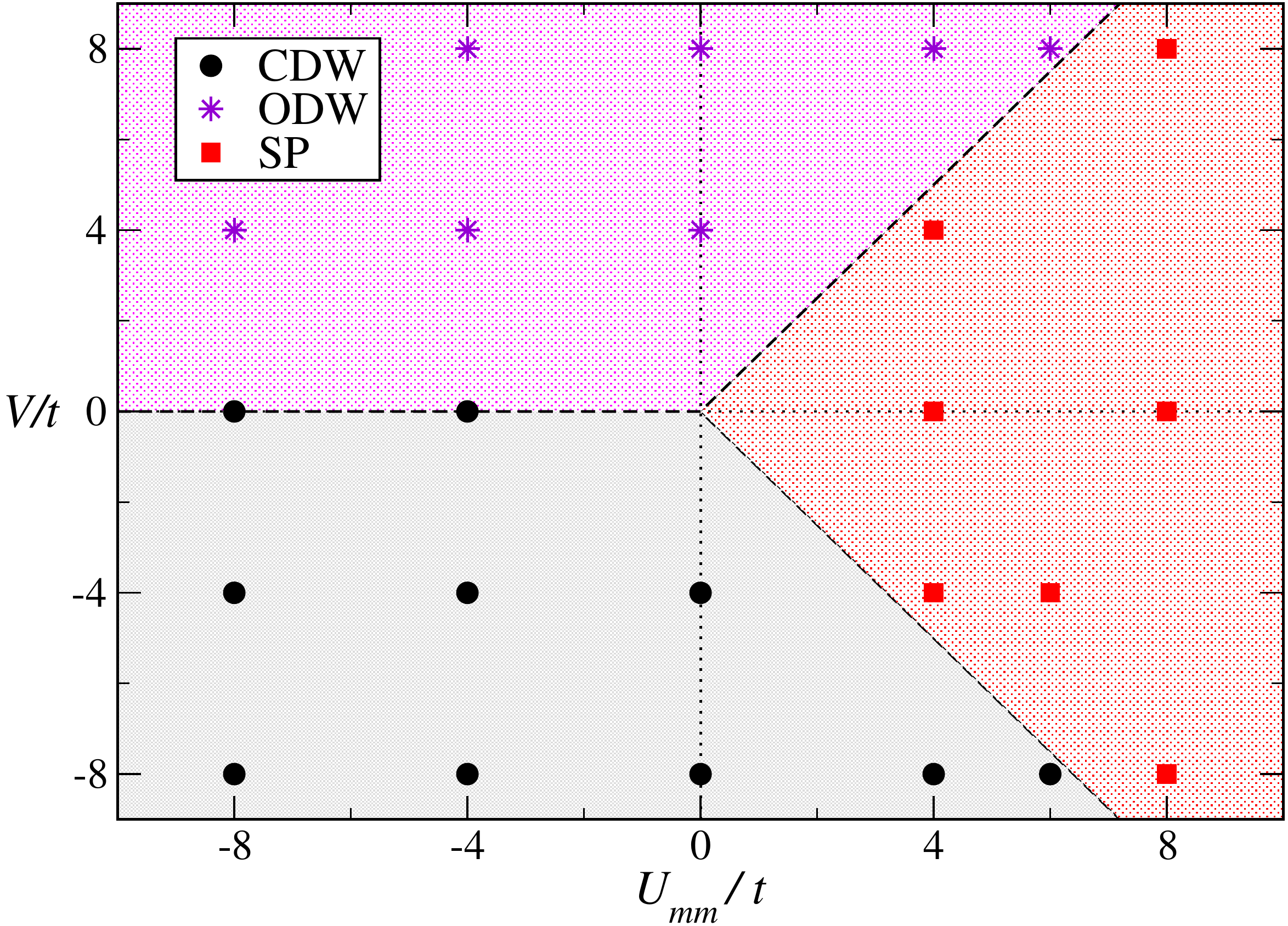}
\includegraphics[width=\columnwidth,clip]{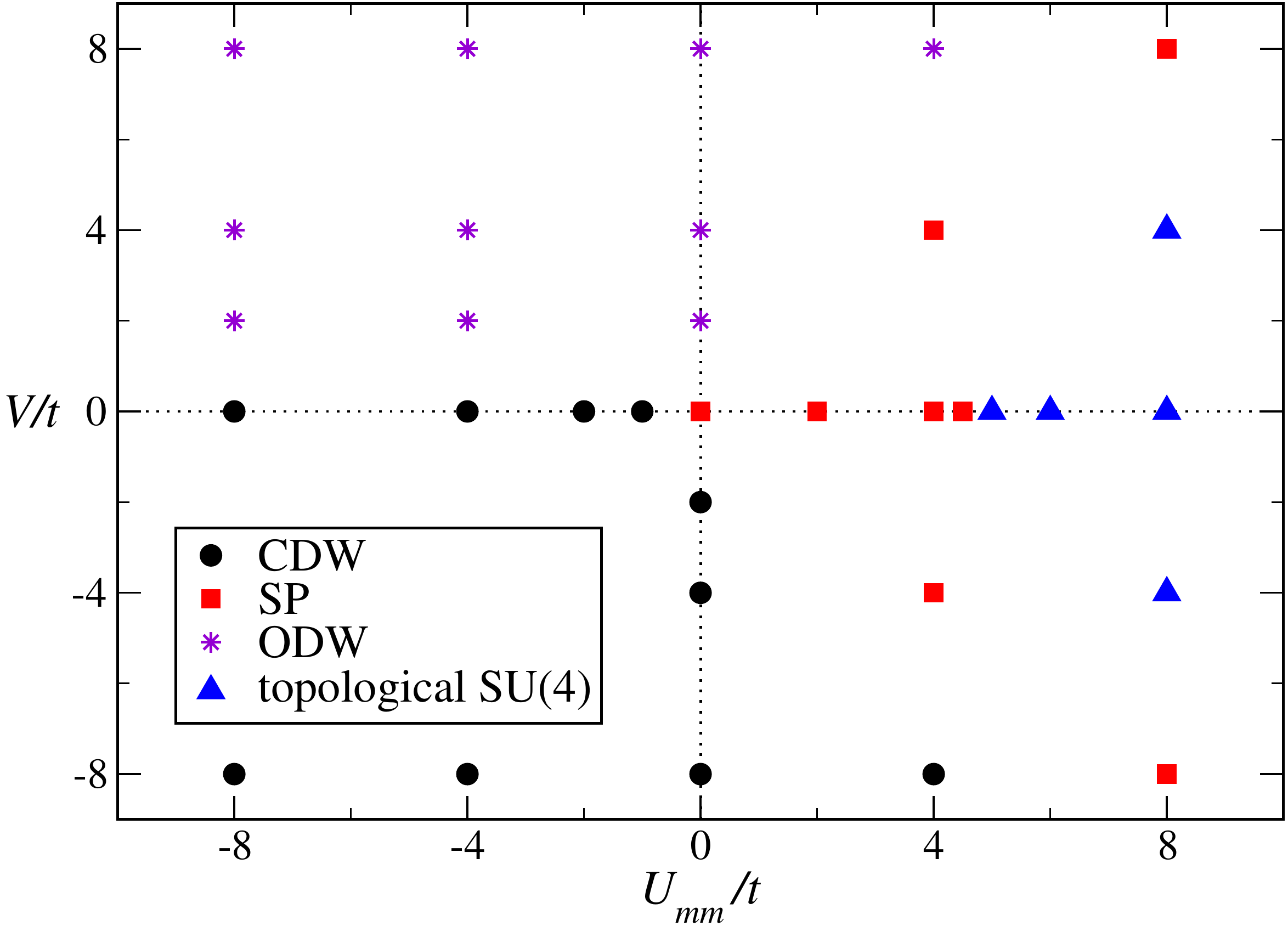}
\caption{(Color online) Phase diagrams for $N=4$ $g$-$e$ model (\ref{eqn:Gorshkov-Ham}) obtained by DMRG. From top to bottom, panels correspond respectively to $V_{\text{ex}}^{g\text{-}e}/t=-1$, $V_{\text{ex}}^{g\text{-}e}/t=0$   and $V_{\text{ex}}^{g\text{-}e}/t=1$. 
Symbols correspond to the numerical data obtained by DMRG with $L=36$ while colored regions and dashed lines indicate the one-loop numerical RG results. 
\label{fig:phasediag_Gorshkov_N4_Vex}}
\end{figure}

%%%%%%%%%%%%%%%%%%%%%%%%%%%%%%%%%%%%%%%%%%%%%%%%%%%%%
Here we consider again model $\mathcal{H}_{g\text{-}e}$
 (\ref{eqn:Gorshkov-Ham}) as in Sec.~\ref{sec:N2-Gorshkov}, but in the
$N=4$ case.  In the low-energy analysis of Sec.~\ref{sec:weak-coupling}, it was argued that, 
in comparison with the rich $N=2$ case, there were no more (symmetry-protected) topological phases for
the nuclear spin degrees of freedom, 
but only degenerate ones (CDW, ODW or
SP) and the non-degenerate OH and RS phases. Our numerical simulations do 
confirm these predictions at weak-coupling as shown in
Fig.~\ref{fig:phasediag_Gorshkov_N4_Vex} for fixed $V_{\text{ex}}^{g\text{-}e}/t=-1$, $0$ and $1$, 
although the one-loop RG results from Sec.~\ref{sec:RG-phase-diag} were
obtained at much smaller $V_{\text{ex}}^{g\text{-}e}/t$ values. 
The phase diagram for $V_{\text{ex}}^{g\text{-}e}/t=0$ clearly shows symmetry with respect to $V=0$ 
(see the middle panel of  Fig.~\ref{fig:phasediag_Gorshkov_N4_Vex}).  
Actually, this is a natural consequence of the orbital-charge interchange symmetry discussed in 
Sec.~\ref{sec:orbita-charge-interchange}; the transformation $V \rightarrow -V$ maps the CDW 
phase on the $V>0$ side to the ODW one on the $V<0$ side (see Table \ref{tab:Gorshkov}).  

Moreover, both CDW and ODW are rather insensitive to the value of $V_{\text{ex}}^{g\text{-}e}$. On the contrary, as was emphasised in the previous sections, the sign of $V_{\text{ex}}^{g\text{-}e}$ plays a major role in the positive $U_{mm}$ region. For  $V_{\text{ex}}^{g\text{-}e}<0$,  the SP phase gives way to the trivial RS phase. 
For $V_{\text{ex}}^{g\text{-}e}>0$, on the other hand, 
the SP phase  remains stable at weak and intermediate coupling as found using RG. 
There is however a crucial difference for $V_{\text{ex}}^{g\text{-}e}/t=1$ at strong coupling since we also find a large region of the topological SU(4) phase discussed in Sec.~\ref{sec:SUN-topological-phase} 
(see the lower panel of Fig.~\ref{fig:phasediag_Gorshkov_N4_Vex}).

%%%%%%%%%%%%%%%%%%%%%%%%%%%%%%%%%%%%%%%%%%%%%%%%%%%%%%%%%%
%%%%%%%%%%%%%%%%%%%%%%%%%%%%%%%%%%%%%%%%%%%%%%%%%%%%%

\begin{figure}[ht]
\centering
\includegraphics[width=\columnwidth,clip]{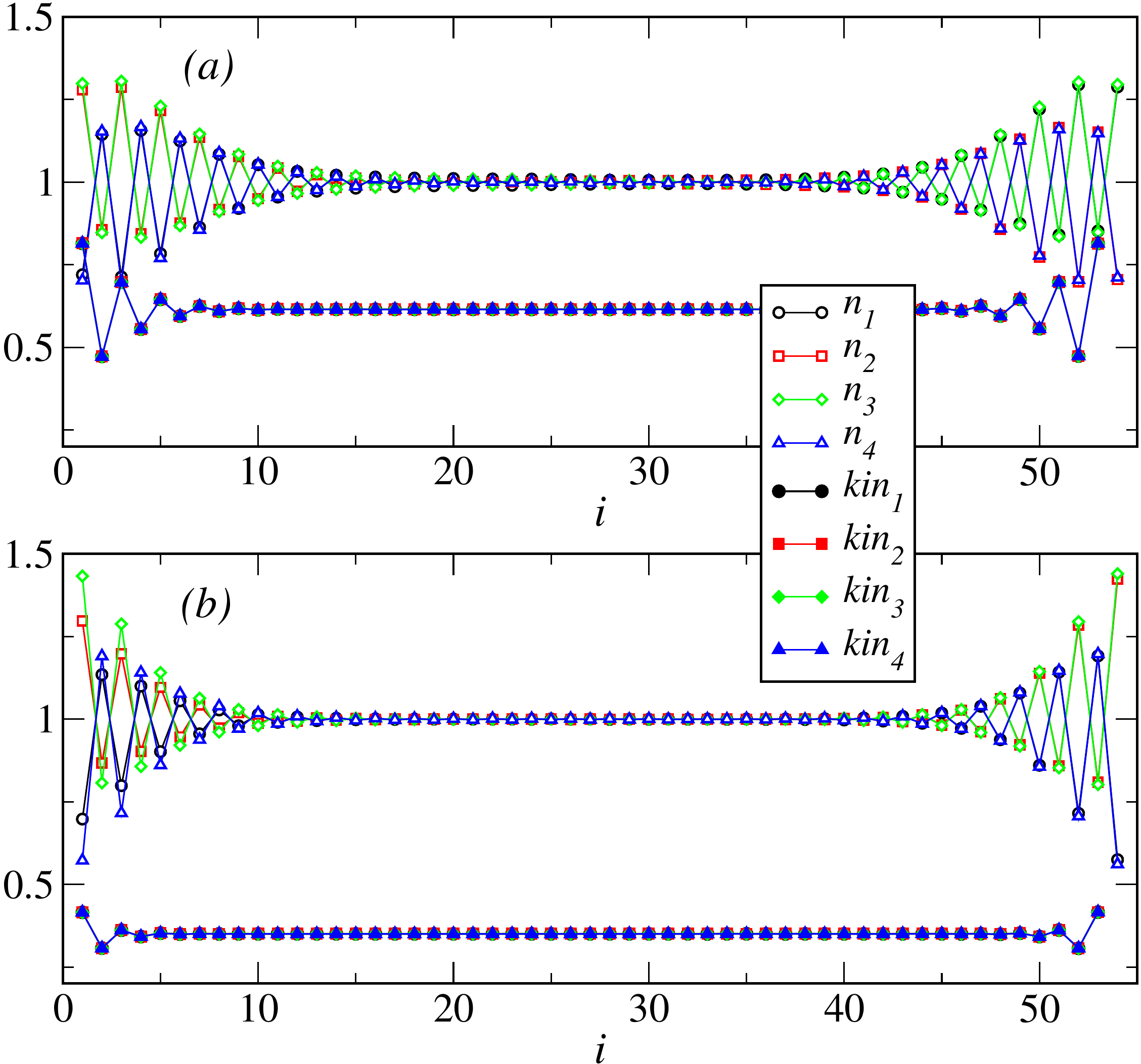}
\caption{(Color online) DMRG results for the local fermion densities 
and kinetic energies for each flavor $\alpha=1,\ldots,4$ in the $N=4$ case. 
Panel (a) corresponds to the $g$-$e$ model with $U/t=8$, $V=0$ and $V_{\text{ex}}/t=1$ on $L=54$ chain. Panel (b) corresponds to the $p$-band model with $U_1/t=12$ and $U_2/t=4$ on $L=54$ chain. The presence 
of localized edge states is clearly visible in both cases. 
\label{fig:edge_DMRG}}
\end{figure}

Clear signatures of the topological SU(4) phase are given by the existence of 6-fold degenerate 
edge states~\cite{Nonne-M-C-L-T-13} (see Fig.~\ref{fig:edge_DMRG}(a)), or by the 6-fold degeneracy of the dominant eigenvalue in the entanglement spectrum of half a system~\cite{Tanimoto-T-14} (data not shown). 
While the edge states should not occur 
in the true ground-state, which is highly entangled but exponentially close in energy to the other low-lying states (similarly to the spin-1 Haldane non-magnetic ground-state which lies very close to the so-called Kennedy triplets), it is known that DMRG will target a minimally entangled state~\cite{Jiang-W-B-12} and thus for a large enough system size (at a fixed number of states $m$), the DMRG algorithm will ultimately lead to one of the quasi-degenerate ground-states with some edge states configurations, as is observed in Fig.~\ref{fig:edge_DMRG}. For $N=4$, a simple physical interpretation of the 6-fold degeneracy is given by the number of ways of choosing two colors among four.   
Using the VBS wave function obtained in Sec.~\ref{sec:SUN-VBS-state}, 
one can explicitly compute the local fermion densities $n_{\alpha,i}$ ($\alpha=1,2,3,4$).  
Near the left edge of a sufficiently large system, two of the four $\{n_{\alpha,i}\}$ decay as $1+3 (-1/5)^r$ 
and the other two as $1-3 (-1/5)^r$ ($r$ being the distance from the left edge).  
The existence of the two different kinds of color-pairing on the left and right edges 
is clearly seen in Fig.\ref{fig:edge_DMRG}(a) and gives another support for the SPT nature  
of the SU(4) phase found here. 

%%%%%%%%%%%%%%%%%%%%%%%%%%%%%%%%%%%%%%%%%%%%%%%%%%%%%%%%%%
%%%%%%%%%%%%%%%%%%%%%%%%%%%%%%%%%%%%%%%%%%%%%%%%%%%%%

%%%%%%%%%%%%%%%%%%%%%%%%%%%%%%%%%%%%%%%%%%%%%%%%%%%%%
\subsection{\texorpdfstring{$\boldsymbol{N=4}$}{N=4} SU(2)$_\text{o}$  \texorpdfstring{$\boldsymbol{g}$}{g}-\texorpdfstring{$\boldsymbol{e}$}{e} model}
\begin{figure}[ht]
\centering
\includegraphics[width=\columnwidth,clip]{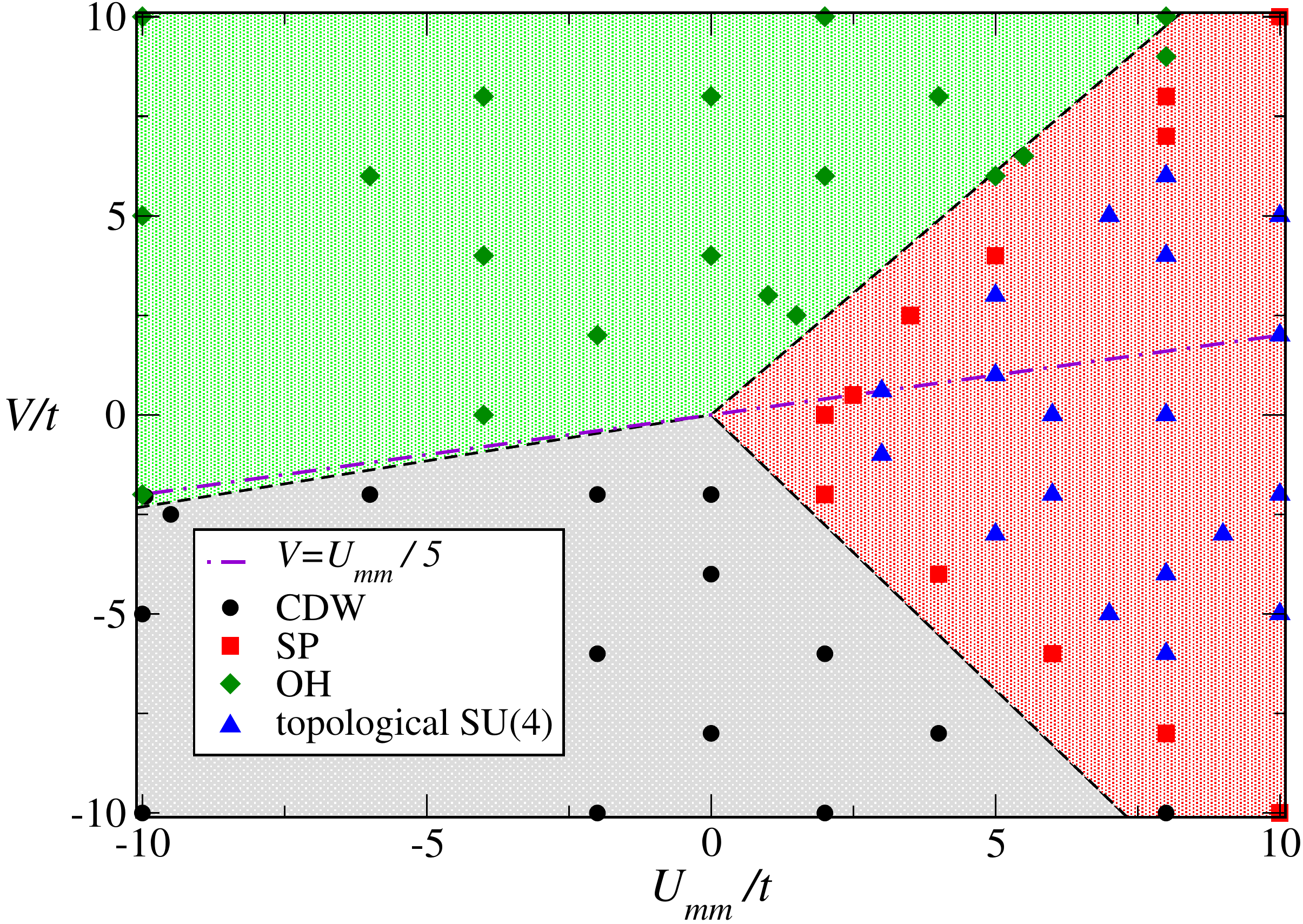}
\caption{(Color online) Phase diagram for the $N=4$ $g$-$e$ model (\ref{eqn:Gorshkov-Ham}) with SU(2)$_\text{o}$ symmetry, i.e. $U_{mm}-V=V_{\text{ex}}^{g\text{-}e}$. Symbols correspond to the numerical data obtained by DMRG with $L=36$ while colored regions and dashed lines indicate the one-loop numerical RG results. We also plot the special line $V=U_{mm}/5$ (see text). 
\label{fig:phasediag_Gorshkov_N4_SU2}}
\end{figure}

%%%%%%%%%%%%%%%%%%%%%%%%%%%%%%%%%%%%%%%%%%%%%%%%%%%%%
We now consider the same $N=4$ model but imposing SU(2)$_\text{o}$ symmetry, 
i.e. $U_{mm}-V=V_{\text{ex}}^{g\text{-}e}$ [$J=J_{z}$; see Eq.~\eqref{eqn:Gorshkov-to-Hund}]. 
The phase diagram as a function of $(U_{mm},V)$ is shown in Fig.~\ref{fig:phasediag_Gorshkov_N4_SU2} together with the one-loop RG result. 
We observe that the agreement is excellent at weak-coupling, and rather good at all couplings 
for the phase boundaries CDW/SP, CDW/OH and OH/SP. Still, we emphasize that the RG results shown as dashed lines are mostly guide to the eyes for these transitions.  Moreover, as expected from our strong-coupling analysis, we do confirm the presence of the SU(4) topological phase  along the special line $V=U_{mm}/5$ at strong $V>0$.\footnote{See also Ref.~\onlinecite{Nonne-M-C-L-T-13} where this line was investigated in strong coupling and numerically in the equivalent formulation of a generalized Hund model} In fact, this topological phase occupies a large fraction of the phase diagram, which in our opinion makes its potential observation quite promising. 
A quantum phase transition  necessarily takes place between the SP and the SU(4) topological phase. A  precise numerical determination of its nature  is beyond the scope of this paper.  

%%%%%%%%%%%%%%%%%%%%%%%%%%%%%%%%%%%%%%%%%%%%%%%%%%%%%%%%%%
%%%%%%%%%%%%%%%%%%%%%%%%%%%%%%%%%%%%%%%%%%%%%%%%%%%%%
%%%%%%%%%%%%%%%%%%%%%%%%%%%%%%%%%%%%%%%%%%%%%%%%%%%%%%%%%
\subsection{\texorpdfstring{$\boldsymbol{N=4}$}{N=4} SU(2)$_\text{o}$ generalized Hund model}
\begin{figure}[htb]
\centering
\includegraphics[width=\columnwidth,clip]{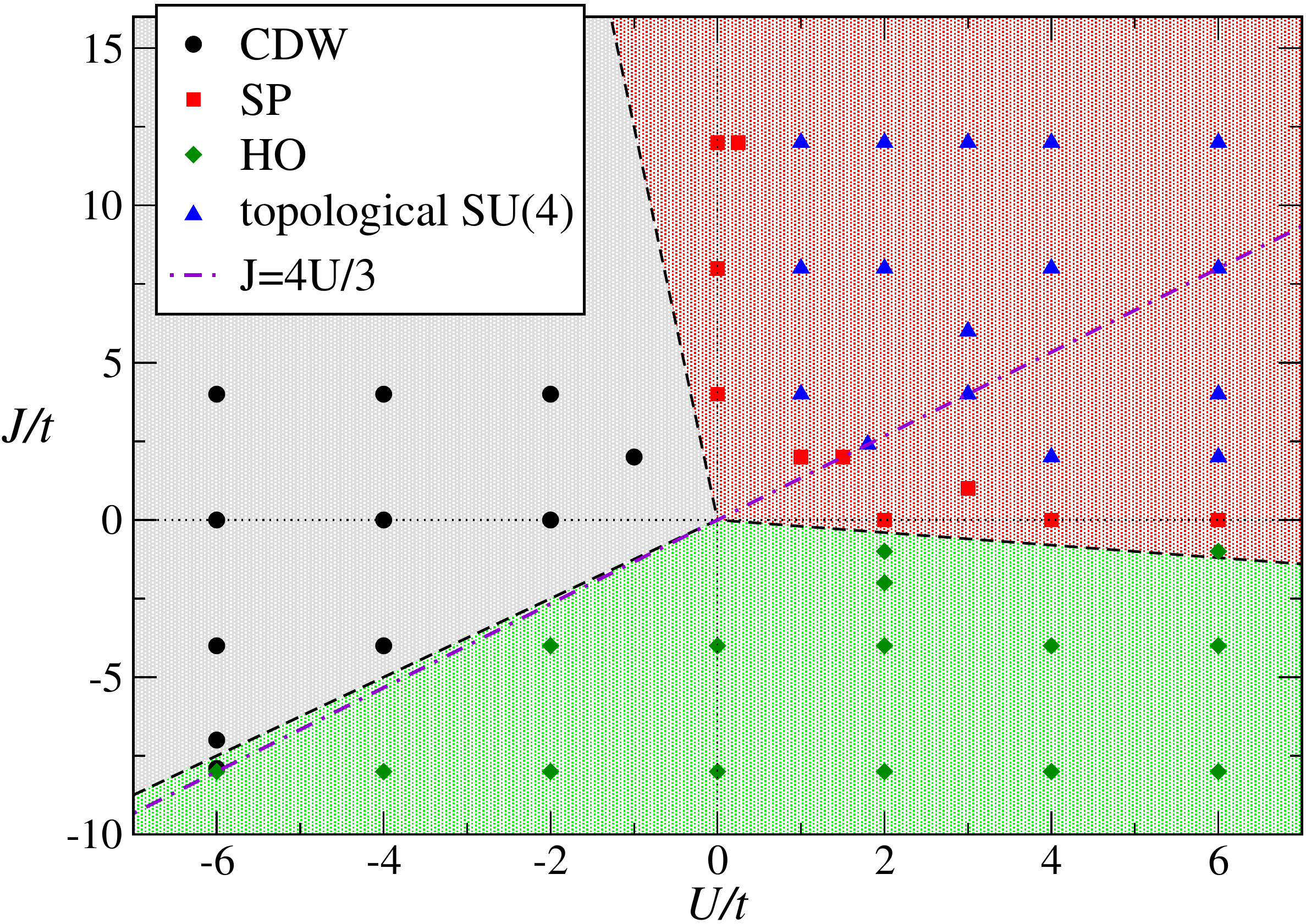}
\caption{(Color online) Phase diagram for the $N=4$ generalized Hund model (\ref{alkaourmodel}) 
with SU(2)$_{\text{o}}$ symmetry.  
Symbols correspond to the numerical data obtained by DMRG with $L=36$ while colored regions and dashed lines indicate the one-loop numerical RG results. We also plot the special line $J=4U/3$ (see text). 
\label{fig:phasediag_alka_N4}}
\end{figure}

%%%%%%%%%%%%%%%%%%%%%%%%%%%%%%%%%%%%%%%%%%%%%%%%%%%%%%%%%
As discussed in Sec.~\ref{sec:Gorshkov-Hamiltonian}, 
the SU(2)$_\text{o}$ model can also be parametrized as a function of $(U,J)$ in the generalized Hund model (\ref{alkaourmodel}). 
This means that we can simply take the data of the previous paragraph and replot them accordingly in Fig.~\ref{fig:phasediag_alka_N4}. Obviously, we obtain the same set of phases, and the same extent of 
agreement with the one-loop RG numerical result as far as the structure in the weak-coupling region 
and the locations of the phase transitions are concerned.  
As already noted in Ref.~\onlinecite{Nonne-M-C-L-T-13}, the topological SU(4) phase is stable along the special line $J=4U/3$ at strong coupling $J>0$, but our numerical results prove that it has an unexpectedly large extent 
in the first quadrant $U,J>0$. 

%%%%%%%%%%%%%%%%%%%%%%%%%%%%%%%%%%%%%%%%%%%%%%%%%%%%%%%%%%
%%%%%%%%%%%%%%%%%%%%%%%%%%%%%%%%%%%%%%%%%%%%%%%%%%%%%
%%%%%%%%%%%%%%%%%%%%%%%%%%%%%%%%%%%%%%%%%%%%%%%%%%%%%%%%%%
\subsection{\texorpdfstring{$\boldsymbol{N=4}$}{N=4} generalized Hund model without SU(2)$_\text{o}$ symmetry}
\label{sec:N4-gen-Hund-wo-SU2}
%%%%%%%%%%%%%%%%%%%%%%%%%%%%%%%%%%%%%%%%%%%%%%%%%%%%%%%%%%
\subsubsection{$J_z=0$}
\begin{figure}[ht]
\centering
\includegraphics[width=\columnwidth,clip]{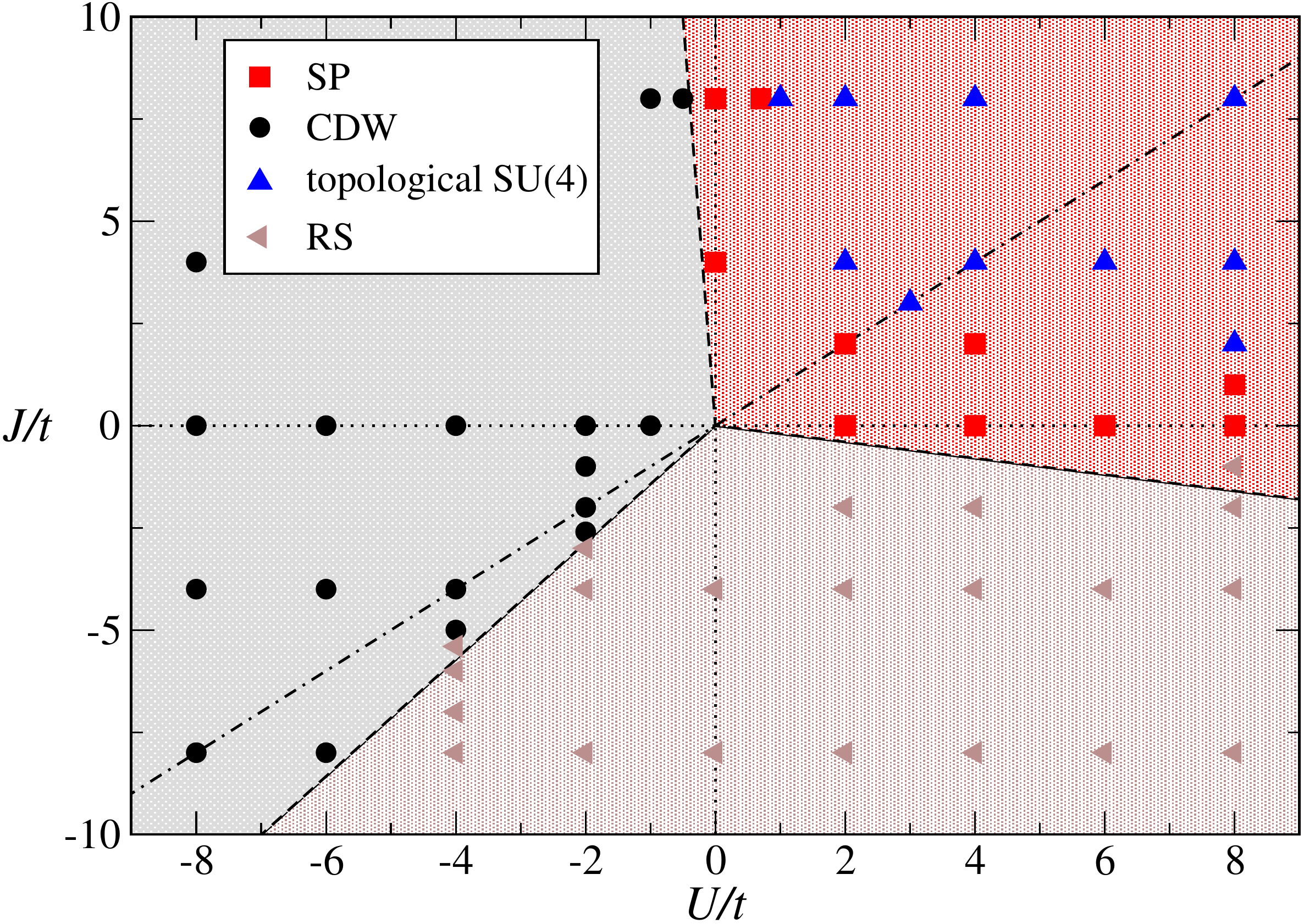}
\caption{(Color online) Phase diagram for the $N=4$ generalized Hund  model (\ref{alkaourmodel}) 
with $J_z=0$. Symbols correspond to the numerical data obtained by DMRG with $L=36$ while colored regions and dashed lines indicate the one-loop numerical RG results. We also plot the special line $J=U$ where the model can be mapped onto the $N=4$ $p$-band model. 
\label{fig:phasediag_alka_N4_Jz_0}}
\end{figure}

%%%%%%%%%%%%%%%%%%%%%%%%%%%%%%%%%%%%%%%%%%%%%%%%%%%%%%%%%%
We can also investigate parameter region \emph{without} SU(2)$_\text{o}$ symmetry ($J\neq J_z$) 
for the generalized Hund model  \eqref{alkaourmodel} in order to check the robustness of the observations 
made for the (fine-tuned) SU(2)-symmetric model. 
In Fig.~\ref{fig:phasediag_alka_N4_Jz_0}, we present our numerical results for $J_z=0$ together with the RG phase boundaries. 
Again, we obtained remarkable agreement at weak coupling 
as well as the semi-quantitative results concerning the phase transitions. 
The main difference from the SU(2)$_\text{o}$ case consists in the disappearance 
of the OH which is replaced by the trivial singlet phase RS. 
In the strong-coupling picture, this result is obvious since the model maps 
onto a large-$D$ spin-2 chain [see eq.\eqref{eqn:eff-Ham-orbital-Haldane}]. 
However, the topological SU(4) phase is scarcely affected by the breaking of SU(2)$_\text{o}$  
and it still occupies a large fraction of the $U,J>0$ region.  
Finally, we have indicated in Fig.~\ref{fig:phasediag_alka_N4_Jz_0} 
the $J=U$ line which can be mapped onto the special line $U_1= 3U_2$  
of the $N=4$ $p$-band model upon the identification $J=U=2U_2$.  We will use this property later 
in Sec~\ref{sec:N4-p-band}. 

%%%%%%%%%%%%%%%%%%%%%%%%%%%%%%%%%%%%%%%%%%%%%%%%%%%%%%%%%%
%%%%%%%%%%%%%%%%%%%%%%%%%%%%%%%%%%%%%%%%%%%%%%%%%%%%%
%%%%%%%%%%%%%%%%%%%%%%%%%%%%%%%%%%%%%%%%%%%%%%%%%%%%%%%%%%
\subsubsection{$J_z/t=\pm 4$}
\label{sec:N4-gen-Hund-woSU2-J-pm4}
%%%%%%%%%%%%%%%%%%%%%%%%%%%%%%%%%%%%%%%%%%%%%%%%%%%%%%%%%%
\begin{figure}[ht]
\centering
\includegraphics[width=\columnwidth,clip]{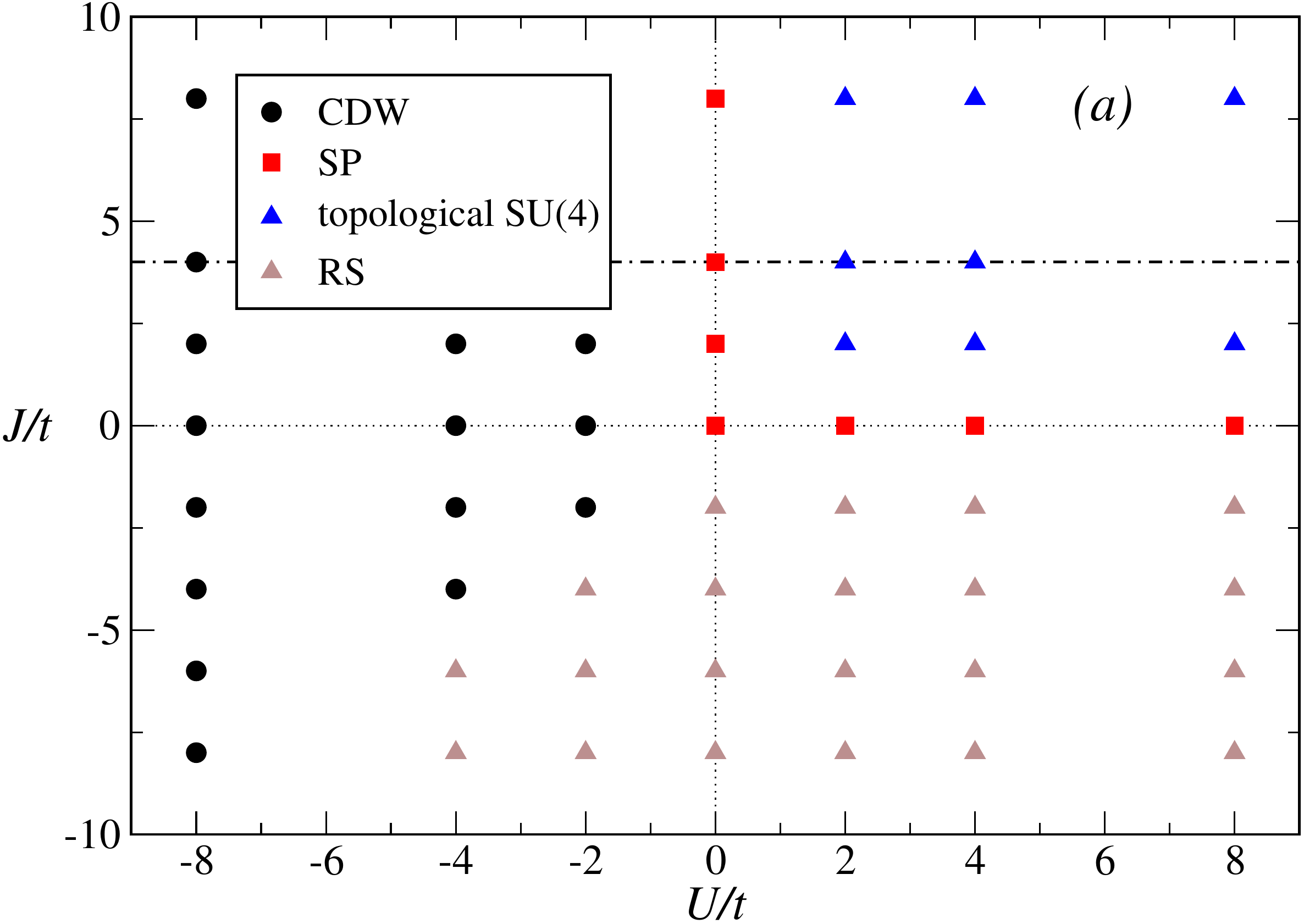}
\includegraphics[width=\columnwidth,clip]{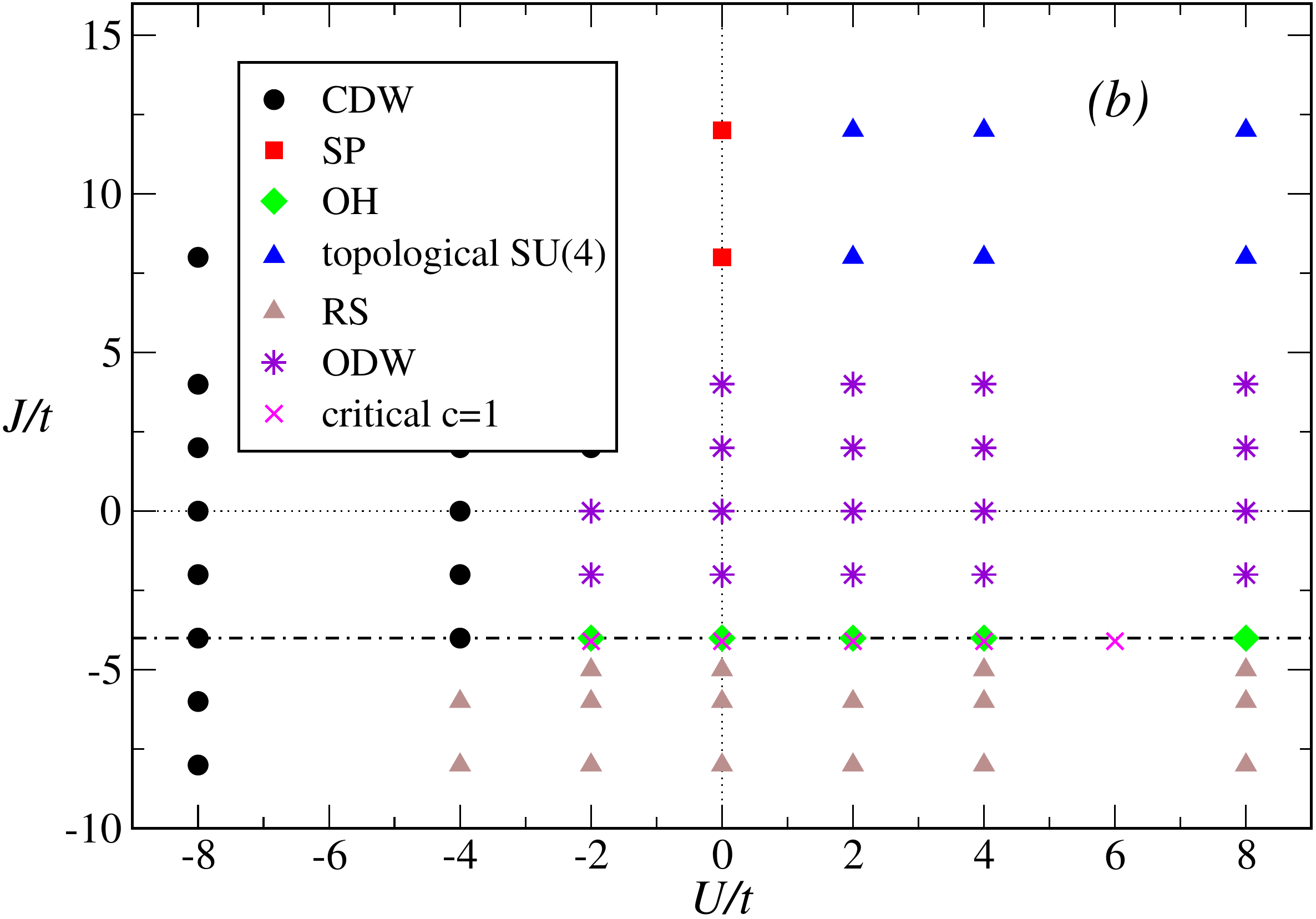}
\caption{(Color online) Phase diagram for the $N=4$ generalized Hund  model (\ref{alkaourmodel})  
with $J_z/t=4$ [(a)] and $J_z/t=-4$ [(b)]. Symbols correspond to the numerical data obtained by DMRG with $L=36$. Dash-dotted lines correspond to SU(2)$_\text{o}$ lines when $J=J_z$. The critical region for $J_z/t=-4$ is almost invisible on this scale (see text).
\label{fig:phasediag_alka_N4_Jz_4}}
\end{figure}
%%%%%%%%%%%%%%%%%%%%%%%%%%%%%%%%%%%%%%%%%%%%%%%%%%%%%

For fixed finite $J_z$, generically there is no SU(2)$_\text{o}$ symmetry.  
Nevertheless, one can understand part of the phase diagram starting from the line $J=J_z$. 

For $J_z/t=4$, as is seen in Fig.~\ref{fig:phasediag_alka_N4}, if we fix $J=J_z$ then the system will evolve 
from the CDW phase to the topological SU(4) one through the SP region with increasing $U$. 
Since these are all gapped phases, they must have a finite extension in the phase diagram. Our numerical results in Fig.~\ref{fig:phasediag_alka_N4_Jz_4}(a) confirm this expectation and moreover prove that these phases occupy a large fraction of the phase diagram. The remaining part of it contains the RS phase  in agreement with strong-coupling picture.

Considering now a fixed $J_z/t=-4$ and our previous results in Fig.~\ref{fig:phasediag_alka_N4} for $J=J_z$, we expect the appearance of CDW and OH by varying $U$. Our numerical phase diagram in Fig.~\ref{fig:phasediag_alka_N4_Jz_4}(b) recovers, of course, this result, but there is a crucial difference from the previous case. 
Indeed, our strong-coupling analysis reveals that starting from the OH phase, deviations from $J=J_z$ will induce an effective $D (T^z)^2$ term with $D=J_z-J$ [see Eq.~(\ref{eqn:eff-Ham-orbital-Haldane})]. 
This on-site anisotropy is well-known for spin-2 chain~\cite{Schollwock-G-J-96}, and it drives the OH phase either 
to the Ising ODW phase for large $D<0$, or to a non-degenerate singlet phase for $D>0$ 
(the so-called large-$D$ phase, which is equivalent to RS here) through an intermediate 
extended gapless $c=1$ phase lying in the interval $0.04 \lesssim D/{\cal J} \lesssim 3.0$, 
where ${\cal J}={\cal J}_{xy}={\cal J}_z$ is the effective spin exchange \eqref{eq:eff_coupling_Hund}.~\footnote{The upper critical value has been estimated more recently~\cite{Tonegawa-O-N-S-N-K-11} to be $D/{\cal J}=2.4$.}
This scenario away from the OH region is confirmed by our numerical phase diagram, 
although the  extent of the intermediate critical region is rather small in Fig.~\ref{fig:phasediag_alka_N4} due to the smallness of ${\cal J}$.  For the same reason, we have not investigated here whether the intermediate-$D$ phase,  
which has been proposed long-time ago by Oshikawa~\cite{Oshikawa1992} and only recently observed numerically 
in anisotropic spin-2 chains~\cite{Tonegawa-O-N-S-N-K-11,Tu2011,Tzeng2012,Kjall2013}, 
could appear in our phase diagram.  

\begin{figure}[ht]
\centering
\includegraphics[width=\columnwidth,clip]{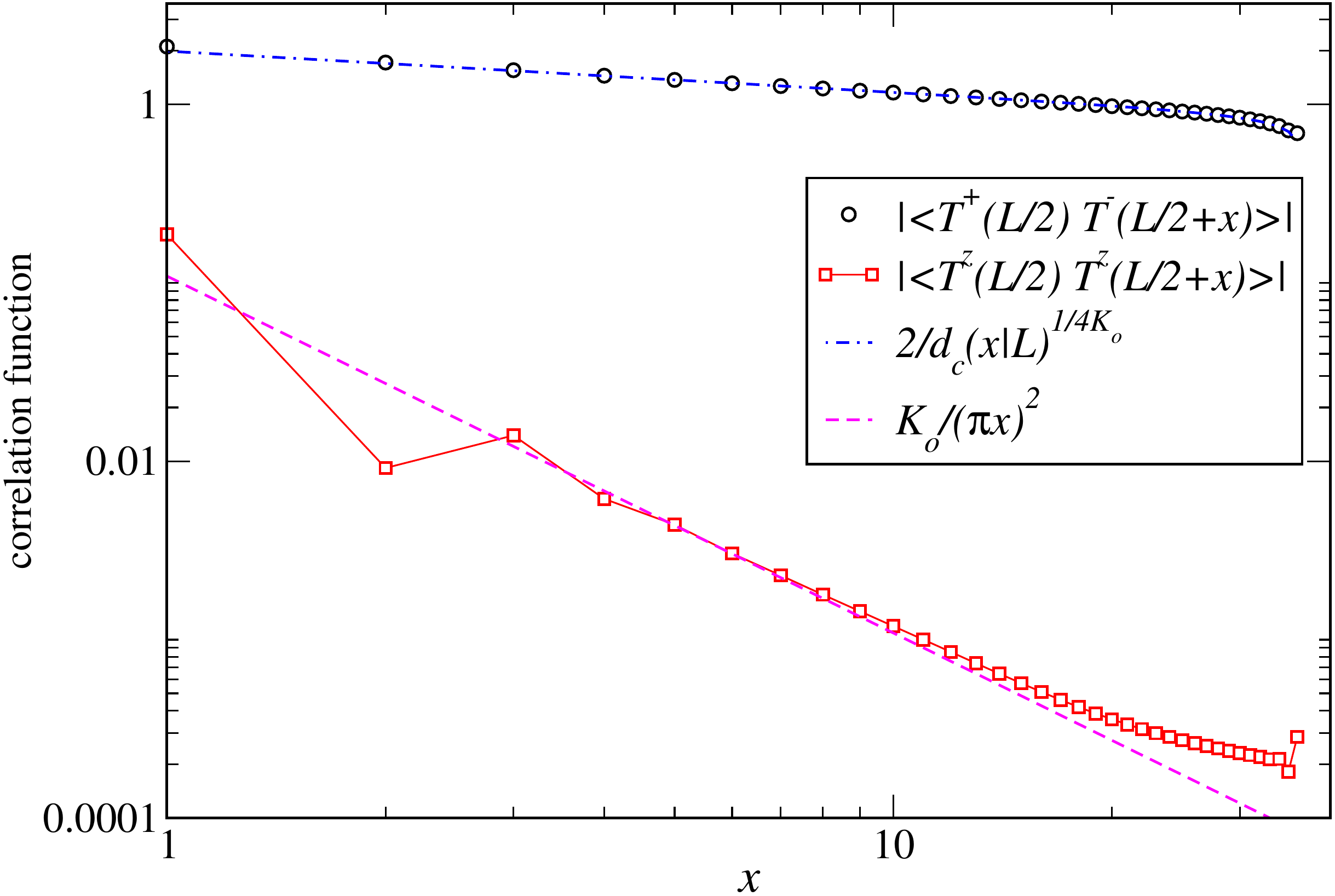}
\caption{(Color online) Absolute values of the transverse and longitudinal pseudo spin correlation functions measured from the middle of the chain on a $L=72$ system with parameters $U=0$, $J/t=-4.1$ and $J_z/t=-4$. Both can be fitted with a similar Luttinger parameter $K_\text{o} \simeq 1.09>1$ and appropriate functional forms (see text). 
\label{fig:corr_critical}}
\end{figure}

The existence of the critical region may be further evidenced by the measurement of 
the pseudo spin correlation functions. Using the low-energy predictions \eqref{correlBCSNeven} for $N=4$, 
and taking into account that we are measuring correlation from the middle of a chain with OBC, 
we use the appropriate functional form for the distance~\cite{Hikihara2004,Cazalilla-04,Roux-C-L-A-09}:
\begin{equation}
\langle T^+(L/2) T^-(L/2+x) \rangle \sim (-1)^x \left(d_c(x)\right)^{-1/4K_o}
\end{equation}
where 
\begin{equation}
d_c(x)=\frac{d(x+L/2|2(L+1)) d(x-L/2|2(L+1))}{\sqrt{d(2x|2(L+1))d(L|2(L+1))}}
\end{equation}
with $d(x|L)=L |\sin(\pi x/L)|/\pi$ is the conformal distance. Thus fitting, we get an excellent agreement (see Fig.~\ref{fig:corr_critical}) with the data and a Luttinger parameter $K_\text{o}=1.09$  
indeed larger than 1 as expected.  
An identical value was obtained when fitting the longitudinal correlations, too. 
This critical phase is thus described by the orbital Luttinger liquid \eqref{hoNeven}.  

%%%%%%%%%%%%%%%%%%%%%%%%%%%%%%%%%%%%%%%%%%%%%%%%%%%%%%%%%%
%%%%%%%%%%%%%%%%%%%%%%%%%%%%%%%%%%%%%%%%%%%%%%%%%%%%%

Another difference from the weak-coupling results lies in the large $J/t>0$ region where we have found 
surprisingly the reentrance of the SP and topological SU(4) phases that were found 
in other parts of the full three-dimensional parameters phase diagram. 
This confirms again that, contrary to the OH phase 
whose stability is limited to the proximity of the SU(2)$_\text{o}$-symmetric points, 
the SU(4) SPT phase could be stabilized for a large variety of parameters and thus could potentially be realized experimentally. 

%%%%%%%%%%%%%%%%%%%%%%%%%%%%%%%%%%%%%%%%%%%%%%%%%%%%%%%%%%
\subsection{Varying $J_z$ in the \texorpdfstring{$\boldsymbol{N=4}$}{N=4}  \texorpdfstring{$\boldsymbol{g}$}{g}-\texorpdfstring{$\boldsymbol{e}$}{e} model}
\begin{figure}[ht]
\centering
\includegraphics[width=\columnwidth,clip]{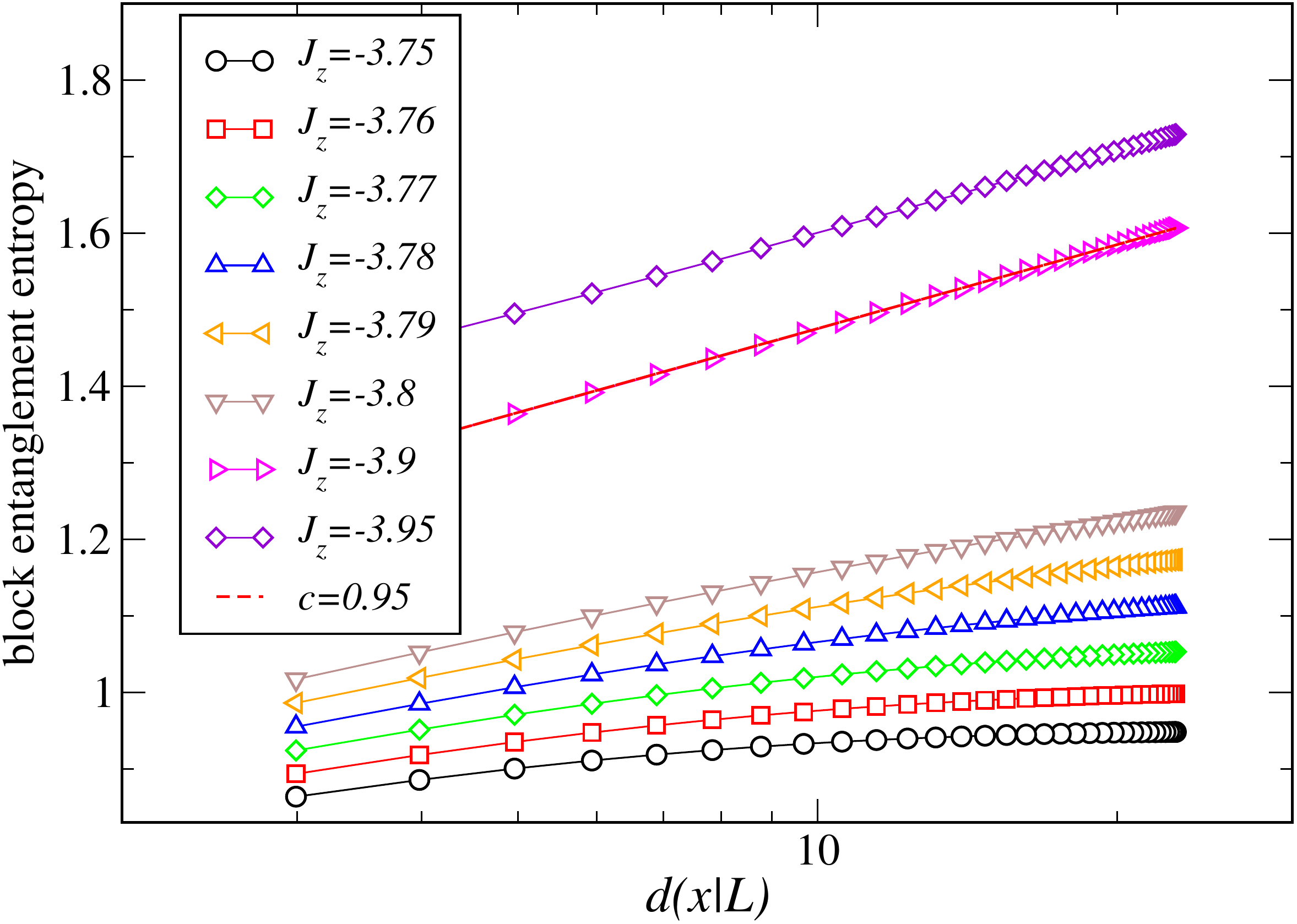}
\caption{(Color online) Von Neumann entanglement entropy $S_{\text{vN}}$  of a block of $x$ sites 
(starting from the left open edge) vs conformal distance $d(x|L)=(L/\pi) \sin (\pi x/L)$ for $U=0$ 
and $J/t=-4$ with varying parameters $J_z$ from the SU(2)$_\text{o}$ point $J_z=J$ 
with OH phase to the $J_z=0$ RS singlet phase. In the intermediate region, there is an extended critical gapless phase compatible 
with $c=1$ central charge.
\label{fig:entropy_N4_L72}}
\end{figure}

%%%%%%%%%%%%%%%%%%%%%%%%%%%%%%%%%%%%%%%%%%%%%%%%%%%%%%%%%%
As was shown before, if one starts from the OH phase in the SU(2)$_\text{o}$ case and then increases $J_z$, the OH phase will ultimately be replaced by the trivial RS phase.  
However, in the strong-coupling, we have an effective spin-$N/2$(=2) chain with some on-site anisotropy $D$ term. 
For such a system, we know that the transition from the Haldane phase to the trivial large-$D$ phase goes through an \emph{extended} gapless region~\cite{Schollwock-G-J-96} with central charge $c=1$. In Fig.~\ref{fig:entropy_N4_L72}, we present measurements of the von Neumann entropy $S_{\text{vN}}$ vs conformal distance $d(x|L)=(L/\pi) \sin (\pi x/L)$ for various parameters ($U=0$, $J/t=-4$ and $J_z>J$) obtained on $L=72$ chains. 
It is known~\cite{Calabrese-C-04} that this quantity will saturate in a gapped phase, and will scale as $S_{\text{vN}}=(c/6) \log d(x|L) + \mathrm{Cst}$ in a critical phase with central charge $c$.
As is expected from our strong-coupling 
results, our numerical data do confirm the presence of an extended critical phase compatible with $c=1$. 

If one uses the expressions from the strong-coupling (\ref{eq:eff_coupling_Hund}) for our choice of parameters, we are thus starting from an SU(2) spin-2 chain with exchange ${\cal J}=1/18$ (using $t=1$ as the unit of energy). As recalled in the previous subsection, an on-site anisotropy $D=J_z-J$ will induce a critical phase when  $0.04 \lesssim D/{\cal J} \lesssim 3$, or assuming that ${\cal J}$ is not changed, 
$ -3.998 \lesssim J_z \lesssim < -3.83$ in good agreement with our numerical data too. 

%%%%%%%%%%%%%%%%%%%%%%%%%%%%%%%%%%%%%%%%%%%%%%%%%%%%%%%%%%
%%%%%%%%%%%%%%%%%%%%%%%%%%%%%%%%%%%%%%%%%%%%%%%%%%%%%
%%%%%%%%%%%%%%%%%%%%%%%%%%%%%%%%%%%%%%%%%%%%%%%%%%%%%
\subsection{\texorpdfstring{$\boldsymbol{N=4}$}{N=4}  \texorpdfstring{$\boldsymbol{p}$}{p}-band model}
\label{sec:N4-p-band}
\begin{figure}[ht]
\centering
\includegraphics[width=\columnwidth,clip]{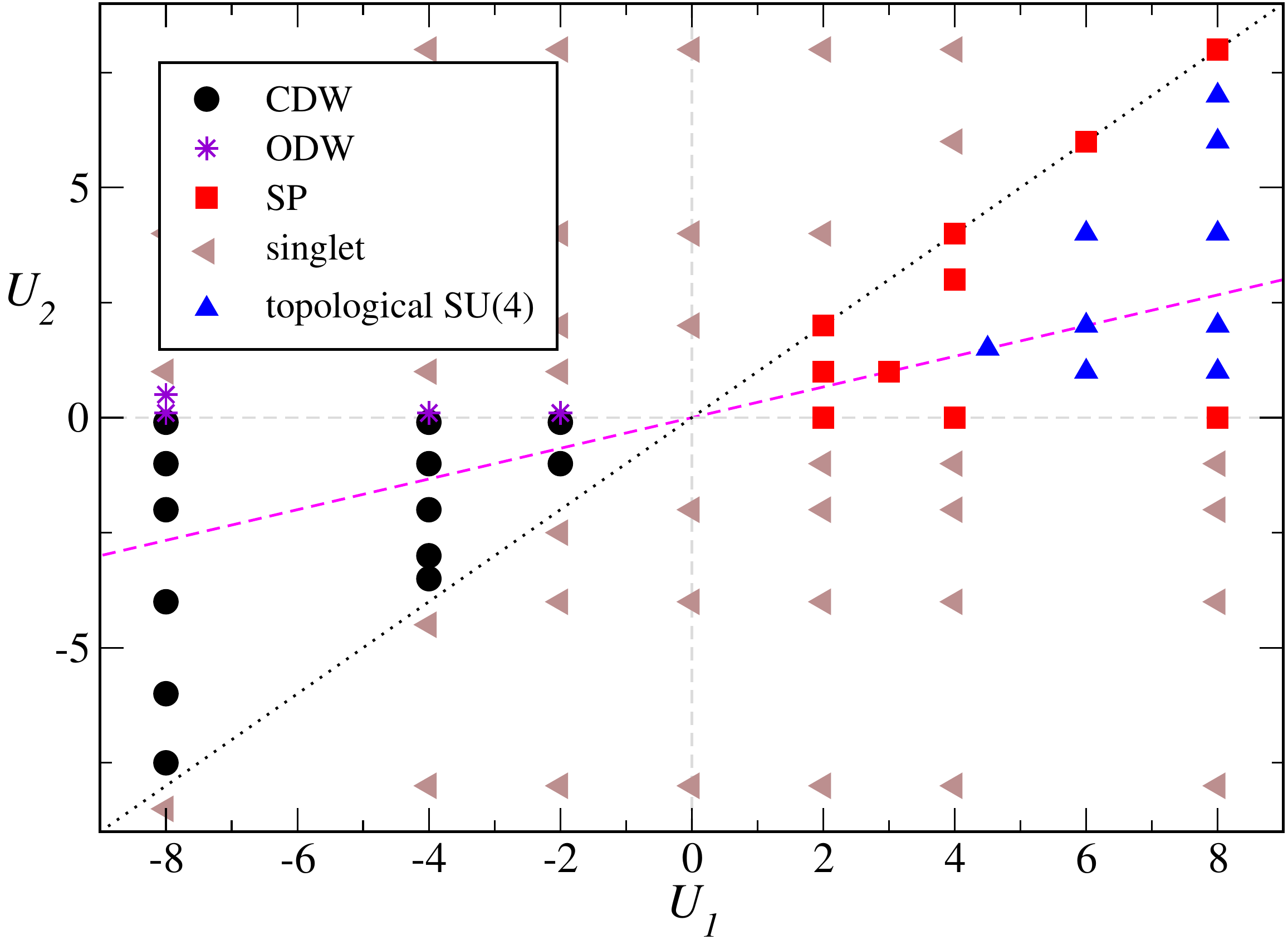}
\caption{(Color online) Phase diagram for half-filled $N=4$ $p$-band model (\ref{eqn:p-band}) obtained by DMRG on $L=32$. Dashed line corresponds to the condition $U_1=3U_2$ satisfied for an axially symmetric trap. $U_2=0$ correspond to two decoupled SU(4) Hubbard chains (see text).
\label{fig:phasediag_pband_N4}}
\end{figure}

%%%%%%%%%%%%%%%%%%%%%%%%%%%%%%%%%%%%%%%%%%%%%%%%%%%%%
Lastly, we investigate the $N=4$  $p$-band model (\ref{eqn:p-band}) which we believe to be quite relevant experimentally. Its phase diagram as  a function of $(U_1/t,U_2/t)$ is depicted in Fig.~\ref{fig:phasediag_pband_N4}. While the physical realization with an axially symmetric trap imposes $U_1=3U_2$, we have already discussed that 
other trapping schemes could remove this constraint. 

%%%%%%%%%%%%%%%%%%%%%%%%%%%%%%%%%%%%%%%%%%%%%%%%%%%%%%%%%%
%%%%%%%%%%%%%%%%%%%%%%%%%%%%%%%%%%%%%%%%%%%%%%%%%%%%%

Starting from this special line and using the equivalence to the generalized Hund model \eqref{alkaourmodel} 
with $J_z=0$ 
(see Sec.~\ref{sec:N4-gen-Hund-wo-SU2}), we obtain identical results as in Fig.~\ref{fig:phasediag_alka_N4_Jz_0}, 
i.e. when increasing $U_1(=3U_2)$, we find respectively the CDW phase (when $U_1<0$), 
the SP phase (for small $U_1>0$, as found in weak-coupling), and our topological SU(4) phase 
(for large $U_1>0)$. Since these are gapped phases, they do have a finite extension 
in the phase diagram. Again, the topological SU(4) phase occupies a rather large portion 
which makes it a good candidate for being realized experimentally. 
As was done for the $g$-$e$ model in the above, 
this topological SU(4) phase can be easily identified numerically thanks to the existence 
of characteristic edge-states in the DMRG simulations, see Fig.~\ref{fig:edge_DMRG}(b) 
and related comments in Sec.~\ref{sec:N4-Gorshkov}.

The rest of the phase diagram is dominated by trivial singlet phases. However, contrary to the $N=2$ $p$-band model (see Fig.~\ref{fig:phasediag_pband_N2}) where two trivial phases could be distinguished with respect to the symmetry $p_x \leftrightarrow p_y$, here we do not have a full picture. For instance, for $U_1=0$ and large $U_2/t \gg 1$, 
the ground-state is a complicated superposition of different $T^z$ eigenstates
(that are neither $T^x$ nor $T^y$ eigenstates) which has thus no special features concerning 
the orbital degrees of freedom.   

Before concluding this section, 
we have to comment about the special line $U_2=0$ where the model decouples into 
two identical (single-band) SU(4) Hubbard chains (one for each orbital). Such a chain is known to be either 
in a CDW (for $U_1<0$) or in a SP (for $U_1>0$) phase, each of which is two-fold degenerate. 
As a consequence, for $U_1>0$ we have four-fold degenerate SP phase depending on how dimerization 
patterns on the two chains are combined; for $U_1<0$, the CDW on each chain can be in-phase or out-of-phase, which in our terminology translates respectively into CDW or ODW for the whole system 
(see Fig.~\ref{fig:2-DWPhases-SU2}), 
again giving four-fold degeneracy.  
Any small finite $U_2$ splits these four degenerate states into two pairs of degenerate states, 
thereby stabilizing either CDW or ODW depending on its sign.\footnote{%
Near the line $U_2=0$, the impact of the deviation from $U_2=0$ is readily estimated.
When $U_2>0$, the two CDWs on different orbitals (`chains') repel each other 
due to the positive $V=U_2$-interaction and favor out-of-phase CDW, i.e., ODW.  
When $U_2<0$, on the other hand, they attract each other and consequently
stabilize usual CDW.}

\section{Concluding remarks}  
\label{sec:conclusion}
%%%%%%%%%%%%%%%%%%%%%%%%%%%%%%%%%%%%%%%%%%%%%%%%%%%%
The possibility to realize SU($N$)-symmetric models in alkaline-earth cold atoms experiments has revived the interest in determining what kind of electronic phases, possibly exotic, can be stabilized in these systems and more generally in establishing their phase diagrams. 
While this remains a challenging problem in general, we have presented a fairly complete study relevant 
for alkaline-earth fermionic atoms loaded into a 1D optical lattice at half-filling.  
The two models we considered take into account two orbitals 
as well as the SU($N$) internal degrees of freedom and we focused on the $N$-even case, 
which seems to harbor various interesting topological phases. 

Working in 1D allows us to use rather powerful analytical and unbiased numerical tools in order to complete this program. Moreover, this strategy has often been used in the past even 
to gain insight on possible phases in higher dimensions. Last but not least, 1D optical lattices are easily created experimentally so that the exotic phases proposed could be investigated in future experiments.

Our choice of working at  half-filling aims at investigating Mott phases, which are presumably simpler in the sense that some (charge) degrees of freedom will be frozen, but may still exhibit a variety of properties 
as exemplified in our phase diagrams 
where several exotic SPT phases have been found and characterized thanks to their nontrivial edge states, for instance. 
Let us remind that precisely in an SPT phase, edge states are protected (and thus cannot be removed without closing a gap) as long as some particular (protecting) symmetry is present.

The addition of the orbital degree of freedom is the key ingredient in our study. 
Indeed, without it, there are no SPT phases for 1D (singe-band) SU($N$) Hubbard models. This additional degree of freedom may be provided either by a metastable $e$ state (on top of the ground-state $g$) 
or by populating the two degenerate first-excited $p_x$ and $p_y$ states forming the $p$-bands of the optical lattice. 
Now, if one considers contact interactions only, the resulting minimal models are respectively the $g$-$e$ model 
[see Eqs. (\ref{eqn:Gorshkov-Ham}) or (\ref{alkaourmodel})] and the $p$-band Hamiltonian (\ref{eqn:p-band-simple}). Depending on their parameters, we have first clarified their symmetries as well as their strong-coupling limits, which provided  
a firm ground for the subsequent analyses and allowed a physical interpretation of some of their phases. 

Combining the strong-coupling approach, a low-energy field-theory and a large-scale unbiased numerical (DMRG) 
simulations, we have obtained a large number of phase diagrams of the two models depending on the value of $N$ 
(specifically, $N=2$ and $4$) 
and its parameters. Our main conclusion is that the interplay between the orbital and the SU($N$) nuclear-spin degrees of freedom gives rise to several interesting phases: in particular, we presented microscopic models whose ground-states realize 
two different kinds of SPT  phases 
(see Sec.~\ref{sec:strong-coupling-positive-J} and \ref{sec:strong-coupling-negative-J}). 

One of these SPT phases concerns the orbital pseudo spins $\mathbf{T}$  
and can be described by an effective (pseudo spin) $T=N/2$ Heisenberg chain, 
possibly with some single-ion anisotropy. If the original model we consider possesses 
the orbital SU(2)$_{\text{o}}$ symmetry (which may require some fine-tuning), 
then there is no such anisotropy so that the physical properties are identical 
to those of the spin-$N/2$ Heisenberg chain  
(see Figs.~\ref{fig:phasediag_Gorshkov_N4_SU2} and \ref{fig:phasediag_alka_N4}). 
Recent studies have shown\cite{Pollmann-B-T-O-12} that this gapped phase, when $N/2$ odd, is topologically 
protected by \emph{any} one of the following symmetries: (i) $\pi$ rotations around two of the three spin axis; (ii) time-reversal; (iii) bond inversion.  Away from the SU($2$)$_{\text{o}}$ regime, the phase diagram is dominated by the trivial rung-singlet (RS) phase 
corresponding to the so-called large-$D$ phase in the spin-chain language, so that the observation 
of the SPT phase remains challenging. Quite interestingly too, in the case of intermediate values of $D$, 
there is an extended critical phase for the integer $N/2$ strictly larger than 1, that we have been able to characterise 
as the Luttinger liquid of this orbital pseudo spin degree of freedom. 

Our main observation is the appearance in a much wider region of parameter space of another SU($N$) topological phase, corresponding in the strong-coupling limit to an SU($N$) Heisenberg chain with a self-conjugate representation 
(Young diagram with $N/2$ rows and 2 columns) at each site. Thanks to the VBS approach, we have been able to show: 
(i) this is a featureless gapped phase in the bulk, (ii) with open boundary conditions, there exist edge states (corresponding to self-conjugate representation with $N/2$ rows and 1 column), (iii) this is an SPT phase protected 
by $\text{PSU}(N) \simeq \text{SU}(N)/\mathbb{Z}_{N}$ (this is the case in the SU($N$) phase of our systems) 
or $\mathbb{Z}_N \times \mathbb{Z}_N$ symmetry for any $N$. Therefore, this provides a microscopic realization of one (among $N$) possible SPT phases for SU($N$) models~\cite{Duivenvoorden-Q-13}, characterised by the number of boxes modulo $N$ in the Young diagram describing the edge state (here $N/2$). Note also 
that even if the SU($N$) symmetry is broken but there remains some bond inversion symmetry, 
then this topological phase remains protected iff $N/2$ is odd as the Haldane one. 

Both our strong-coupling approach and our numerical simulations have confirmed the existence of this phase in a large regime of parameters, which make its potential observation more realistic.  
Nevertheless, the detection of our topological phases is still a real challenge given 
that the edge states may be substantially suppressed or even absent 
if one takes into accout a harmonic trap~\cite{Kobayashi-O-O-Y-M-12, Kobayashi-O-O-Y-M-14} 
and it appears difficult, though not hopeless\cite{Endres-etal-stringOP-11}, at the moment 
to directly measure the rather involved non-local order parameters.  
An exciting possibility would be to use a box trapping scheme~\cite{Gaunt2013} 
where presumably edge states should be more visible. 

Quite remarkably, this topological SU($N$) phase is not found in the weak-coupling regime, both in  the low-energy approach as well as in the numerical simulations, but instead is replaced by the spin-Peierls-like ground state 
with bond-strength modulations.  
As discussed in Ref.~\onlinecite{Nonne-M-C-L-T-13}, we expect that the quantum phase transition between
the topological SU($N$) phase and the dimerized one is described by a SU($N$)$_2$ CFT with central
charge $c= 2(N^2 -1)/(N+2)$. Since this prediction is independent on the microscopic model, 
we are looking forward to checking it using simpler Hamiltonians with less degrees of freedom, 
which will be easier from the numerical point of view.    

In this paper, we did not consider the case of odd $N$, which can also be realized 
in the systems of alkaline-earth fermions by trapping only a subset of $N$(=even) nuclear multiplet.  
In fact, already in the strong-coupling limit, one can see that the systems with even-$N$ considered here 
and those with odd-$N$ behave quite differently.  
For instance, as the orbital pseudo spin can never be quenched even in the Mott region when $N$ odd, 
one obtains an SU($N$)-orbital-coupled effective Hamiltonian for the region 
that was described by the SU($N$) Heisenberg model \eqref{eqn:2nd-order-effective-Ham-Gorshkov} 
or \eqref{eqn:2nd-order-effective-Ham-p-band} when $N$ even.  
Mapping out the phases in the odd-$N$ system would be an interesting future problem.  

%%%%%%%%%%%%%%%%%%%%%%%%%%%%%%%%%%%%%%%%%%%%%%%%%%%%%%%%%%%%%%%%%%%%%%%%%%%%%%%%
\section*{Acknowledgements}
The authors would like to thank H. Nonne for collaboration on this project. 
Numerical simulations have been performed using HPC resources from GENCI--TGCC, GENCI--IDRIS (Grant 2014050225) and CALMIP. 
S.C. would like to thank IUF for financial support. 
K.T. has benefited from stimulating discussions with 
A.~Bolens, K.~Penc, and K.~Tanimoto on related projects. 
He was also supported in part by JSPS Grant-in-Aid for Scientific Research(C) No. 24540402.
%%%%%%%%%%%%%%%%%%%%%%%%%%%%%%%%%%%%%%%%%%%%%%%%%%%%%%%%%%%%%%%%%%%%%%%%%%%%%%%%
\appendix
%%%%%%%%%%%%%%%%%%%%%%%%%%%%%%%%%%%%%%%%%%%%%%%%%%%%%%%
\section{Decomposition of SU({\em 2N}) in terms of SU({\em N})$\boldsymbol{\times}$SU(2)}
\label{sec:decomp-SUN-SU2}
%%%%%%%%%%%%%%%%%%%%%%%%%%%%%%%%%%%%%%%%%%%
As we have seen in Sec. \ref{sec:models-strong-coupling}, 
the largest symmetry of the system is U($2N$) since 
we deal with fermions with two different types of indices: $\alpha=1,\ldots,N$ for SU($N$) and 
$m=g,e$ for orbitals (or $p_x$ and $p_y$ for the $p$-band model).   
The Mott state with the fixed number of fermions at each site corresponds to one of 
the irreducible representations of SU($2N$).  
In the presence of interactions, the symmetry of the system changes as Eq. \eqref{eqn:symmetry-change}.  
Therefore, it is helpful to know how a given irreducible representation of SU($2N$) decomposes 
into those of SU($N$) and SU(2) (orbital).   

As a warming-up, we begin with the $N=2$ case.  Then, we have four species of fermions 
$c_{g\uparrow}$, $c_{g\downarrow}$, $c_{e\uparrow}$ and $c_{e\downarrow}$ and 
the largest symmetry is SU(4) [U(4), precisely].   
Let us consider the Mott-insulating state where we have an integer number ($n$) of fermions at each site. 
Then, the fermionic property restricts the possible representations at each site to 
the following four:
\begin{equation}
\yng(1) \; (n=1), \;\; \yng(1,1) \; (n=2), \;\; \yng(1,1,1) \; (n=3), \;\; \yng(1,1,1,1) \; (n=4).
\end{equation}
These on-site states correspond respectively to SU(4) irreducible representations 
with dimensions 4, 6, 4 and 1.  

It is easy to see that the four states in the $n=1$ (${\tiny \yng(1)}$) case are grouped into two 
\begin{subequations}
\begin{equation}
\left\{ c_{g\uparrow}^{\dagger}|0\rangle \; , \;
c_{g\downarrow}^{\dagger}|0\rangle  \right\} \; , \;\;
\left\{ c_{e\uparrow}^{\dagger}|0\rangle \; , \;
c_{e\downarrow}^{\dagger}|0\rangle  \right\} \; ,
\end{equation}
which span the two independent ($g$ and $e$) sets of the two-dimensional ($S=1/2$) representations of spin-SU(2).  
Note that the spin operators $\bolS_{g}+\bolS_{e}$ does not see the orbital indices.  
For the orbital SU(2), we see that another grouping 
\begin{equation}
\left\{ c_{g\uparrow}^{\dagger}|0\rangle \; , \;
c_{e\uparrow}^{\dagger}|0\rangle  \right\} \; , \;\;
\left\{ c_{g\downarrow}^{\dagger}|0\rangle \; , \;
c_{e\downarrow}^{\dagger}|0\rangle  \right\}
\end{equation}
gives the two ($\uparrow$ and $\downarrow$) basis sets for the two-dimensional ($T=1/2$) 
representations of orbital-SU(2). 
\end{subequations}
We write these results as
\begin{equation}
\underbrace{\yng(1)}_{\text{SU}(4)} \sim (\underbrace{\yng(1)}_{\text{SU}(2)_{\text{s}}} , 
 \underbrace{\yng(1)}_{\text{SU}(2)_{\text{o}}})  \; .
 \end{equation}
 
There are six states with two fermions at each site ($n=2$; half-filled) and these six states can be grouped into
\begin{equation}
\left\{
c_{g\uparrow}^{\dagger}c_{e\uparrow}^{\dagger}|0\rangle \; , \;\;
\frac{1}{\sqrt{2}}\left(
c_{g\uparrow}^{\dagger}c_{e\downarrow}^{\dagger}|0\rangle 
+ c_{g\downarrow}^{\dagger}c_{e\uparrow}^{\dagger}|0\rangle \right) \; , \;\;
c_{g\downarrow}^{\dagger}c_{e\downarrow}^{\dagger}|0\rangle 
\right\}
\end{equation}
and
\begin{equation}
\left\{
c_{g\uparrow}^{\dagger}c_{g\downarrow}^{\dagger}|0\rangle \; , \;\;
\frac{1}{\sqrt{2}}\left(
c_{g\uparrow}^{\dagger}c_{e\downarrow}^{\dagger}|0\rangle 
- c_{g\downarrow}^{\dagger}c_{e\uparrow}^{\dagger}|0\rangle \right) \; , \;\;
c_{e\uparrow}^{\dagger}c_{e\downarrow}^{\dagger}|0\rangle 
\right\}  \; .
\end{equation}
One can easily see that the above two respectively correspond to 
\begin{equation}
(S=1){\otimes}(T=0)  \;\; \text{and} \;\; (S=0){\otimes}(T=1) \; .
\end{equation}
Therefore, the spin-SU(2) and the orbital-SU(2) are entangled and when the former 
is in a triplet (singlet), the latter should be in a singlet (triplet).  
Again, in terms of Young diagrams, this may be written as
\begin{equation}
\yng(1,1) \sim 
 (\underbrace{\yng(2)}_{\text{SU}(2)_{\text{s}}} , \underbrace{\yng(1,1)}_{\text{SU}(2)_{\text{o}}}) \oplus 
(\underbrace{\yng(1,1)}_{\text{SU}(2)_{\text{s}}}, \underbrace{\yng(2)}_{\text{SU}(2)_{\text{o}}})
\Rightarrow 
 (\yng(2) , \bullet) \oplus (\bullet, \yng(2)) \; ,
 \end{equation}
 where $\bullet$ denotes the singlet.  
 
 For general $N$, we use the rules described in Refs.~\onlinecite{Georgi-book-99} (chapter 15) 
 and \onlinecite{Itzykson-N-66} (in particular, Table C of Ref.~\onlinecite{Itzykson-N-66} is quite useful).  
 The decomposition of fermionic states reads, for various local fermion number $n$ ($n_{c}\leq 2N$), as 
 \begin{subequations}
 \begin{align}
 & \yng(1) \sim (\underbrace{\yng(1)}_{\text{SU}(N)},\underbrace{\yng(1)}_{\text{SU(2)}}) 
 \quad (n=1) \\
 & \yng(1,1) \sim \left(\yng(2),\bullet \right) \oplus \left(\yng(1,1), \yng(2) \right)  \quad (n=2) \\
 & \yng(1,1,1) \sim \left( \yng(2,1), \yng(1) \right) \oplus \left(\yng(1,1,1),\yng(3) \right)  \quad (n=3) 
 \end{align}
 \begin{equation}
 \begin{split}
 & \yng(1,1,1,1) \sim \left( \yng(2,2), \bullet \right) \oplus \left( \yng(2,1,1),\yng(2) \right) 
 \oplus \left(\yng(1,1,1,1), \yng(4) \right)  \\
 &    \qquad \qquad (n=4) 
 \end{split}
 \label{eqn:decomp-SU2-SUN-4} 
 \end{equation}
 \begin{equation}
 \begin{split}
 & \yng(1,1,1,1,1) \sim \left( \yng(2,2,1), \yng(1) \right) \oplus \left( \yng(2,1,1,1),\yng(3) \right)  \\
& \qquad \oplus \left(\yng(1,1,1,1,1), \yng(5) \right)  \quad (n=5) 
 \end{split}
\end{equation}
\end{subequations}
It is easy to check that the dimensions on the both sides match.  
Consider the decomposition \eqref{eqn:decomp-SU2-SUN-4} for $N=4$.   
Apparently, the dimensions of the left-hand side is $8!/(4!4!)=70$.  
The sum of the dimensions appearing on the right-hand side is given by 
\begin{equation}
20{\times}1 + 15{\times}3 + 1{\times}5=70 \; ,
\end{equation}
which coincides with the one on the left-hand side.  
From these results, it is obvious that the SU($N$) irreducible representations contained in  
the fermionic states of the form $\prod c_{m\alpha}^{\dagger}|0\rangle$ are represented 
by Young diagrams with {\em at most two columns}.  
If we denote the lengths of the two columns by $p$ and $q$ ($p+q=n$, $p \geq q$), 
\begin{equation}
\text{\scriptsize $p$} \left\{ 
\yng(2,2,2,1,1)
\right.
\raisebox{2.2ex}{%
$\left.
\vphantom{\yng(2,2,2)}
 \right\} \text{\scriptsize $q$}
$ }
\end{equation}
the `spin' $T$ of the orbital SU(2) is given by 
\begin{equation}
\underbrace{\yng(2)}_{p-q} \; , \quad 
T = \frac{1}{2}(p-q) \; .
\end{equation} 
%%%%%%%%%%%%%%%%%%%%%%%%%%%%%%%%%%%%%%%%%%%%%%%%%%%%%
\section{$p$-band Hamiltonian}
\label{sec:p-band-hamiltonian}
\begin{figure}[H]
\begin{center}
\includegraphics[scale=0.9]{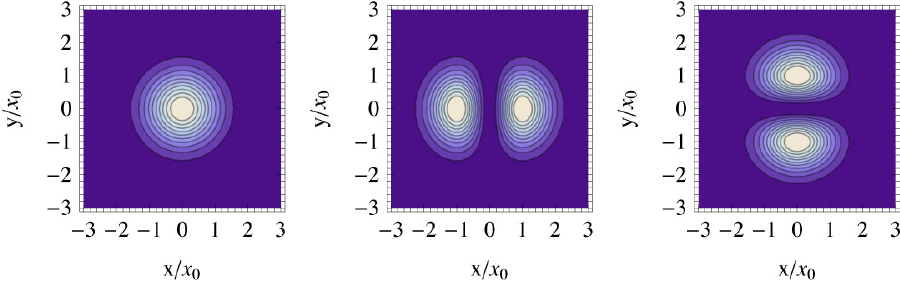}
\caption{Contour plots of squared wave functions $| \phi_{n_x,n_y}|^{2}$ 
for three orbitals $(n_x,n_y)=(0,0)$, $(1,0)$ and $(0,1)$.     
\label{fig:pxpy-orbitals}}
\end{center}
\end{figure}

%%%%%%%%%%%%%%%%%%%%%%%%%%%%%%%%%%%%%%%%%%%%%%%%%%%%%
In this appendix, we sketch the derivation of the $p$-band Hamiltonian \eqref{eqn:p-band}.  
The eigenfunctions of the single-particle part $\mathcal{H}_{0}$ \eqref{eqn:single-particle-Ham} 
is given by the Bloch function:
\begin{equation} 
\psi^{(n)}_{n_x,n_y,k_z}(x,y,z) = \phi_{n_x,n_y}(x,y) \varphi^{(n)}_{k_z}(z) \; ,
\end{equation}
where $\phi_{n_x,n_y}(x,y)$ and $\varphi^{(n)}_{k_z}(z)$ are the eigenfunctions of 
$\mathcal{H}_{\perp}(x,y)$ and $\mathcal{H}_{/\!/}(z)$, respectively.  
Since $\mathcal{H}_{\perp}$ is the two-dimensional harmonic oscillator, we can obtain the explicit 
form of $\phi_{n_x,n_y}(x,y)$.   
First three (normalized) eigenfunctions are given as (see Fig. \ref{fig:pxpy-orbitals})
\begin{subequations}
\begin{align}
&  \phi_{0,0}(x,y) = \frac{1}{\sqrt{\pi } x_{0}}\be^{-\frac{x^2+y^2}{2 x_{0}^2}} \\
&  \phi_{1,0}(x,y) = \frac{\sqrt{2}}{\sqrt{\pi}x_{0}} 
\left(\frac{x}{x_{0}} \right)  \be^{-\frac{x^2+y^2}{2 x_{0}^2}} \, , \\
& \phi_{0,1}(x,y) =  \frac{\sqrt{2}}{\sqrt{\pi}x_{0}} 
\left(\frac{y}{x_{0}} \right)  \be^{-\frac{x^2+y^2}{2 x_{0}^2}} 
\label{eqn:pxpy-orbitals}
\end{align}
\end{subequations}
with $x_{0} = \sqrt{\hbar/(m\omega_{xy})}$.  
We call the levels with $(n_x,n_y)=(0,0)$, $(1,0)$ and $(0,1)$ `$s$', `$p_x$' and `$p_y$', 
respectively.   
%%%%%%%%%%%%%%%%%%%%%%%%%%%%%%%%%%%%%%%%%%%%%%%%%%%%%%%%%%
%%%%%%%%%%%%%%%%%%%%%%%%%%%%%%%%%%%%%%%%%%%%%%%%%%%%%%%

To derive an effective Hubbard-like Hamiltonian\cite{Jaksch-Z-05}, it is convenient to move from 
the Bloch function $\psi^{(n)}_{n_x,n_y,k_z}(x,y,z)$ to the Wannier function defined as
\begin{equation}
W^{(n)}_{n_x,n_y;R}(x,y,z) \equiv 
\frac{1}{\sqrt{N_{\text{cell}}}} \phi_{n_x,n_y}(x,y) \sum_{k_z}
\be^{- i k_z R} \varphi^{(n)}_{k_z}(z)
%\label{eqn:Wannier-pxpy}
\end{equation}
($R$ labels the center of the Wannier function and $N_{\text{cell}}$ is then number of unit cells) 
and introduce the corresponding creation/annihilation operators 
\begin{equation}
\begin{split}
& c_{a\alpha}(\bolr) 
= \sum_{R}\sum_{n=\text{bands}} W_{a;R}^{(n)}(x,y,z) c_{a\alpha,R}^{(n)}  \\
& c^{\dagger}_{a\alpha}(\bolr) 
= \sum_{R}\sum_{n=\text{bands}} W_{a;R}^{(n)\ast}(x,y,z) c_{a\alpha,R}^{(n)\dagger} \\
& \quad (a=p_x,p_y, \; \alpha=1,\ldots, N) \; ,
\end{split}
\end{equation}
As in Sec. \ref{sec:models-strong-coupling}, we have used the short-hand notation $a=p_{x},p_{y}$ 
meaning $p_{x}=(n_x,n_y)=(1,0)$ and $p_{y}=(n_x,n_y)=(0,1)$.  

Following the standard procedure\cite{Jaksch-Z-05}, we can derive the Hubbard-type interactions from 
the original contact interaction $g \delta^{3}(\bolr)$:
\begin{equation}
\begin{split}
& \frac{1}{2} \sum_{R}  \sum_{a=p_x,p_y} U_{aaaa} 
\hat{V}_{aaaa}(R) \\
& + \frac{1}{2} \sum_{R} \sum_{\substack{a\neq b\\=p_x,p_y}} \Biggl\{ 
U_{aabb} \hat{V}_{aabb}(R) + U_{abba} \hat{V}_{abba}(R)  \\
& \phantom{+ \frac{1}{2} \sum_{R} \sum_{\substack{a\neq b\\=p_x,p_y}} \Biggl\{ }
+ U_{abab} \hat{V}_{abab}(R)   \Biggr\}   
\end{split}
\label{eqn:contact-int-by-Wannier}
\end{equation}
where the superscript `$(0)$' for the fermion operators of the lowest Bloch band has been suppressed 
and $U_{abcd}$ is defined by
\begin{equation}
\begin{split}
& U_{abcd} \equiv g 
\int\! d\bolr 
W_{a;R}^{(0)\ast}(\bolr)W_{b;R}^{(0)\ast}(\bolr)
W_{c;R}^{(0)}(\bolr)W_{d;R}^{(0)}(\bolr)  \\
& \hat{V}_{abcd}(R) \equiv 
c_{a\alpha,R}^{\dagger} c_{b\beta,R}^{\dagger}c_{c\beta,R}c_{d\alpha,R}  \\
& \quad
(a,b,c,d=p_{x}, p_{y}) \; .
\end{split}
\label{eqn:def-Uabcd-pxpy}
\end{equation}
Since the Wannier functions are real and the two orbitals $W_{p_x/p_y;R}^{(0)}(\bolr)$ 
are symmetry-related ($C_4$), there are only two independent couplings:
\begin{equation}
\begin{split}
& U_{1} \equiv U_{p_x p_x p_x p_x} = U_{p_y p_y p_y p_y}  \\
& U_{2} \equiv U_{p_x p_x p_y p_y} = U_{p_y p_y p_x p_x}
= U_{p_x p_y p_y p_x} \\
& \phantom{U_{2}} = U_{p_y p_x p_x p_y}
= U_{p_x p_y p_x p_y}= U_{p_y p_x p_y p_x} \; .
\end{split}
\label{eqn:def-U1-U2}
\end{equation}
Using the explicit forms \eqref{eqn:pxpy-orbitals}, one can readily verify that 
the above two coupling constants $U_1$ and $U_2$ actually are {\em not} independent and 
satisfy $U_1 = 3 U_2$.   In fact, this ratio is constant for {\em any} axially symmetric potential 
$V_{\perp}(x,y)$.  

Plugging the above into Eq. \eqref{eqn:contact-int-by-Wannier}, 
we obtain the Hamiltonian of the $p$-band model \eqref{eqn:p-band}:
\begin{equation}
\begin{split}
\mathcal{H}_{p\text{-band}} =& - t \sum_{i}
( c_{a\alpha,i}^{\dagger}c_{a\alpha,i+1}  + \text{H.c.} ) \\
& + \sum_{i} \sum_{a=p_x,p_y} 
(\epsilon_{a} - t_{0}) \, n_{a,i} \\
&+  \frac{1}{2}U_{1}  \sum_{i}
n_{a,i}(n_{a,i}-1) +  U_{2}\,  \sum_{i} n_{p_x,i} n_{p_y,i} \\
& + U_{2} \sum_{i}
c_{p_x\alpha,i}^{\dagger} c_{p_y \beta,i}^{\dagger}c_{p_x \beta,i}c_{p_y \alpha,i} \\
& + \frac{1}{2}U_{2}\sum_{i} \sum_{\substack{a\neq b\\=p_x,p_y}} 
c_{a,\alpha,i}^{\dagger} c_{a,\beta,i}^{\dagger}c_{b,\beta,i}c_{b,\alpha,i}  \; ,
\end{split}
\label{eqn:p-band-in-appendix}
\end{equation}
where $\epsilon_{p_x}=\epsilon_{p_y}=3\hbar \omega_{xy}/2$ and 
the hopping amplitude $t$ is defined as
\begin{equation}
\begin{split}
& t = t^{(0)}(\pm 1)  \\
& \int\! d\bolr \, 
W_{a;R_1}^{(n_1)\ast}(\bolr)\mathcal{H}_{/\!/}(z) W_{b;R_2}^{(n_2)}(\bolr) 
\equiv - \delta_{ab} \delta_{n_1 n_2} t^{(n_1)}(R_1-R_2)   \;.
\end{split}
\end{equation}   
When the last term in \eqref{eqn:p-band-in-appendix} ({\em pair-hopping}) is rewritten in terms of 
the orbital pseudo spin $\bolT$, eq.\eqref{eqn:p-band} is recovered.  
%%%%%%%%%%%%%%%%%%%%%%%%%%%%%%%%%%%%%%%%%%%%%%%%%%%%
\section{Conformal field theory data}
\label{sec:CFT-data}
%%%%%%%%%%%%%%%%%%%%%%%%%%%%%%%%%%%%%%%%%%%%%%%%%%%%
In this Appendix, we recall some useful formula of SU($N$)$_k$  CFT which are useful in the
low-energy approach of two-orbital SU($N$) models (Sec.~\ref{sec:SUN-topological-phase}).

Let us first consider the SU(2)$_N$ CFT which is generated by the orbital current
$\mathbf{j}_{\text{L,R}}$ in our problem.  The left chiral current satisfies the SU(2)$_N$ Kac-Moody algebra 
which reads as follows within our conventions:
\begin{eqnarray}
j^{i}_{\text{L}}\left(z\right) j^{j}_{\text{L}}\left(0\right) \sim \frac{N \delta^{ij}}{8 \pi^2 z^2} 
+ \frac{ i \epsilon^{ijk}}{2 \pi z}  j^{k}_{\text{L}}\left(0\right),
\label{KacMoodyApp}
\end{eqnarray}
with a similar result for the right current.
The   SU(2)$_N$  primary operators with spin $j = 0,\ldots, N/2$  is an 
SU(2) $\times$ SU(2) tensor with $(2j+1)^2$ components which are denoted by 
$\Phi^{j}_{m, \bar m}$  ($|m, \bar m| \le j$). They transform in the spin-j representation of
SU(2) and have scaling dimension $d_j = 2 j (j+1)/(N+2)$. \cite{Knizhnik-Z-84,DiFrancesco-M-S-book}
They are defined through the OPE: \cite{DiFrancesco-M-S-book}
\begin{eqnarray}
j^{i}_{\text{L}}\left(z\right) \Phi_{m, \bar m}^{j}\left(\omega, \bar \omega\right) 
 &\sim & -\frac{1}{z-\omega}
T^i_{ms} \Phi_{s, \bar m}^{j}\left(\omega, \bar \omega\right)
\nonumber \\
j^{i}_{\text{R}}\left(\bar z\right) \Phi_{m, \bar m}^{j}\left(\omega, \bar \omega\right) 
 &\sim & \frac{1}{\bar z- \bar \omega} 
\Phi_{m, \bar s}^{j}\left(\omega, \bar \omega\right)
T^i_{\bar s \bar m} ,
\label{theopeApp}
\end{eqnarray}
where $T^i$ are the usual spin-$j$ matrices. The conjugate of $\Phi^{j}_{m, \bar m}$ is defined by:
\begin{equation}
\Phi^{j \dagger}_{m, \bar m} = \left(-1\right)^{2j - m - \bar m} \Phi^{j}_{-m, -\bar m} .
\label{Su2conj}
\end{equation}

We need also the SU(2)$_N$  fusion rule which describes the product between two primary operators
with spin $j_1$ and $j_2$: \cite{Zamolodchikov-F-86}
\begin{eqnarray}
j_1 \otimes j_2 &=& | j_1 -  j_2 |, | j_1 -  j_2 | +1, \ldots, \nonumber \\
&& {\rm min} \left( j_1 +  j_2, N/2 - j_1 -  j_2 \right) .
\label{Su2fusion}
\end{eqnarray}
Related to this decomposition is the SU(2)$_N$  operator algebra:\cite{Zamolodchikov-F-86}
\begin{eqnarray}
 \Phi^{j_1}_{m_1, \bar m_1} \left( z, \bar z\right) && \Phi^{j_2}_{m_2, \bar m_2} \left( 0, 0 \right)  \sim
\sum_{j=0}^{N/2} \sum_{m, \bar m=-j}^{j} | z |^{d_j - d_{j_1} - d_{j_2}} \nonumber \\
 && C \left(\begin{array}{cccccc}
  j  & m &  \bar m \\
  j_1  & m_1  &  \bar m_1 \\
 j_2  & m_2  &  \bar m_2
\end{array}\right) \Phi^{j}_{m, \bar m} \left( 0, 0 \right) ,
\label{SU2Nfusionop}
\end{eqnarray}
where $C$ are the structure constants of the operator algebra which are related to the Wigner $3j$ symbols as:
\begin{eqnarray}
&& C \left(\begin{array}{cccccc}
  j  & m &  \bar m \\
  j_1  & m_1  &  \bar m_1 \\
 j_2  & m_2  &  \bar m_2
\end{array}\right)  = \rho_{ j, j_1, j_2 }
\nonumber \\
&& \left(\begin{array}{cccccc}
  j  & j_1 &  j_2 \\
  -m  & m_1  &  m_2  \\
\end{array}\right)
\left(\begin{array}{cccccc}
j  & j_1 &  j_2 \\
  -  \bar m   & \bar m_1  &  \bar m_2  \\
\end{array}\right),
\label{structconst}
\end{eqnarray}
where $ \rho_{ j, j_1, j_2 }$ is a constant which can be found in Ref. \onlinecite{Zamolodchikov-F-86}
and we have the constraints: $m = m_1 + m_2$, $\bar m = \bar m_1 + \bar m_2$ which stem from the properties
of $3j$ symbols.
The explicit application of the operator algebra  (\ref{SU2Nfusionop}) for $j_1=j_2 = 1/2$ leads to
\begin{eqnarray}
 \Phi^{1/2}_{1/2, 1/2} \left( z, \bar z\right) &&   \Phi^{1/2}_{1/2, 1/2} \left( 0, 0 \right)  \sim 
 \nonumber \\
&& \frac{1}{3} | z |^{1/(N+2)} \rho_{ 1, 1/2, 1/2 }  \Phi^{1}_{1, 1} \left( 0, 0 \right) 
 \nonumber \\
 \Phi^{1/2}_{-1/2, -1/2} \left( z, \bar z\right) &&   \Phi^{1/2}_{-1/2, -1/2} \left( 0, 0 \right)  \sim  \nonumber \\
&&  \frac{1}{3} | z |^{1/(N+2)}  \rho_{ 1, 1/2, 1/2 }  \Phi^{1}_{-1, -1} \left( 0, 0 \right) 
\nonumber \\
  \Phi^{1/2}_{1/2, 1/2} \left( z, \bar z\right)  &&  \Phi^{1/2}_{-1/2, -1/2} \left( 0, 0 \right)  \sim 
 \nonumber \\
 && \frac{1}{2} | z |^{-3/(N+2)}  \rho_{ 0, 1/2, 1/2 } 
  \nonumber \\
  &+& \frac{1}{6} | z |^{1/(N+2)}  \rho_{ 1, 1/2, 1/2 } \Phi^{1}_{0, 0} \left( 0, 0 \right)
  \nonumber \\
   \Phi^{1/2}_{1/2, -1/2} \left( z, \bar z\right)  &&  \Phi^{1/2}_{-1/2, 1/2} \left( 0, 0 \right)  \sim 
 \nonumber \\
 && - \frac{1}{2} | z |^{-3/(N+2)}  \rho_{ 0, 1/2, 1/2 } 
  \nonumber \\
  &+& \frac{1}{6} | z |^{1/(N+2)}  \rho_{ 1, 1/2, 1/2 } \Phi^{1}_{0, 0} \left( 0, 0 \right) .
 \label{SU2Nfusionopspin1demi}
\end{eqnarray}

At this stage, we introduce another parametrization of the spin-1/2  SU(2)$_N$ field which
will be used in Sec.~\ref{sec:SUN-topological-phase}: $g_{pl} \equiv  \Phi^{1/2}_{m, \bar m}$ where $p=g,e   \; ({\rm or}  \; p_x, p_y)  
 \rightarrow m=1/2,-1/2$
and $l=g,e  \; ({\rm or}  \; p_x, p_y) \rightarrow  \bar m =1/2,-1/2$. With this definition and 
the OPEs (\ref{SU2Nfusionopspin1demi}),
we deduce that the trace of the SU(2)$_N$ primary field which transforms in the spin-1 representation, 
reads as follows:
\begin{equation}
{\rm Tr} \left(  \Phi^{\text{SU(2)}_N}_{\rm j=1} \right) \sim
{\rm Tr} \left( g \right) {\rm Tr} \left( g^{\dagger} \right) 
-   \frac{1}{2} g_{p l}  g^{\dagger}_{p l} .
\label{adjointSU2App}
\end{equation}

The  SU(2)$_N$  primary operators can also be related to that of the ${\mathbb{Z}}_N$  CFT ($f^{2j}_{2m, 2\bar m}$)
through the coset construction ${\mathbb{Z}}_N$ $\sim$  SU(2)$_N$ / U(1)$_\text{o}$. \cite{Zamolodchikov-F-JETP-85,Gepner-Q-87}
In the paper, the U(1)$_\text{o}$ CFT is described by a bosonic field which is the orbital field $\Phi_{o}$
with chiral components $\Phi_{o L,R}$. Within our conventions, the relationship between
the primary fields is:
\begin{equation}
\Phi^{j}_{m, \bar m} = f^{2j}_{-2m, 2\bar m}  \exp \left( - i m \sqrt{8 \pi/N} \; \Phi_{o L} 
- i \bar m \sqrt{8 \pi/N} \; \Phi_{o R} \right),
\label{primariesApp}
\end{equation}
where the ${\mathbb{Z}}_N$ primary operators have
scaling dimension $ \Delta^{j}_{m, \bar m} = 2 j (j+1)/(N+2) - (m^2 + {\bar m}^2)/N$.
The most important one for our purpose are the ${\mathbb{Z}}_N$ ordered spin operators $\sigma_k \sim
f^{k}_{k, k}$ and the disordered ones $\mu_k \sim f^{k}_{-k, k}$ ($k=1,\ldots, N-1$).
The relation (\ref{primariesApp})  gives in particular the following identifications:
\begin{eqnarray}
 \Phi^{1/2}_{1/2, 1/2} &\simeq& \mu_1 \exp \left( - i  \sqrt{2 \pi/N} \; \Phi_{o } \right) 
\nonumber \\
\Phi^{1/2}_{-1/2, -1/2} &\simeq& \mu^{\dagger}_1 \exp \left(  i  \sqrt{2 \pi/N} \; \Phi_{o } \right)
\nonumber \\
\Phi^{1/2}_{-1/2, 1/2} &\simeq&  \sigma_1 \exp \left( i  \sqrt{2 \pi/N} \; \Theta_{o } \right)
\nonumber \\
 \Phi^{1}_{1, 1} &\simeq& \mu_2 \exp \left(- i \sqrt{8 \pi/N} \; \Phi_{o} \right)
 \nonumber \\
  \Phi^{1}_{-1, -1} &\simeq& \mu^{\dagger}_2 \exp \left(i \sqrt{8 \pi/N} \; \Phi_{o} \right),
 \label{identiprimZnApp}
\end{eqnarray}
where $\Theta_{o }$ is the dual field associated with $\Phi_\text{o}$.
The last identification that we need is the ${\mathbb{Z}}_N$ description of $\Phi^{1}_{0, 0}$ which 
can be determined by the SU(2)$_N$ fusion rule $\Phi^{1/2}_{1/2, 1/2} \Phi^{1/2}_{-1/2, -1/2}$
(see Eq. (\ref{SU2Nfusionopspin1demi})).
Using the identification (\ref{identiprimZnApp}) for $\Phi^{1/2}_{\pm 1/2, \pm 1/2}$ and the following 
OPE for the  ${\mathbb{Z}}_N$  CFT ($C$ being an unimportant positive constant)
\begin{eqnarray}
\mu_1  \left(z, \bar z \right) \mu^{\dagger}_1  \left(0, 0 \right)  &\sim& |z|^{-\frac{2(N-1)}{N(N+2)}}
\nonumber \\
&- C&
|z|^{\frac{2(N+1)}{N(N+2)}} \epsilon_1  \left(0, 0 \right) ,
\label{ZnOPEApp}
\end{eqnarray}
we get:
\begin{equation}
\Phi^{1}_{0, 0} \simeq - \epsilon_1 ,
\label{spin1primaryparaApp}
\end{equation}
where $\epsilon_1$   is the thermal operator of the $Z_N$  CFT  with scaling dimension $4/(N+2)$.
In our convention, $\langle \epsilon_1 \rangle > 0 $ in a phase where the ${\mathbb{Z}}_N$ is broken so that
the disorder parameters cannot condense.

These results generalize in the SU($N$) case. We will only need for our purpose the 
values of scaling dimensions of SU($N$)$_2$ primary fields.
The SU($N$)$_k$ primary field  transforms in some representation $R$ 
of the SU($N$) group and its scaling dimension is given by \cite{Knizhnik-Z-84}:
\begin{equation}
\Delta_{R} = \frac{2 C_{R}}{N+k},
\label{dimensioprimary}
\end{equation}
where $C_{R}$ is the quadratic Casimir in the representation $R$.
Its expression can be obtained from the general formula where $R$ is written as a Young diagram:
\begin{equation}
C_{R} = T^aT^a = \frac{1}{2} \left\{
l(N-l/N) +\sum_{i=1}^{n_{\text{row}}} b_i^2 - \sum_{i=1}^{n_{\text{col}}}a_i^2 
\right\}
\label{Cas}
\end{equation}
for Young diagram of $l$ boxes consisting of $n_{\text{row}}$ rows of length $b_i$ each 
and $n_{\text{col}}$ columns of length $a_i$ each.
For instance, we get $C_{R} = (N^2 -1)/2N$ for the fundamental representation, 
$C_{R} = N$ for the adjoint representation, $C_{R} (k) = k(N+1) (N - k)/2N$ 
for the $k$-th basic antisymmetric representation
made of a Young diagram with a single column and $k$ boxes, and $C_{R} = N - 2/N +1$ for the symmetric 
representation with dimension $N(N+1)/2$.
In particular, in the SU($N$)$_2$ case, i.e. the CFT which describes the nuclear spin degrees
of freedom in our paper, the scaling dimensions of various primary operators needed in Sec.~\ref{sec:SUN-topological-phase} are:  
\begin{eqnarray}
\Delta_G &=& \frac{N^2-1}{N(N+2)} \nonumber 
\\
\Delta_{\rm adj} &=& \frac{2N}{N+2} \nonumber
\\
 \Delta_{S} &=& \frac{2(N - 2/N +1)}{N+2} \nonumber
 \\
\Delta_{A} &=& \frac{2(N+1)(N-2)}{N(N+2)},
 \label{scalingdimSUN2App}
\end{eqnarray}
which describes respectively the scaling dimension of the SU($N$)$_2$ primary field which
transforms in the fundamental, adjoint,  symmetric representation with dimension $N(N+1)/2$,  
and antisymmetric representation with dimension $N(N-1)/2$ of SU($N$).
%%%%%%%%%%%%%%%%%%%%%%%%%%%%%%%%%%%%%%%%%%%%%%%%%%%%
\section{Majorana-fermionization of the  half-filled \texorpdfstring{$\boldsymbol{N=2}$}{N=2} 
\texorpdfstring{$\boldsymbol{p}$}{p}-band model}
\label{sec:N2-p-band-continuum}
%%%%%%%%%%%%%%%%%%%%%%%%%%%%%%%%%%%%%%%%%%%%%%%%%%%%
In this Appendix, we investigate the zero-temperature phase diagram of the 
half-filled $N=2$ $p$-band model in the general case with two different coupling
constants $U_{1,2}$ by means of the low-energy approach.
As seen in Sec.~\ref{sec:models-strong-coupling}, the U(1)$_\text{o}$ continuous orbital symmetry is explicitly broken when $U_1 \ne 3 U_2$
and the low-energy effective Hamiltonian is no longer parametrized by nine coupling constants
as in Eq.~(\ref{lowenergyham}).
In the special $N=2$ case, one can use the standard field-theoretical methods 
based on bosonization and refermionization techniques as in the two-leg ladders. \cite{Gogolin-N-T-book}
In the context of the $N=2$ generalized Hund model at half-filling, we have described extensively this approach
in Ref. \onlinecite{Nonne-B-C-L-10}.

Using the Abelian bosonization, one can define four chiral bosonic field
$\Phi_{m \sigma \text{R,L}}$ ($m= p_x,p_y;\sigma= \uparrow, \downarrow$) from the four 
left-right moving Dirac fermions of the continuum limit for $N=2$.
The next step of the approach is to introduce a bosonic basis which singles out
the different degrees of freedom for $N=2$, i.e. charge, (nuclear) spin, orbital, and
spin-orbital degrees of freedom:
\begin{eqnarray}
  &&\Phi_{p_x\uparrow \text{L,R}}=\frac{1}{2}
  (\Phi_\text{c}+\Phi_\text{s}+\Phi_\text{o}+\Phi_{\text{so}})_{\text{L,R}} \nonumber\\
  &&\Phi_{p_x\downarrow \text{L,R}}=\frac{1}{2}
  (\Phi_\text{c}-\Phi_\text{s}+\Phi_\text{o}-\Phi_{\text{so}})_{\text{L,R}} \nonumber\\
  &&\Phi_{p_y\uparrow \text{L,R}}=\frac{1}{2}
  (\Phi_\text{c}+\Phi_\text{s}-\Phi_\text{o}-\Phi_{\text{so}})_{\text{L,R}} \nonumber\\
  &&\Phi_{p_y\downarrow \text{L,R}}=\frac{1}{2}
  (\Phi_\text{c}-\Phi_\text{s}-\Phi_\text{o}+\Phi_{\text{so}})_{\text{L,R}}.
  \label{basebosons}
\end{eqnarray}

From these new bosonic fields, one can now consider a refermionization procedure
by introducing eight left and right moving Majorana fermions through:
\begin{eqnarray}
  &&\xi_{\text{L}}^2+i\xi_{\text{L}}^1=
  \frac{\eta_1}{\sqrt{\pi a_0}}
  \exp{(-i\sqrt{4\pi}\Phi_{sL})}\nonumber\\
  &&\xi_{\text{R}}^2+i\xi_{\text{R}}^1=
  \frac{\eta_1}{\sqrt{\pi a_0}}
  \exp{(i\sqrt{4\pi}\Phi_{sR})}\nonumber\\
  &&\xi_{\text{L}}^4-i\xi_{\text{L}}^5=
  \frac{\eta_2}{\sqrt{\pi a_0}}
  \exp{(-i\sqrt{4\pi}\Phi_{oL})}\nonumber\\
  &&\xi_{\text{R}}^4-i\xi_{\text{R}}^5=
  \frac{\eta_2}{\sqrt{\pi a_0}}
  \exp{(i\sqrt{4\pi}\Phi_{oR})}\nonumber\\
  &&\xi_{\text{L}}^6+i\xi_{\text{L}}^3=
  \frac{\eta_3}{\sqrt{\pi a_0}}
  \exp{(-i\sqrt{4\pi}\Phi_{soL})}\nonumber\\
  &&\xi_{\text{R}}^6+i\xi_{\text{R}}^3=
  \frac{\eta_3}{\sqrt{\pi a_0}}
  \exp{(i\sqrt{4\pi}\Phi_{soR})} \nonumber\\
  &&\xi_{\text{L}}^8+i\xi_{\text{L}}^7=
  \frac{\eta_4}{\sqrt{\pi a_0}}
  \exp{(- i\sqrt{4\pi}\Phi_{cL})} \nonumber\\
&&\xi_{\text{R}}^8+i\xi_{\text{R}}^7=
  \frac{\eta_4}{\sqrt{\pi a_0}}
  \exp{(i\sqrt{4\pi}\Phi_{cR})} ,
 \label{refer}
\end{eqnarray}
where the Klein factors $\eta_{1,2,3,4}$ ensure the anti-commutation rules for the Majorana fermions. 
 
 With these definitions, the continuum Hamiltonian of the
  half-filled $N=2$ $p$-band model can then be expressed in terms of these eight Majorana fermions:
   \begin{eqnarray}
  \mathcal{H}&=& - \frac{i v_\text{F}}{2}\sum_{a=1}^8
  (\xi_{\text{R}}^a \partial_x \xi_{\text{R}}^a
  - \xi_{\text{L}}^a \partial_x \xi_{\text{L}}^a)\nonumber\\
    &+&\frac{g_1}{2}\left(\sum_{a=1}^3
  \xi_{\text{R}}^a\xi_{\text{L}}^a\right)^2+g_2\left(\sum_{a=1}^3
  \xi_{\text{R}}^a\xi_{\text{L}}^a\right) \xi_{\text{R}}^4\xi_{\text{L}}^4\nonumber\\
  &+&  \xi_{\text{R}}^6\xi_{\text{L}}^6\left[ g_3 \sum_{a=1}^3
  \xi_{\text{R}}^a\xi_{\text{L}}^a + g_4 \xi_{\text{R}}^4 \xi_{\text{L}}^4 \right] +\frac{g_5}{2}\left(
   \xi_{\text{R}}^5\xi_{\text{L}}^5 + \sum_{a=7}^8   \xi_{\text{R}}^a\xi_{\text{L}}^a \right)^2
  \nonumber\\
  &+& \left(\xi_{\text{R}}^5\xi_{\text{L}}^5 + \sum_{a=7}^8   \xi_{\text{R}}^a\xi_{\text{L}}^a \right) 
   \Biggl( g_6 \sum_{a=1}^3 \xi_{\text{R}}^a\xi_{\text{L}}^a + g_7  \xi_{\text{R}}^4\xi_{\text{L}}^4 
  \nonumber\\  
 &+&   
  g_8  \xi_{\text{R}}^6\xi_{\text{L}}^6  \Biggl),
  \label{Hamiltonian_tperp0_majoranapp}
\end{eqnarray}
where we have neglected the velocity-anisotropy terms for the sake of simplicity.
The different coupling constants of the continuum limit are given by:
\begin{eqnarray}
   &&g_1= - g_5 = -a_0\left(U_1 + U_2 \right) \nonumber\\
  &&g_2= - g_8 = -2a_0 U_2\nonumber\\
  &&g_3= -g_7 = a_0\left(U_2 - U_1 \right) \nonumber\\
  &&g_4= g_6 = 0  ,
\label{majocouplingsapp}
\end{eqnarray}
where we have included the operators with coupling constants $g_{4,6}$ since they will be generated
in the one-loop RG calculation.

From Eq. (\ref{Hamiltonian_tperp0_majoranapp}), one observes that the three Majorana fermions 
$\xi_{\text{R,L}}^a, (a = 1,2,3)$, which accounts for the physical properties of the (nuclear) spin degrees of freedom,
play a symmetric role as the result of the SU(2)$_s$ spin-symmetry of the lattice model. In addition, the two 
Majorana fermions $\xi_{\text{R,L}}^a, (a = 7,8)$, associated to the charge degrees of freedom, are unified
with one Majorana fermion $\xi_{\text{R,L}}^5$ of the orbital sector. This signals the emergence of a new independent
SU(2) symmetry for all $U_1$ and $U_2$ that we have revealed on the lattice from the charge pseudo spin operator 
(\ref{chargepseudospin}). The continuous symmetry of model (\ref{Hamiltonian_tperp0_majoranapp}) is actually
SU(2) $\times$ SU(2) $\sim$ SO(4).

The one-loop RG of model (\ref{Hamiltonian_tperp0_majoranapp}) can be easily determined within the 
Majorana formalism and
we find:
\begin{eqnarray}
  \dot{g_1} &=& \frac{1}{2\pi} g^2_1 + \frac{1}{2\pi} g^2_2 + \frac{1}{2\pi} g^2_3
  + \frac{3}{2\pi} g^2_6 
     \nonumber\\
    \dot{g_2} &=&  \frac{1}{\pi} g_1 g_2 +  \frac{1}{2\pi} g_3 g_4  + \frac{3}{2\pi} g_6 g_7
      \nonumber\\
       \dot{g_3} &=&  \frac{1}{\pi} g_1 g_3 +  \frac{1}{2\pi} g_2 g_4  + \frac{3}{2\pi} g_6 g_8
           \nonumber\\
            \dot{g_4} &=&  \frac{3}{2\pi} g_2 g_3 +  \frac{3}{2\pi} g_7 g_8  
        \nonumber\\
        \dot{g_5} &=& \frac{1}{2\pi} g^2_5 + \frac{3}{2\pi} g^2_6 + \frac{1}{2\pi} g^2_7
  + \frac{1}{2\pi} g^2_8  
      \nonumber\\
       \dot{g_6} &=&  \frac{1}{\pi} g_1 g_6 +  \frac{1}{2\pi} g_2 g_7  + \frac{1}{2\pi} g_3 g_8
       + \frac{1}{\pi} g_5 g_6
             \nonumber\\
              \dot{g_7} &=&  \frac{3}{2\pi} g_2 g_6 +  \frac{1}{2\pi} g_4 g_8  + \frac{1}{\pi} g_5 g_7
             \nonumber\\
              \dot{g_8} &=&  \frac{3}{2\pi} g_3 g_6 +  \frac{1}{2\pi} g_4 g_7  + \frac{1}{\pi} g_5 g_8 .
    \label{RGN2}
\end{eqnarray}

These RG equations enjoy some hidden symmetries:
 \begin{eqnarray}
 \Omega_1 &:& g_{2,3,6}  \to - g_{2,3,6} \nonumber  \\
\Omega_2 &:& g_{3,4,8}  \to - g_{3,4,8}  \nonumber \\
\Omega_3 &:& g_{6,7,8} \to - g_{6,7,8}  \nonumber \\
\Omega_4 &:& g_{2,4,7} \to - g_{2,4,7},
\label{eq:symmRG2}
\end{eqnarray}
which correspond to duality symmetries on the Majorana fermions
\begin{eqnarray}
\Omega_1 &:& \xi^{1,2,3}_\text{L} \to -\xi^{1,2,3}_\text{L} \nonumber  \\
\Omega_2 &:& ~~ \xi^6_\text{L} ~ \to -\xi^6_\text{L} \nonumber  \\
\Omega_3 &:& \xi^{5,7,8}_\text{L} \to -\xi^{5,7,8}_\text{L} \nonumber  \\
\Omega_4 &:& ~~ \xi^4_\text{L} ~ \to -\xi^4_\text{L}  ,
\label{dualitiespband}
\end{eqnarray}
while the right-moving Majorana fermions remain invariant. 
The four dualities (\ref{dualitiespband}), together with the trivial one 
$\Omega_0$, gives five possible $\mathrm{SO}(8)$-symmetric  rays  which attract the  one-loop RG  (\ref{RGN2}) flows in 
the far IR regime. Along  these rays, the interacting part of the effective Hamiltonian (\ref{Hamiltonian_tperp0_majoranapp}) simplifie as follows:
\begin{eqnarray}
\Omega_0 &:& \mathcal{H}^{\Omega_0}_{\text{int}} = \frac{g}{2}\left(\sum_{a=1}^8\xi^a_\text{R}\xi^a_\text{L}\right)^2 
\nonumber  \\
\Omega_1 &:& \mathcal{H}^{\Omega_1}_{\text{int}} = \frac{g}{2}\left(\sum_{a=4}^8\xi^a_\text{R}\xi^a_\text{L} 
- \sum_{a=1}^3\xi^a_\text{R}\xi^a_\text{L}\right)^2 \nonumber  \\
\Omega_2 &:& \mathcal{H}^{\Omega_2}_{\text{int}} 
= \frac{g}{2}\left(\sum_{a\neq6}\xi^a_\text{R}\xi^a_\text{L} - \xi^6_\text{R}\xi^6_\text{L}\right)^2 \nonumber  \\
\Omega_3 &:& \mathcal{H}^{\Omega_3}_{\text{int}} 
= \frac{g}{2}\left(\sum_{a\neq5,7,8}\xi^a_\text{R}\xi^a_\text{L} - 
\sum_{b=5,7,8}\xi^b_\text{R}\xi^b_\text{L}\right)^2 \nonumber  \\
\Omega_4 &:& \mathcal{H}^{\Omega_4}_{\text{int}} 
= \frac{g}{2}\left(\sum_{a\neq4}\xi^a_\text{R}\xi^a_\text{L} - \xi^4_\text{R}\xi^4_\text{L}\right)^2 .
\label{SO8raysapp}
\end{eqnarray}
with $g>0$. The nature of the underlying electronic phase can then be inferred by a straightforward semiclassical
approach on the bosonic representation of the different models in Eqs. (\ref{SO8raysapp}) by means 
of the identification (\ref{refer}). 
The following five different fully gapped Mott-insulating phases are found in this analysis.

\bigskip
\noindent ~ \textbf{Spin-Peierls phase:}
\vspace{0.25cm}

The trivial duality $\Omega_0$ correspond to the SO(8) GN model. As seen in Sec.~\ref{sec:Phases_with_dynamical_symmetry_enlargement}
in the general SO($4N$) case, the underlying Mott-insulating phase is a SP one with spontaneous dimerization.

\bigskip

\noindent ~ \textbf{Spin Haldane phase:}

\vspace{0.25cm}

\noindent For the first non-trivial duality symmetry $\Omega_1$, the semi-classical approach leads 
to a non-degenerate phase where the bosonic fields are pinned as follows:
\begin{equation}
\langle \Phi_{\text{s}} \rangle = \langle \Theta_{\text{so}} \rangle  = \frac{\sqrt{\pi}}{2} ~ ; ~~ 
\langle \Phi_{\text{c,o}} \rangle  = 0 ~~  \mbox{(SH phase)},
\label{HSsemi}
\end{equation}
where $\Phi_{a} = \Phi_{a\text{L}} +  \Phi_{a\text{R}}$ and  
$\Theta_{a} = \Phi_{a\text{L}} -  \Phi_{a\text{R}}$ 
($a =\text{c},\text{s},\text{o},\text{so}$) are respectively
the total bosonic field and the dual field. The field configurations (\ref{HSsemi}) correspond to the SH phase. 
\cite{Nonne-B-C-L-10}

\bigskip

\noindent ~ \textbf{Rung-Singlet phase:}

\vspace{0.25cm}

The duality symmetry $\Omega_2$ leads to a non-degenerate phase with field configurations:
\begin{equation}
\langle \Phi_{\text{c,s,o}} \rangle = \langle \Theta_{\text{so}} \rangle  = 0 ~~  \mbox{(RS phase)}.
\label{RSso}
\end{equation}
The physical picture of the corresponding phase is a singlet formed between the orbital and nuclear spins:
\begin{equation}
\ket{\mathrm{RS}} = \prod_{i}\frac{1}{\sqrt{2}}\big(c^\dagger_{p_x\uparrow,i}c^\dagger_{p_y\downarrow,i}-c^\dagger_{p_x\downarrow,i}c^\dagger_{p_y\uparrow,i}\big)\ket{0}.
\label{RSWF}
\end{equation}
Such phase is similar to the RS phase of the two-leg spin-1/2 ladder where a singlet is formed on 
each rung of the ladder. \cite{Gogolin-N-T-book}
Since $T^z_i \ket{\mathrm{RS}} = 0$, the RS phase can also be interpreted as an orbital large-$D$ (OLD) phase along
the $z$-axis.

\bigskip

\noindent ~ \textbf{Charge Haldane phase:}

\vspace{0.25cm}

For the duality symmetry $\Omega_3$, we obtain again a non-degenerate phase 
with the following pinning:
\begin{equation}
\langle \Phi_{\text{c}} \rangle = \langle \Theta_{\text{o}} \rangle = \frac{\sqrt{\pi}}{2} ~ ; ~~ 
\langle \Phi_{\text{s,so}} \rangle = 0 ~~  \mbox{(CH phase)}.
\end{equation}
Such field configurations signal the emergence of a Haldane phase for the charge degrees of 
freedom, which has been dubbed  charge Haldane  (CH) phase (or equivalently Haldane charge) in Refs.~\onlinecite{Nonne-L-C-R-B-10,Nonne-L-C-R-B-11}.
The spin degrees of freedom of this phase are described by the pseudo spin operator (\ref{chargepseudospin}),
which is a spin-singlet that carries charge. This CH phase is deduced from the usual SH phase by 
the Shiba transformation (\ref{eqn:Shiba-tr-SU2-2band}).
 
\bigskip

\noindent ~ \textbf{Orbital large-\texorpdfstring{$\boldsymbol{D}$}{D} phase:}

\vspace{0.25cm}

\noindent For the last duality  symmetry, \textit{i.e.} $\Omega_4$, 
the semi-classical approach gives the following vacuum expectation values:
\begin{equation}
\langle \Phi_{\text{c,s,so}} \rangle = \langle \Theta_{\text{o}} \rangle = 0 .
\end{equation}
The corresponding Mott insulating phase is non-degenerate and featureless. 
In the strong-coupling regime, a ground state for that phase is the singlet state:
\begin{equation}
\ket{\text{OLD}_x} = \prod_{i}\frac{1}{\sqrt{2}}\left(
c^\dagger_{p_x\uparrow,i}c^\dagger_{p_x\downarrow,i} 
-c^\dagger_{p_y\uparrow,i}c^\dagger_{p_y\downarrow,i}\right)\ket{0} ,
\label{oldx}
\end{equation}
which is characterized by $T^x_i \ket{\text{OLD}_x} = 0$. The resulting spin-singlet phase 
is an orbital large-$D$ (OLD) phase along the $x$-axis.   
We can also think of a similar state along the $y$-axis:
\begin{equation}
\ket{\text{OLD}_y} = \prod_{i} \frac{1}{\sqrt{2}}\left(  
c^{\dagger}_{p_x\uparrow,i}c^{\dagger}_{p_x\downarrow,i}
+ c^{\dagger}_{p_y\uparrow,i}c^{\dagger}_{p_y\downarrow,i}  \right) |0\rangle  \; .
\end{equation}
The latter is different from the RS phase (\ref{RSWF}) since (\ref{oldx}) (respectively (\ref{RSWF}))  
is  antisymmetric (respectively symmetric) under the $p_x \leftrightarrow p_y$ exchange.

\bigskip

\noindent ~ \textbf{Phase diagram in the weak coupling regime:}

\vspace{0.25cm}
%%%%%%%%%%%%%%%%%%%%%%%%%%%%%%%%%%%%%%%%%%%%%%%%%%%%%%%%%%
\begin{figure}[!ht]
\centering
\includegraphics[width=0.5\columnwidth,clip]{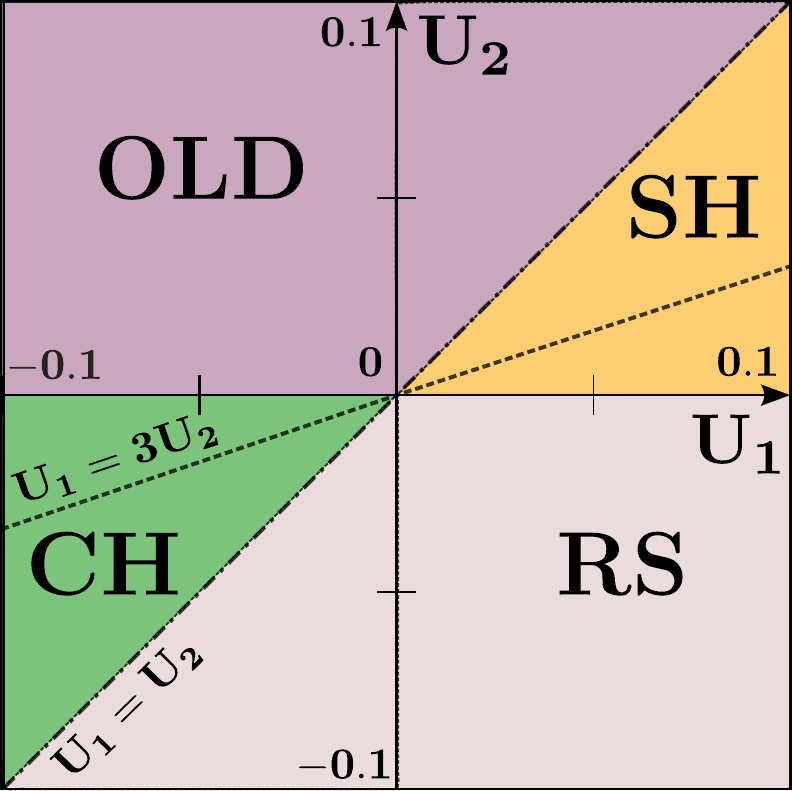}
\caption{(Color online) Phase diagram for $N=2$ $p$-band model \eqref{eqn:p-band-simple} obtained by solving numerically the one-loop RG Eqs (\ref{RGN2}) with initial conditions (\ref{majocouplingsapp}).
The line $U_1=3U_2$ corresponds to the axially symmetric trapping scheme. 
\label{fig:RG_phasediag_pband_N2}}
\end{figure}
%%%%%%%%%%%%%%%%%%%%%%%%%%%%%%%%%%%%%%%%%%%%%%%%%%%%%

Following the same procedure as described in Sec.~\ref{sec:RG-phase-diag}, we solve numerically the RG equations (\ref{RGN2}) with initial conditions (\ref{majocouplingsapp}) to obtain the low-energy phase diagram of the $N=2$ $p$-band model \eqref{eqn:p-band-simple} in the ($U_1, U_2$) plane.
We identify four out of the five regions discussed above. Indeed, the SP phase is not realized. 
These fours regions are readily identified as SH, RS, CH and OLD by the flows of the couplings 
$g_i(l_{\text{max}}) = \pm g_{\text{max}}$ that are in agreement with the symmetries \eqref{eq:symmRG2}.
The phase diagram in the low energy limit Fig.~\ref{fig:RG_phasediag_pband_N2} is equivalent with the one obtained with the DMRG technique in Fig.~\ref{fig:phasediag_pband_N2} (see discussion in Sec.~\ref{sec:DMRG_N2pband}). 

\bigskip

From the duality symmetries (\ref{dualitiespband}), we can, as well, discuss the nature of the quantum
phase transitions that occur in Fig.~\ref{fig:RG_phasediag_pband_N2} by investigating the self-dual manifolds. 

\bigskip

\noindent ~ \textbf{CH/RS or SH/OLD transition:}

\vspace{0.25cm}

The transition between the CH and RS phases, or between SH and OLD, is  governed by the self-dual manifold
of the duality $\Omega_2 \Omega_3$ where $\xi^{5,6,7,8}_\text{L} \to -\xi^{5,6,7,8}_\text{L}$. 
The self-dual manifold is then described by $g_3=g_4=g_6=g_7 =0$. From the initial conditions 
(\ref{majocouplingsapp}), we observed that the line $U_1 =U_2$ of the $p$-band model belongs to that manifold.
The interacting part of the effective Hamiltonian (\ref{Hamiltonian_tperp0_majoranapp}) simplifies as follows
along that line:
\begin{eqnarray}
\mathcal{H}^{\mathrm{CH/RS}}_{\tiny\mbox{int}} 
= \frac{g_1}{2}\left(\sum_{a=1}^4\xi^a_\text{R}\xi^a_\text{L} \right)^2
-  \frac{g_1}{2}\left(\sum_{a=5}^8\xi^a_\text{R}\xi^a_\text{L} \right)^2 ,
\label{transHCRS}
\end{eqnarray}
which takes the form of two decoupled SO(4) GN models. Due to the particular structure of model (\ref{transHCRS}),
one of this SO(4) GN displays a critical behavior while the other is massive. We thus conclude that 
the quantum phase transition CH/RS or SH/OLD belongs to the SO(4)$_1$ universality class with central charge $c=2$.

\bigskip

\noindent ~ \textbf{SH/RS or CH/OLD transition:}

\vspace{0.25cm}

One can repeat the analysis for  the transition between the SH and RS phases, or between CH and OLD.  In that case,
the relevant duality is $\Omega_1 \Omega_2$ with $\xi^{1,2,3,6}_L \to -\xi^{1,2,3,6}_L$. 
The resulting self-dual manifold is $g_2=g_4=g_6=g_8 =0$. From the initial conditions 
(\ref{majocouplingsapp}), we observed that the line $U_2 = 0$ of the $p$-band model belongs to that manifold.
The interacting part of the effective Hamiltonian (\ref{Hamiltonian_tperp0_majoranapp}) simplifies as follows
along that line:
\begin{equation}
\begin{split}
\mathcal{H}^{\mathrm{SH/RS}}_{\text{int}} 
=& \frac{g_1}{2}\left(\sum_{a=1,2,3,6}\xi^a_\text{R}\xi^a_\text{L} \right)^2  %\\& 
-  \frac{g_1}{2}\left(\sum_{a=4,5,7,8}\xi^a_\text{R}\xi^a_\text{L} \right)^2 ,
\end{split}
\label{transHSRS}
\end{equation}
which takes also the form of two decoupled SO(4) GN models with an emerging SO(4)$_1$ quantum criticality with
$c=2$. This last result can be easily understood since when $U_2 = 0$  the $p$-band model (\ref{eqn:p-band}) is equivalent
to two decoupled half-filled Hubbard chains and therefore a critical behavior with central charge $c=1+1=2$ occurs.

\bigskip

\noindent ~ \textbf{SH/CH or RS/OLD transition:}

\vspace{0.25cm}

In this last case, the quantum phase transition is described by the duality $\Omega_1 \Omega_3$ 
with $\xi^{1,2,3,5,7,8}_\text{L} \to -\xi^{1,2,3,5,7,8}_\text{L}$. The self-dual manifold is $g_2 = g_3 = g_7 = g_8 =0$.
Using the initial conditions of the $p$-band model (\ref{majocouplingsapp}), the non-interacting point belongs to that
manifold and we expect thus that the SH/CH and RS/OLD transitions occur for $U_1 =U_2 =0$.

%%%%%%%%% BIB-files %%%%%%%%%%%%%%%%%%%%%%%%%%%%%%%%%%%%%%%%
\bibliographystyle{apsrev4-1}
\end{document}